\documentclass[classic, hyperref,watermark,histinit,frontispiece,plain]{teipel-thesis-en} 

\graphicspath{{figures/}}
\usepackage[scientific-notation=true]{siunitx}
\usepackage[english]{babel}
\usepackage{graphicx, color}
\graphicspath{{figures/}}
\usepackage{amsmath}
\usepackage{amscd}
\usepackage{youngtab}
\usepackage{titlesec}
\usepackage{listings}
\usepackage{enumerate}
\usepackage{amssymb}
\usepackage{amsthm}
\usepackage{syntonly}
\usepackage{fancyhdr}
\usepackage{cancel}
\usepackage{siunitx}
\usepackage{slashed}
\usepackage{bbm}
\usepackage{dsfont}
 \usepackage{comment} 
\usepackage{multirow}
\usepackage{gensymb}
\usepackage{cite} 
\usepackage{fancyhdr}
\usepackage{dsfont}
\usepackage{float}

\usepackage{fancyhdr}
	\UniversityName{Universidad del Valle}
	\SchoolName{Departamento de Física}
	\DepartmentName{Facultad de Ciencias Naturales y Exactas }
	\UniversityLocation{Cali}
	\ThesisType{A thesis submitted in partial fulfillment of the requirements for the degree of Master of Science in Physics}
	\title{Cosmology on the Generalized Proca Theory}
	\authorName{Santiago García Serna} 
	\supervisor{Prof. César Alonso Valenzuela Toledo.}
	\cosupervisor{Dr. John Bayron Orjuela Quintana.}
	\supervisorTitle{}
	\thesisPlaceDate{Cali, 2025}
	\examinationDate{ Date:}
	\approvalStatement{Approved by \hspace{2.0 cm}}
	\firstExaminer{\hspace{5.0 cm}}
	\firstExaminerTitle{}
	\secondExaminer{\hspace{5.0 cm}}
	\secondExaminerTitle{}
	\chaptercolor{brown!75!magenta}
	\appendixcolor{brown!80!magenta}
    	\hyperlinkcolor{red!55!brown}
    	\titlecolor{white}
    	\titlebackgroundcolor{brown!85!magenta}  
     \coverpageimage{0.3}{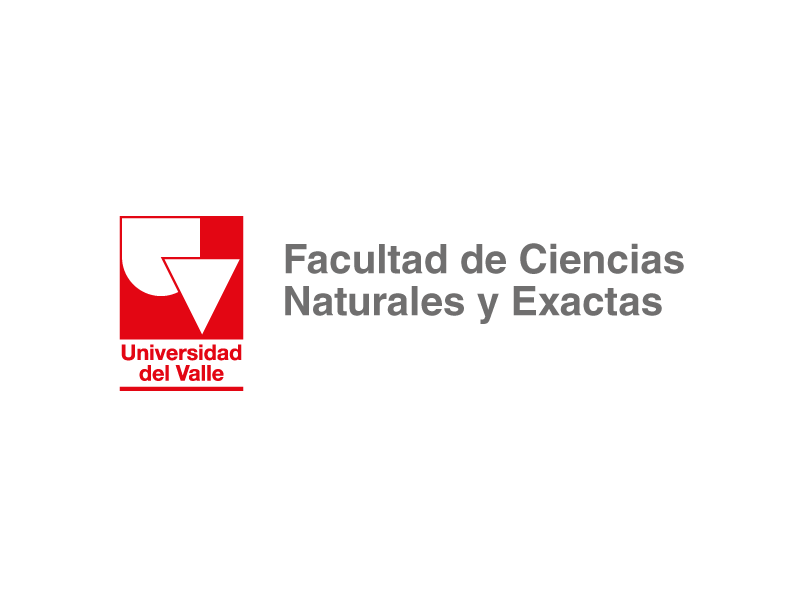}
     \vspace{2.0 cm}
	\UniversityLogoPath{logoUV}
	\UniversityBWLogoPath{logoUV}
	\LogoScaleFactor{1}
	\draftIndication{}
	\setFrontispiece[0.4]{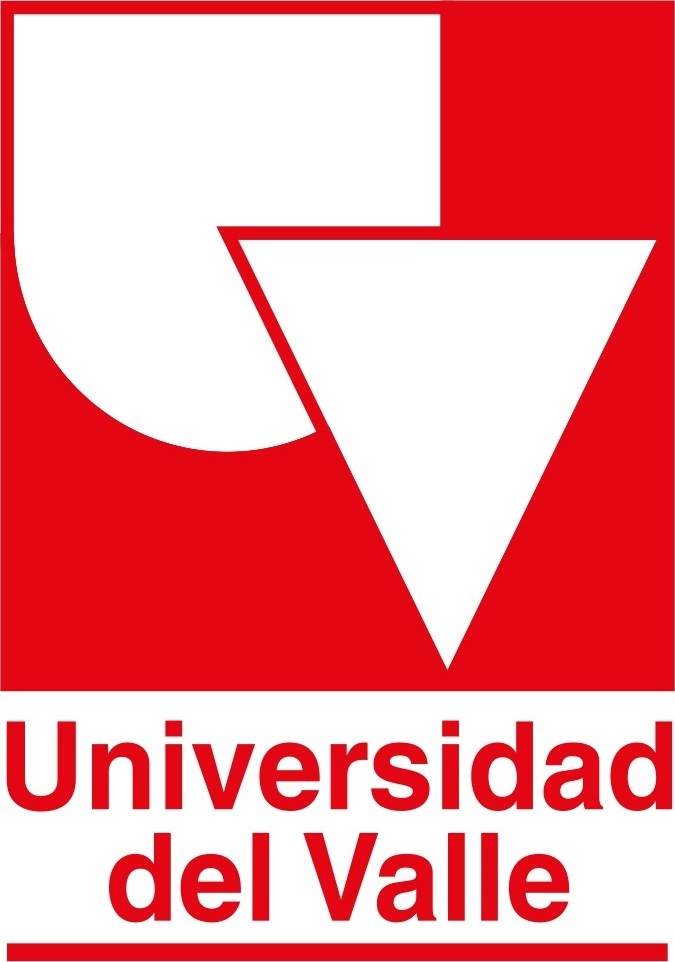}{} 
\begin{document}

\maketitle

\beginfrontmatter
	
	\begin{abstract}

In this thesis, we employ the dynamical system approach to investigate two cosmological models: the first one based on a dark energy scenario in an anisotropic Bianchi-I background, and the other exploring the viability of dark energy or inflation in an isotropic background. Both models address key challenges in understanding the expansion history of the Universe. In the first part, the complexity introduced by shear prevents the analytical identification of certain fixed points in the model. To address this, we develop a numerical framework to study anisotropic dark energy scenarios involving the interaction between a scalar tachyon field and a vector field in a Bianchi-I background. This approach overcomes the limitations of analytical fixed-point methods, allowing us to identify parameter regions that support anisotropic accelerated attractor solutions and providing a versatile tool for exploring dark energy models.\\

The second part focuses on the Generalized SU(2) Proca theory, an extension of vector-tensor theories which has recently garnered attention for its potential to describe key phases of cosmic evolution, including primordial inflation and late-time accelerated expansion. However, its full cosmological implications remain unexplored. In this part, we perform a comprehensive analysis of the dynamical properties of the GSU2P theory in a flat, homogeneous, and isotropic spacetime, through a dynamical-system approach. Our analysis reveals the presence of three pairs of fixed points, one of them corresponding to de-Sitter expansion which may represent either a stable or unstable phase in the evolution of the universe. These points, nonetheless, give rise to an indeterminate or infinite Hubble parameter, which renders them cosmologically unviable. Additionally, we find two key pseudostationary states: the attractor lines, along which the system exhibits constant-roll dynamics, and the central zone, characterized by oscillatory radiation-like behaviour of the field. The dynamics within the central zone could represent a graceful exit from the primordial inflationary phase to a radiation dominated phase, or a state of the dark energy component prior to the late-time cosmic acceleration. However, within the central zone, the dynamics of the vector field leads to recurrent instances of a nonphysical expansion rate. The absence of a limit cycle in the central zone further exacerbates the issue, as the system may follow unbounded phase-space trajectories, and the expansion rate becomes complex once it escapes the region. Collectively, these challenges undermine the viability of the GSU2P theory as a cosmological model for cosmic acceleration.\\

Overall, this work employs dynamical systems techniques to assess the viability of alternative cosmological models. Our numerical framework circumvents analytical limitations in anisotropic dark energy scenarios, identifying parameter regions that allow for accelerated expansion. For the Generalized SU(2) Proca theory, we demonstrate that despite the presence of pseudostationary structures suggesting possible cosmological transitions, the model ultimately fails due to recurrent nonphysical expansion rates and unstable trajectories.



\end{abstract}

\begin{acknowledgements}
I would like to express my deepest gratitude to my family — my father Santiago García Restrepo, my mother Maria Teresa Serna Díaz, and my sister Maria Camila García Serna — for their unwavering support, patience, and encouragement throughout these intense years of academic training. Their readiness to assist me and provide everything I needed has been invaluable.

I would like to extend a special thank you to J. Bayron Orjuela-Quintana for his invaluable support and insightful discussions throughout the research process. His contributions were instrumental to the development of this project. I also wish to express my sincere gratitude to my professors, especially Prof. César Alonso Valenzuela, for his unwavering academic and professional guidance, and to Prof. Yeinzon Rodriguez for his continuous support and encouragement.

Finally, I am deeply grateful to Universidad del Valle and the Department of Physics for their financial support over the past two years. Their assistance made it possible for me to pursue my post-graduate studies.

\end{acknowledgements}
	\tableofcontents
	
\beginmainmatter

        

\chapter{Introduction to Modern Cosmology\label{Ch: ModernCosmology}}
\InitialCharacter{E}instein's General Relativity (GR) revolutionized our comprehension of the Universe, replacing centuries of philosophical speculation with a mathematically rigorous framework for describing the dynamics of spacetime. This groundbreaking theory redefined gravity not as a force acting at a distance, but as the geometric manifestation of spacetime curvature induced by the presence of matter and energy. With this profound conceptual shift, General Relativity laid the cornerstone for modern cosmology. The Einstein field equations encode the fundamental interplay between the geometry of spacetime and the distribution of matter-energy within the Universe, expressed as:

\begin{equation}
    R_{\mu\nu} - \frac{1}{2} R g_{\mu\nu} + \Lambda g_{\mu\nu} = 8\pi G T_{\mu\nu},
    \label{ec:FieldEquations}
\end{equation}

where $R_{\mu\nu}$ represents the Ricci tensor, $R$ is the Ricci scalar, $g_{\mu\nu}$ denotes the metric tensor, $T_{\mu\nu}$ is the energy-momentum tensor, $\Lambda$ is the cosmological constant, and $m_{\text{P}}$ is the reduced Planck mass. Throughout this work, we adopt natural units where $c = \hbar = k_B = 1$.

For the first time in history, the most fundamental cosmological inquiries could be addressed through the rigorous lens of physical law: What are the origins of the light elements? Why does the Universe exhibit such remarkable homogeneity and isotropy on large scales? What mechanisms were responsible for generating the primordial density fluctuations that subsequently evolved into galaxies and the intricate cosmic web? These profound questions, once confined to the realm of metaphysical speculation, have found compelling, albeit partial, answers through the synergistic interplay between theoretical physics and increasingly precise astronomical observations.

The modern cosmological framework is firmly rooted in the paradigm of an expanding Universe, a picture robustly supported by a wealth of empirical evidence~\cite{Perlmutter_1999, SupernovaSearchTeam:1998fmf,Pan-STARRS1:2017jku}. This model traces the history of the cosmos back to an exceedingly hot and dense primordial state. The subsequent evolution of the Universe is described with remarkable accuracy by the synthesis of General Relativity and the Standard Model of particle physics, augmented by three crucial components: cold dark matter, invoked to explain the formation of large-scale structures; dark energy, introduced to account for the observed accelerated expansion at late times; and an early inflationary epoch, postulated to resolve the horizon and flatness problems and to provide a mechanism for the generation of primordial perturbations.

\section{Measuring Distances in the Universe}

\begin{figure}[ht!] 
    \centering
    \includegraphics[width=0.7\linewidth]{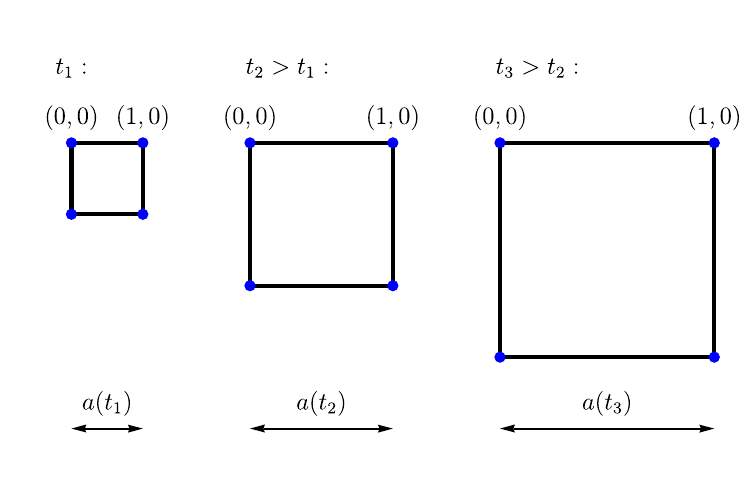} 
    \caption{Schematic representation of cosmic expansion. Two points with fixed comoving coordinates maintain a constant comoving separation, while their physical distance grows proportionally to $a(t)$, illustrating the metric expansion of spacetime.}
    \label{fig:ExpansionEsquema}
\end{figure}
\subsection{Scale Factor}
\label{subsec:ScaleFactor_v2} 

The current Universe is characterized by an accelerated expansion, a phenomenon robustly confirmed through observations of Type Ia supernovae~\cite{Perlmutter_1999, SupernovaSearchTeam:1998fmf, Pan-STARRS1:2017jku}. This expansion implies that the average physical separation between gravitationally unbound structures, such as galaxies or galaxy clusters, increases over cosmic time. To describe this large-scale dynamic without being tied to the stretching fabric of spacetime itself, cosmologists employ \emph{comoving coordinates}, $\mathbf{x}$. This system factors out the universal expansion by introducing a dimensionless, time-dependent \emph{scale factor}, $a(t)$. Within this framework, the physical position $\mathbf{r}(t)$ of an object is related to its fixed comoving coordinate $\mathbf{x}$ by:
\begin{equation}
    \mathbf{r}(t) = a(t) \mathbf{x}.
\end{equation}
By convention, the scale factor is normalized such that $a(t_0) = 1$ at the present cosmic time $t_0$. Figure~\ref{fig:ExpansionEsquema} provides a schematic illustration: as the Universe evolves, commoving coordinates (grid points) remain fixed, while the physical distances between them scale proportionally with $a(t)$, reflecting the metric expansion.

The kinematics of cosmic expansion can obtain examining the time derivative of the physical position $\mathbf{r}(t)$. Defining the Hubble parameter $H(t) \equiv \dot{a}(t)/a(t)$ as the rate of expansion, the total velocity of a particle relative to an observer at the origin is given by:
\begin{align}
    \dot{\mathbf{r}} =\dot{a}\mathbf{x}+a\dot{x}= H(t)\mathbf{r} + \mathbf{v}_p.
    \label{eq:velocity_decomposition_concise}
\end{align}
This expression decomposes the total velocity into two components:
\begin{enumerate} 
    \item The \emph{Hubble flow}, $H(t)\mathbf{r}$, which represents the velocity due to the overall expansion of space.
    \item The \emph{peculiar velocity}, $\mathbf{v}_p \equiv a(t)\dot{\mathbf{x}}$, which describes the motion of the object \emph{relative} to the comoving grid, typically driven by local gravitational interactions.
\end{enumerate}
\begin{figure}[t!] 
    \centering
    \includegraphics[width=0.7\linewidth]{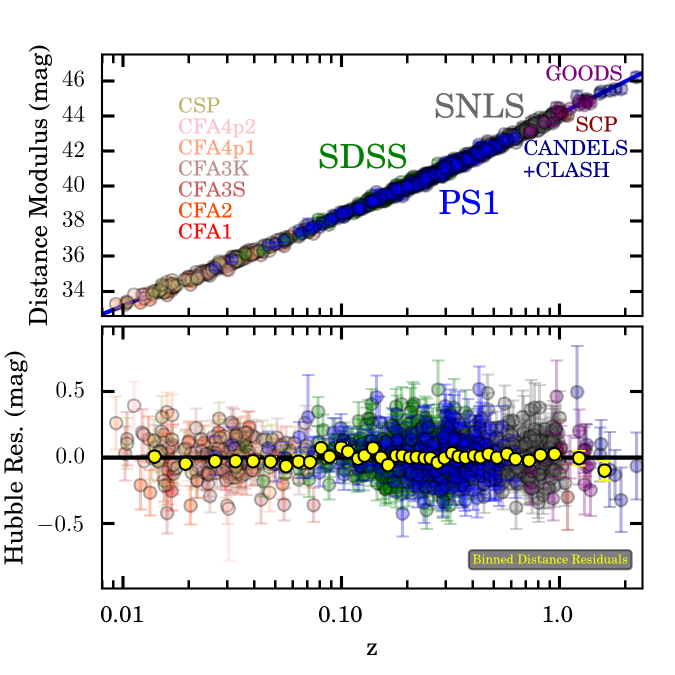} 
    \caption{A modern Hubble diagram illustrating the linear relationship between the recession velocities (often inferred from redshift) and distances of galaxies. The slope of the best-fit line corresponds to the Hubble constant, $H_0$. Data shown are representative of modern measurements, such as those from~\cite{Pan-STARRS1:2017jku}, reinforcing this fundamental cosmological principle.}
    \label{fig:HubbleDiagram}
\end{figure}

On sufficiently large scales, the effects of local structures average out, making peculiar velocities statistically negligible ($\langle \mathbf{v}_p \rangle \approx 0$). In this regime, the motions of distant objects are dominated by the homogeneous Hubble flow. This theoretical picture directly leads to the empirically observed Hubble-Lemaître law:
\begin{equation}
    v = H_0 r,
\end{equation}
where $v$ is the recession velocity of a distant galaxy, $r$ is its proper distance, and $H_0 = H(t_0)$ is the Hubble constant, the value of the Hubble parameter today~\cite{Hubble168}. The linearity of this relationship is confirmed by redshift-distance measurements like those shown in Figure~\ref{fig:HubbleDiagram}.

\subsection{Redshift as a Consequence of Expansion}

While the scale factor $a(t)$ provides the theoretical framework for cosmic expansion, its most direct and crucial observable consequence is the effect it has on electromagnetic radiation traveling across cosmological distances. As photons propagate through the expanding Universe from a distant source to an observer, the stretching of spacetime itself causes their wavelengths to increase. This phenomenon is known as cosmological redshift~\cite{dodelson2020modern}.

Consider light emitted by a source at cosmic time $t_\text{em}$ with a wavelength $\lambda_\text{em}$. As this light travels towards us, the Universe continues to expand. By the time it is detected at the present time $t_0$, its wavelength will have been stretched to $\lambda_\text{obs}$. The amount of stretching is directly proportional to the change in the scale factor during the light's journey:
\begin{equation}
    \frac{\lambda_\text{obs}}{\lambda_\text{em}} = \frac{a(t_0)}{a(t_\text{em})}.
\end{equation}
Cosmological redshift, denoted by $z$, is defined as the fractional change in wavelength:
\begin{equation}
    z \equiv \frac{\lambda_\text{obs} - \lambda_\text{em}}{\lambda_\text{em}} = \frac{\lambda_\text{obs}}{\lambda_\text{em}} - 1.
\end{equation}
Combining these equations and using the convention $a(t_0) = 1$, we arrive at the fundamental relationship between redshift and the scale factor at the time of emission:
\begin{equation}
    1+z = \frac{a(t_0)}{a(t_\text{em})} = \frac{1}{a(t_\text{em})}.
    \label{eq:redshift_scale_factor}
\end{equation}
Therefore, measuring the redshift $z$ of a distant object provides a direct measurement of the scale factor of the Universe when the light was emitted ($a(t_\text{em}) = 1/(1+z)$). Since $\lambda_\text{obs} > \lambda_\text{em}$ for an expanding universe ($a(t_0) > a(t_\text{em})$), $z$ is positive, indicating a shift towards longer (redder) wavelengths—hence the term ``redshift''.

It is important to distinguish this cosmological redshift, caused by the expansion of space itself, from the Doppler redshift caused by the peculiar velocity of the source relative to the observer. While both effects can alter observed wavelengths, for distant objects ($z \gtrsim 0.1$), the cosmological component typically dominates and serves as the primary indicator of distance (via the Hubble-Lemaître law) and cosmic epoch~\cite{dodelson2020modern,liddle2015introduction}. Redshift measurements are thus fundamental not only for mapping the large-scale structure of the Universe but also for probing its expansion history and composition.

\subsection{The Geometric Framework: Isotropy and the FLRW Metric}
\label{sec:FLRW}

The scale factor $a(t)$ and cosmological redshift $z$, discussed previously, describe the dynamics of cosmic expansion. However, a complete cosmological model requires specifying the geometric stage upon which this expansion unfolds. The standard model of cosmology is built upon the \emph{Cosmological Principle}, a foundational assumption postulating that the Universe is statistically homogeneous and isotropic on sufficiently large scales. This principle is strongly supported by observations, particularly of the Cosmic Microwave Background (CMB), which reveals a remarkably uniform temperature across the sky, with anisotropies only at the level of $\Delta T / T \sim 10^{-5}$~\cite{Planck:2018vyg,Planck:2018jri,COBE:1992syq,WMAP:2003elm}.

Symmetries, like those encapsulated in the Cosmological Principle, are useful in physics; they simplify descriptions and often reveal fundamental properties. For a spacetime exhibiting spatial homogeneity and isotropy, the most general line element describing its geometry is the Friedman-Lemaître-Robertson-Walker (FLRW) metric. In the limit of no gravity or expansion ($a(t)=1, k=0$), it reduces to the Minkowski metric $g_{\mu\nu} = \text{diag}(-1,1,1,1)$ of special relativity.

\begin{figure}[t!]
    \centering
    \includegraphics[width=0.9\textwidth]{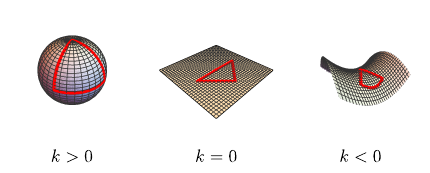} 
    \caption{Diagram illustrating the three possible spatial geometries in the FLRW model, determined by the curvature parameter $k$. Left ($k > 0$): A closed, spherical geometry where parallel lines eventually converge. Center ($k = 0$): A flat, Euclidean geometry where parallel lines remain parallel. Right ($k < 0$): An open, hyperbolic geometry where parallel lines diverge.}
    \label{fig:curvedgeometry}
\end{figure}

The spatial part of the FLRW metric can be derived by considering a 3-dimensional hyperspace embedded in a 4-dimensional Euclidean or pseudo-Euclidean space. Depending on whether the embedding space has a positive, zero, or negative curvature signature ($\sigma = +1, 0, -1$, respectively), the geometry of the 3D slice changes. Let $R$ be the characteristic radius of curvature (infinite for $\sigma=0$). A point in the embedding space satisfies:
\begin{align}
    \sigma R^2 &= x^2 + y^2 + z^2 + w^2, \\
    w^2 &= \sigma R^2 - r^2, \\
    dw &= -\frac{r \, dr}{w} = -\frac{r \, dr}{\sqrt{\sigma R^2 - r^2}}.
\end{align}
The line element $d\Sigma^2$ in the 3D spatial slice is inherited from the 4D embedding space's metric $d\Sigma^2 = dx^2+dy^2+dz^2+dw^2$. Using spherical coordinates for $x, y, z$ ($dr^2 + r^2 d\Omega^2 = dx^2+dy^2+dz^2$), we get:
\begin{align}
    d\Sigma^2 &= dr^2 + r^2 d\Omega^2 + dw^2 \\
             &= dr^2 + r^2 d\Omega^2 + \frac{r^2 \, dr^2}{\sigma R^2 - r^2} \\
             &= \frac{\sigma R^2}{\sigma R^2 - r^2} dr^2 + r^2 d\Omega^2.
\end{align}
Defining the curvature parameter $k = \sigma / R^2$ where $k$ can take values $\{k<0, k=0, k>0\}$, we can write:
\begin{equation}
    d\Sigma^2 = \frac{dr^2}{1 - k r^2} + r^2 d\Omega^2.
\end{equation}
Incorporating the time dimension and the expansion via the scale factor $a(t)$, we arrive at the full FLRW line element:
\begin{equation}
    ds^2 = -dt^2 + a^2(t) \left[ \frac{dr^2}{1 - k r^2} + r^2 \left( d\theta^2 + \sin^2 \theta \, d\phi^2 \right) \right],
    \label{eq:FLRWMetric}
\end{equation}
where $t$ is the cosmic time, $r, \theta, \phi$ are comoving spherical coordinates and $d\Omega^2 = d\theta^2 + \sin^2 \theta \, d\phi^2$. The parameter $k$ dictates the global geometry of the spatial sections:
\begin{itemize}
    \item $k > 0$: Positive curvature, corresponding to a closed, finite Universe (like a 3-sphere).
    \item $k = 0$: Zero curvature, corresponding to a flat, infinite Euclidean Universe.
    \item $k < 0$: Negative curvature, corresponding to an open, infinite hyperbolic Universe.
\end{itemize}
These possibilities are illustrated schematically in Figure~\ref{fig:curvedgeometry}. Current observations, primarily from the CMB, strongly favour a spatially flat Universe ($k \approx 0$)~\cite{Planck:2018vyg,WMAP:2003elm}.

\subsection{The Cosmic Fluid: Energy-Momentum Tensor and Equations of State}
\label{sec:EnergyMomentum}

Having established the standard geometric framework for cosmology – the FLRW metric (\ref{eq:FLRWMetric}) based on the Cosmological Principle – we now turn to the source term in Einstein's field equations: the energy and matter content of the Universe. This content dictates the expansion dynamics $a(t)$ within the given geometry. The distribution of energy and momentum is described by the \emph{energy-momentum tensor}, $T_{\mu\nu}$.

In its most general form for any continuous medium or field configuration, $T_{\mu\nu}$ is a symmetric rank-2 tensor whose components represent densities and fluxes of energy and momentum. In a local inertial frame, its components have direct physical interpretations:
\begin{equation}
T_{\alpha \beta} = \left[
\begin{array}{cccc}
 \color{red}T_{00} & \color{blue}T_{01} & \color{blue}T_{02} & \color{blue}T_{03} \\
 \color{blue}T_{10} & \color{black}T_{11} & \color{orange}T_{12} & \color{orange}T_{13} \\
 \color{blue}T_{20} & \color{orange}T_{21} & \color{black}T_{22} & \color{orange}T_{23} \\
 \color{blue}T_{30} & \color{orange}T_{31} & \color{orange}T_{32} & \color{black}T_{33}
\end{array}
\right]
\label{eq:TmunuComponents}
\end{equation}
Specifically:
\begin{itemize}
    \item $\color{red}T_{00}$ is the energy density.
    \item $\color{blue}T_{0i} = \color{blue}T_{i0}$ is the energy flux across the $x^i$ surface, equivalent to the momentum density in the $i$-th component.
    \item $\color{black}T_{ii}$  represents normal stress in the $i$-th coordinate direction, which is called ``pressure'' when it is the same in every direction, $x^i$.
    \item $\color{orange}T_{ij}$ ($i \neq j$) represents the shear stress, or the flux of the $i$-th component of momentum across the $x^j$ surface.
\end{itemize}
For a closed system like the Universe as a whole, the energy-momentum tensor is conserved, satisfying the covariant conservation equation $\nabla_{\mu} T^{\mu\nu} = 0$. This equation leads to the fundamental laws governing the evolution of energy and momentum density in an expanding universe.

\subsubsection{The Perfect Fluid Approximation} \label{Section:PerfectFluid}

In cosmology, the large-scale distribution of matter and radiation is remarkably well-approximated by a \emph{perfect fluid}. A perfect fluid is characterized by being isotropic in its rest frame, meaning it has no viscosity (shear stress) and no heat conduction. Its energy-momentum tensor takes a simpler form, completely determined by its energy density $\rho$ and isotropic pressure $p$:
\begin{equation}
    T^{\mu\nu} = (\rho + p) u^{\mu} u^{\nu} + p g^{\mu\nu},
    \label{eq:PerfectFluidTmunu}
\end{equation}
where $u^{\mu}$ is the fluid's four-velocity ($u^{\mu} u_{\mu} = -1$), which in the comoving frame of the FLRW metric is simply $u^{\mu} = (1, 0, 0, 0)$, and $g^{\mu\nu}$ is the metric tensor (from Eq.~\eqref{eq:FLRWMetric}). In this frame, $T^{00} = \rho$ and $T^{i}_j = p \, \delta^{i}_j$.

The thermodynamic properties of the cosmic fluid components are often summarized by their \emph{equation of state}, which relates pressure to energy density, typically expressed as a constant parameter $w$:
\begin{equation}
    w \equiv \frac{p}{\rho}.
    \label{eq:EquationOfState}
\end{equation}
The primary components making up the cosmic inventory within the standard $\Lambda$CDM model are modeled as perfect fluids with distinct equations of state:
\begin{itemize}
    \item \textbf{Non-relativistic Matter:} This includes baryonic matter and cold dark matter (CDM). These components have negligible random velocities compared to their rest mass energy, exerting negligible pressure. Thus, they are characterized by $w_m = 0$.
    \item \textbf{Radiation:} This includes photons (like the CMB) and relativistic neutrinos. These particles move at or near the speed of light and exert significant pressure. Relativistic kinematic theory shows $p_r = \rho_r/3$, so they are characterized by $w_r = 1/3$. (See e.g.,~\cite{dodelson2020modern, liddle2015introduction}).
    \item \textbf{Dark Energy/Cosmological Constant:} The component driving the observed accelerated expansion behaves like a fluid with negative pressure. A cosmological constant $\Lambda$ corresponds to a fluid with $w_{\Lambda} = -1$. More general dark energy models allow for different, possibly time-varying, $w$.
\end{itemize}
Understanding these components and their equations of state is essential for applying Einstein's equations to determine the expansion history of the Universe.

\section{The Standard Cosmological Model: $\Lambda$CDM Dynamics} 
\label{sec:LCDM_Dynamics_v2} 

The standard cosmological model, commonly known as the $\Lambda$CDM model, rests upon the cosmological principle. Consequently, the geometry of spacetime is described by the FLRW line element, Eq.~\eqref{eq:FLRWMetric}, and the governing dynamics at cosmological scales are dictated by Einstein's general relativity, expressed in Eq.~\eqref{ec:FieldEquations}. By considering the cosmic medium as a perfect fluid in its rest frame ($u^i = 0$), the Einstein field equations yield two fundamental dynamical relations, often referred to as the Friedmann equations:

\begin{align}
 3m_{\text{P}}^2 H^2 &= \rho_{\text{total}} - 3m_{\text{P}}^2 \frac{k}{a^2}, \label{ec:PrimeraFLRW}\\
 -2m_{\text{P}}^2\dot{H} - 3m_{\text{P}}^2H^2 &= p + m_{\text{P}}^2\frac{k}{a^2} - m_{\text{P}}^2\Lambda. \label{ec:SegundaFRLW} 
\end{align}
In the first Friedmann equation \eqref{ec:PrimeraFLRW}, the left-hand side represents the geometric aspect, while the right-hand side incorporates the total energy of the sources (where $\rho_{\text{total}} = \rho_m+\rho_r + m_{\text{P}}^2 \Lambda$), and the spatial curvature term proportional to $k/a^2$. The primary unknown functions in Eqs.~\eqref{ec:PrimeraFLRW} and \eqref{ec:SegundaFRLW} are the scale factor $a(t)$ (via $H$ and $\dot{H}$), the density $\rho(t)$, and the pressure $p(t)$. The pressure $p(t)$ is related to the density via the equation of state, $p = w\rho$ (Eq.~\eqref{eq:EquationOfState}). Furthermore, the requirement of energy-momentum conservation, $\nabla_{\mu} T^{\mu\nu} = 0$, provides the fluid continuity equation, which governs the evolution of the density:
\begin{equation}
 \dot{\rho} + 3H(\rho + p) = 0.
 \label{ec:EcuacionDeContinuidad}
\end{equation}

For analyzing the cosmic dynamics, it is often convenient to introduce dimensionless density parameters by normalizing the energy densities by the factor $3m_{\text{P}}^2 H^2$. We define the density parameters for the fluid component $\rho$, the cosmological constant $\Lambda$, and the spatial curvature $k$ as:
\begin{equation}
 \Omega = \frac{\rho}{3m_{\text{P}}^2 H^2}, \quad \Omega_{\Lambda} = \frac{\Lambda}{3H^2} \quad\text{ and } \quad\Omega_k = -\frac{k}{a^2H^2},
 \label{ec:DynamicalVariablesLCDM}
\end{equation}
respectively\footnote{Note that $\Omega$ here refers specifically to the matter/radiation fluid density parameter, distinct from $\Omega_\Lambda$.}. With these definitions, the first Friedman equation \eqref{ec:PrimeraFLRW} simplifies to the well-known constraint equation:
\begin{equation}
 1 = \Omega + \Omega_{\Lambda} + \Omega_k.
 \label{ec:FriedmannequationLambda}
\end{equation}

The second Friedman equation \eqref{ec:SegundaFRLW} can be related to the cosmic acceleration via the deceleration parameter $q \equiv -a\ddot{a}/\dot{a}^2 = -(1 + \dot{H}/H^2)$. Expressed in terms of the density parameters, it yields:

\begin{equation}
 q =\frac{1}{2}\sum_i\Omega_i(1+3w_i) =\frac{1}{2} \Omega_{\text{eff}}(1 + 3w_{\text{eff}}),
 \label{ec:FactorDeDesaceleracion}
\end{equation}
where the effective density parameter is $\Omega_{\text{eff}} = \Omega + \Omega_\Lambda$ and the effective equation of state $w_{\text{eff}}$ satisfies $p_{\text{total}} = w_{\text{eff}} \rho_{\text{total}}$. From either perspective, accelerated expansion ($q < 0$) requires a significant contribution from a component with negative pressure. For a single dominant fluid with equation of state $w$, acceleration occurs when $w < -1/3$. Any fluid capable of generating this effect ($q<0$) is broadly categorized as dark energy (DE).

\subsection{Continuity and Energy Decay Equations for Different Cosmic Fluids}
\label{Ch:EnergyContent} 

The continuity equation \eqref{ec:EcuacionDeContinuidad} dictates the evolution of the energy density $\rho_i$ for any component with a specific equation of state $w_i = p_i/\rho_i$. Let us analyze the behavior of the key components within $\Lambda$CDM as a function of the scale factor $a$, assuming a spatially flat ($k=0$) universe for simplicity in deriving the $a(t)$ dependencies.

\subsubsection{Relativistic Components }
Relativistic components, such as photons and neutrinos, are characterized by $w_r = 1/3$. The continuity equation \eqref{ec:EcuacionDeContinuidad} becomes:
\begin{equation}
 \dot{\rho}_r + 3H(1 + 1/3)\rho_r = \dot{\rho}_r + 4 H \rho_r = 0.
 \label{ec:RadiationMotionEquation}
\end{equation}
Integrating this equation yields the solution:
\begin{equation}
 \rho_r \propto a^{-4}.
\end{equation}
Thus, the energy density of relativistic components decays rapidly as $\propto a^{-4}$ during cosmic expansion. If radiation dominates the energy budget ($\rho_{\text{total}} \approx \rho_r$), the first Friedman equation \eqref{ec:PrimeraFLRW} implies $H^2 \propto a^{-4}$, leading to $a(t) \propto t^{1/2}$.

\subsubsection{Non-Relativistic Components (Matter)}
Non-relativistic components (baryons and cold dark matter) exert negligible pressure, so $w_m = 0$. The continuity equation then reads:
\begin{equation}
 \dot{\rho}_m + 3H(1 + 0)\rho_m = \dot{\rho}_m + 3 H \rho_m = 0.
 \label{ec:MatterMotionEquation}
\end{equation}
The solution is:
\begin{equation}
 \rho_m \propto a^{-3}.
\end{equation}
This indicates that the energy density of non-relativistic matter dilutes proportionally to the volume ($a^3$) as the Universe expands. If matter dominates, Eq.~\eqref{ec:PrimeraFLRW} implies $H^2 \propto a^{-3}$, resulting in $a(t) \propto t^{2/3}$.

\subsubsection{Dark Energy with a Constant Equation of State $w$ ($w$CDM)}
Dark energy broadly refers to any component capable of driving cosmic acceleration ($q<0$), requiring $w < -1/3$. The simplest model beyond the cosmological constant ($\Lambda$, where $w=-1$) is one where dark energy has a constant equation of state parameter $w$. For such a component, the continuity equation is:
\begin{equation}
 \dot{\rho}_{\text{DE}} + 3H(1 + w) \rho_{\text{DE}} = 0.
\end{equation}
The solution is:
\begin{equation}
 \rho_{\text{DE}} \propto a^{-3(1 + w)}.
\end{equation}
This shows that the energy density evolution depends critically on $w$:
\begin{itemize}
    \item If $w > -1$, $\rho_{\text{DE}}$ decreases as the Universe expands (though slower than matter if $w<0$).
    \item If $w = -1$ (the $\Lambda$CDM case), $\rho_{\text{DE}} = \rho_\Lambda$ remains constant.
    \item If $w < -1$ (phantom energy), $\rho_{\text{DE}}$ increases as the Universe expands.
\end{itemize}
Consequently, if a dark energy component with $w < 0$ exists, it dilutes slower than matter and radiation and is expected to eventually dominate the energy content of the Universe\footnote{Henceforth, the terms ``radiation'' and ``relativistic matter'' will be used interchangeably, as well ``non-relativistic matter'' and ``matter''.}. Furthermore, solving the first Friedman equation \eqref{ec:PrimeraFLRW} under the dominance of such a component yields specific expansion laws: for $w = -1$, $a(t) \propto e^{Ht}$ where $H = \sqrt{\Lambda/3}$; for $w \neq -1$, $a(t) \propto t^{2/(3(1 + w))}$~\cite{liddle2015introduction}.

\section{Challenges for the Standard Cosmological Model}
\label{sec:LCDM_Problems_Intro}

Despite the remarkable success of the $\Lambda$CDM model in describing a wide array of cosmological observations, it faces several significant challenges. These include conceptual fine-tuning issues related to the initial conditions of the Universe, as well as emerging tensions between different observational datasets~\cite{DiValentino:2020vnx, Weinberg:2013agg, Planck:2018vyg, Riess:2021jrx,
DES:2017myr, Kazantzidis:2018rnb, Benevento:2020fev, Skara:2019usd,Planck:2019evm, Schwarz:2015cma,Bowman:2018yin, Perivolaropoulos:2021jda, Banerjee:2020bjq}. Such discrepancies hint that $\Lambda$CDM, while a powerful effective theory, might not represent the complete picture of cosmic evolution, potentially requiring extensions involving new physics. Alternative frameworks, such as modified gravity theories or models incorporating dynamical DE, aim to address these outstanding issues
~\cite{Copeland:2006wr, Zlatev:1998tr, Garcia-Serna:2023xfw, Alvarez:2019ues, Guarnizo:2020pkj, Orjuela-Quintana:2021zoe,DeFelice:2010aj, Sotiriou:2008rp, Cardona:2022lcz, Orjuela-Quintana:2023zjm,Horndeski:1974wa,Koyama:2015vza,Allys:2015sht,Allys:2016jaq,Heisenberg:2014rta,Garnica:2021fuu,Rodriguez:2017ckc,DeFelice:2016yws,GallegoCadavid:2020dho,Frusciante:2018aew,Cardona:2023gzq}. In the following subsections, we outline the key problems that motivate the exploration of these alternative approaches.

\subsection{The Flatness Problem}
\label{subsec:FlatnessProblem}

The flatness problem originates from the relationship between the total energy density and the spatial curvature of the Universe, as expressed by the first Friedmann equation in terms of density parameters, Eq.~\eqref{ec:FriedmannequationLambda}. Defining the total density parameter excluding curvature as $\Omega_{\text{total}} = \Omega + \Omega_{\Lambda}$ (where $\Omega$ represents matter and radiation), the equation becomes:
\begin{equation}
 |\Omega_{\text{total}}(t)-1|=|\Omega_k(t)| = \left| -\frac{k}{a^2(t)H^2(t)} \right|.
 \label{ec:FlatnessProblem}
\end{equation}
If the Universe begins perfectly flat ($k=0$), then $\Omega_k(t_i)=0$, implying $\Omega_{\text{total}}(t_i)=1$. Since $a(t)H(t)$ remains finite, this condition $k=0$ ensures $\Omega_k(t)=0$ for all times; a flat universe remains flat. However, if there is any initial deviation from flatness ($k \neq 0$), the quantity $a^2 H^2$ evolves differently depending on the dominant energy component. During the radiation and matter-dominated eras, specifically:
\begin{itemize}
 \item For radiation dominance ($a \propto t^{1/2}$, $H \propto t^{-1}$): 
 \begin{equation}
 a^{2} H^{2} \propto t^{-1} \quad \Longrightarrow \quad |\Omega_{\text{total}}(t)-1| = |\Omega_k(t)| \propto \frac{1}{a^2 H^2} \propto t. \label{ec:CurvatureRadiation}
 \end{equation}
 \item For matter dominance ($a \propto t^{2/3}$, $H \propto t^{-1}$):
 \begin{equation}
 a^{2} H^{2} \propto t^{-2 / 3} \quad \Longrightarrow \quad |\Omega_{\text{total}}(t)-1| = |\Omega_k(t)| \propto \frac{1}{a^2 H^2} \propto t^{2 / 3}. \label{ec:CurvatureMatter}
 \end{equation}
\end{itemize}
In both scenarios, the deviation from flatness, $|\Omega_{\text{total}}(t)-1|$, grows with time. This implies that $\Omega_{\text{total}}=1$ is an unstable equilibrium. For the Universe to be observed as nearly flat today ($\Omega_{\text{total}}(t_0) \approx 1$, implying $|\Omega_k(t_0)| \ll 1$~\cite{Planck:2018vyg}), its initial value must have been extraordinarily close to unity. For example, during Big Bang Nucleosynthesis (BBN)\footnote{Nucleosynthesis is the early epoch in the Universe, roughly $t\sim 1$\,s to minutes, where nuclear reactions formed the first light elements as photons had energies around $\sim 1$\,MeV.}, the total density parameter must have satisfied $|\Omega_{\text{total}}(t_{\text{BBN}})-1| \lesssim 10^{-16}$~\cite{liddle2015introduction}.

The essence of the flatness problem is the lack of a compelling mechanism within the standard Big Bang framework (without inflation) to explain why the initial conditions were so precisely fine-tuned to produce the near-flatness observed today. Why did the Universe start with $\Omega_{\text{total}}$ so incredibly close to 1?

\subsection{The Horizon Problem}
\label{subsec:HorizonProblem}

The horizon problem arises from the remarkable large-scale isotropy observed in the Cosmic Microwave Background (CMB). The CMB radiation, emitted roughly 380,000 years after the Big Bang, exhibits an almost perfectly uniform temperature of $T_0 \approx 2.725\,\text{K}$ across the entire sky, with fluctuations only at the level of $\Delta T / T \sim 10^{-5}$~\cite{Planck:2018vyg}.

\begin{figure}[t!]
 \centering
 \includegraphics[width=10cm]{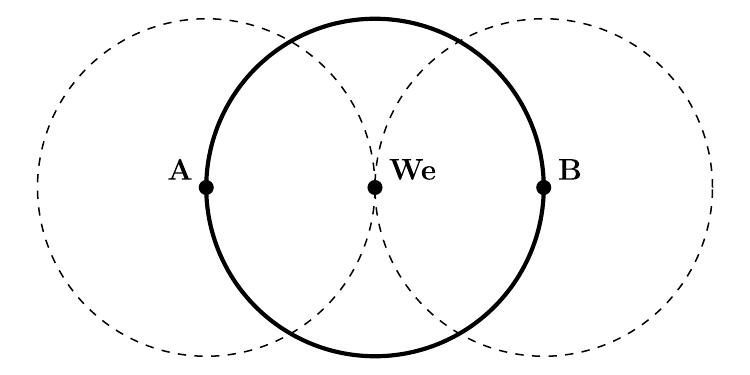}
 \caption{Schematic illustration of the horizon problem. An observer (center) sees CMB radiation from points A and B, which are widely separated on the surface of last scattering. In the standard Big Bang model without inflation, the particle horizons of A and B (dashed circles) do not overlap, meaning they were causally disconnected when the CMB was emitted, yet they exhibit nearly identical temperatures.}
 \label{fig:Horizon} 
\end{figure}

In the standard Big Bang model (without inflation), the Universe has a finite age, implying a finite particle horizon – the maximum distance light could have traveled since the beginning. Calculations show that at the time of CMB emission (last scattering), regions on the sky separated by more than $\sim 1^\circ$ were outside each other's particle horizons\cite{dodelson2020modern}. They were causally disconnected, meaning they could not have exchanged signals or interacted to reach thermal equilibrium. Yet, these causally separated regions are observed to have the same temperature to extremely high precision. How did regions that never communicated achieve such thermal homogeneity? This conundrum is illustrated schematically in Figure~\ref{fig:Horizon}.

\subsection{Inflationary Solution to Primordial Puzzles}
\label{subsec:InflationSolution}

The flatness and horizon problems find an elegant resolution within the paradigm of cosmic inflation, proposed by Guth and Linde~\cite{Guth:1980zm,liddle2015introduction}. Inflation postulates a period of quasi-exponential, accelerated expansion ($\ddot{a} > 0$) in the very early Universe. As established earlier (see Eq.~\eqref{ec:FactorDeDesaceleracion}), such acceleration requires a dominant component with an equation of state $w < -1/3$. A simple example is cosmological constant $\Lambda$, which drives $a(t) \propto e^{\sqrt{\Lambda/3}t}$.

\begin{figure}[t!]
 \centering
 \includegraphics[width=10cm]{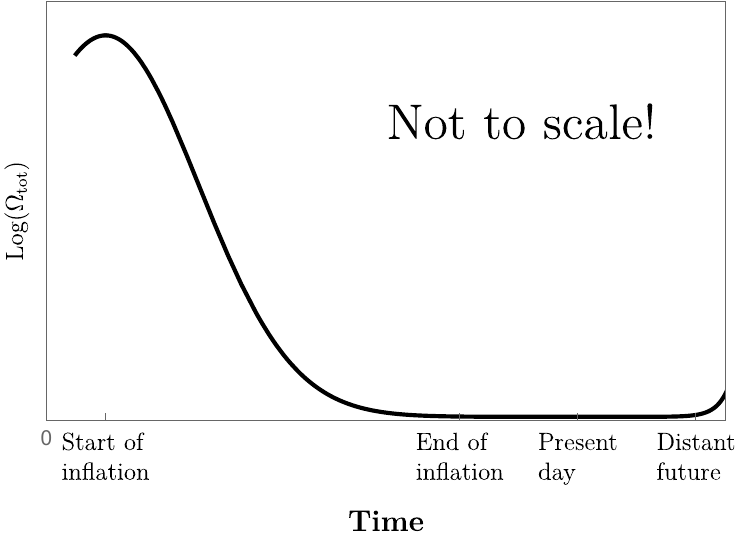}
 \caption{Evolution of the logarithm of the total density parameter $\text{Log}\left({\Omega_{\text{total}}}\right)$. The first part of the curve is an arbitrary initial condition, the inflation process eliminates these conditions and makes the Universe highly flat. $\Omega_{\text{total}}\sim1$, yet, in the future, the Universe could tend to deviate from 1 as observed at the end of the curve~\cite{liddle2015introduction}.}
 \label{fig:OmegaTotal} 
\end{figure}

\begin{figure}[t!]
 \centering
 \includegraphics[width=10cm]{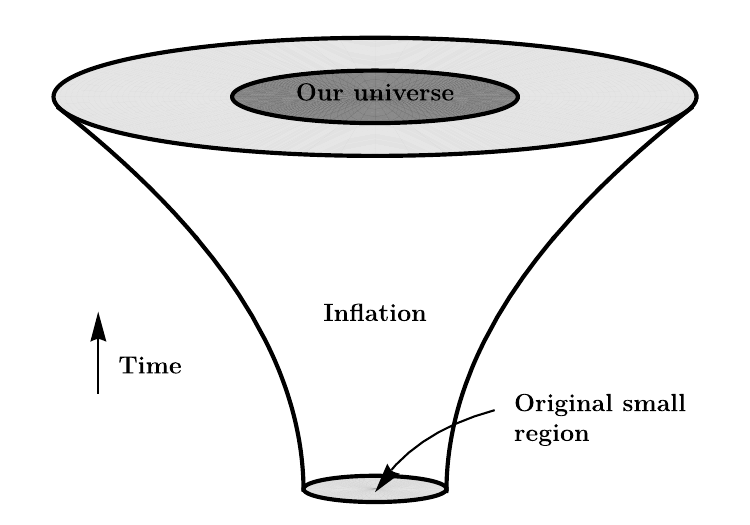}
 \caption{Schematic illustrating the solution to the horizon problem. A small, causally connected region before inflation (left) is expanded exponentially (center) to become vastly larger than the post-inflationary Hubble radius (right). Regions A and B, though causally disconnected today, originated from within the same causally connected pre-inflationary patch~\cite{liddle2015introduction}.}
 \label{fig:EsquemaDeInflacion} 
\end{figure}

Inflation addresses the primordial puzzles as follows:
\begin{itemize}
    \item \textbf{Solving the Flatness Problem:} During a period of accelerated expansion, $\ddot{a}>0$. This implies $\frac{d}{dt}(aH) = \frac{d}{dt}(\dot{a}) > 0$. Consequently, the denominator $a^2 H^2$ in the expression for $|\Omega_k(t)| = |k|/(a^2 H^2)$ grows enormously. Any initial curvature is stretched to near-zero values, driving $|\Omega_{\text{total}}(t)-1|$ exponentially towards zero. Inflation dynamically forces the Universe towards flatness, alleviating the need for extreme fine-tuning of initial conditions (see Figure~\ref{fig:OmegaTotal}).
    \item \textbf{Solving the Horizon Problem:} Prior to inflation, the region corresponding to our entire observable Universe today was microscopically small and causally connected, allowing it to reach thermal equilibrium. Inflation then stretched this tiny, homogeneous patch by an enormous factor (at least $e^{60} \approx 10^{26}$), making it encompass scales vastly larger than the post-inflationary particle horizon. Thus, the large-scale uniformity of the CMB reflects the equilibrium established \emph{before} inflation expanded the region beyond causal contact (see Figure~\ref{fig:EsquemaDeInflacion}).
    \item \textbf{Seeding Structure:} Beyond solving these problems, inflation also provides a natural mechanism for generating the primordial density fluctuations needed to seed galaxy formation. Quantum fluctuations inherent in the inflating field are stretched to cosmological scales, creating the nearly scale-invariant, adiabatic, Gaussian perturbations observed in the CMB and large-scale structure~\cite{Planck:2018jri}.
\end{itemize}

\subsection{Remaining Challenges and Tensions}
\label{subsec:RemainingChallenges}

While inflation resolves key issues of the standard hot Big Bang model, the combined $\Lambda$CDM $+$ Inflation paradigm still faces significant hurdles:

\subsubsection{The Cosmological Constant Problem}
\label{subsubsec:LambdaProblem}
The observed value of the cosmological constant $\Lambda$, which drives the current accelerated expansion, is extremely small compared to theoretical expectations from quantum field theory. Interpreting $\Lambda$ as the vacuum energy density leads to predictions that are famously $60-120$ orders of magnitude larger than observed~\cite{Weinberg, ErrorDELambda}.

\subsubsection{Observational Tensions: $H_0$ and $S_8$}
\label{subsubsec:Tensions}
Recent, precise measurements have revealed statistically significant tensions between cosmological parameters inferred from early-universe probes (like the CMB by Planck) assuming $\Lambda$CDM, and those measured via late-universe probes:
\begin{itemize}
    \item \textbf{$H_0$ Tension:} Measurements of the Hubble constant today ($H_0$) using local distance ladder methods (e.g., Cepheids and SNe Ia via SH0ES) yield a value around $H_0 \approx (73.04 \pm 1.04)\,km/s \text{Mpc}$~\cite{Riess:2021jrx, Freedman:2021ahq, Abdalla:2022yfr}, which is in $\sim 5\sigma$ tension with the value inferred from Planck CMB data assuming $\Lambda$CDM, $H_0 = (67.27 \pm 0.60)\,km/s \text{Mpc}$~\cite{Planck:2018vyg, Planck:2019evm, Abdalla:2022yfr}. This discrepancy suggests a potential breakdown in the standard model connecting the early and late universe~\cite{DiValentino:2021izs, Perivolaropoulos:2021jda}.
    \item \textbf{$S_8$ Tension:} Measurements of the amplitude of matter clustering at late times, often parameterized by $S_8 \equiv \sigma_8 \sqrt{\Omega_m/0.3}$ (where $\sigma_8$ is the root mean square fluctuation amplitude on 8 \text{Mpc} scales), derived from weak lensing and galaxy clustering surveys tend to be lower than the value predicted by Planck CMB data extrapolated forward using $\Lambda$CDM. This tension is typically at the $2-3\sigma$ level and might indicate suppressed structure growth~\cite{DES:2021wwk, KiDS:2020suj, Abdalla:2022yfr}. 
\end{itemize}
\begin{figure}[t!]
    \centering
    \includegraphics[width=8cm]{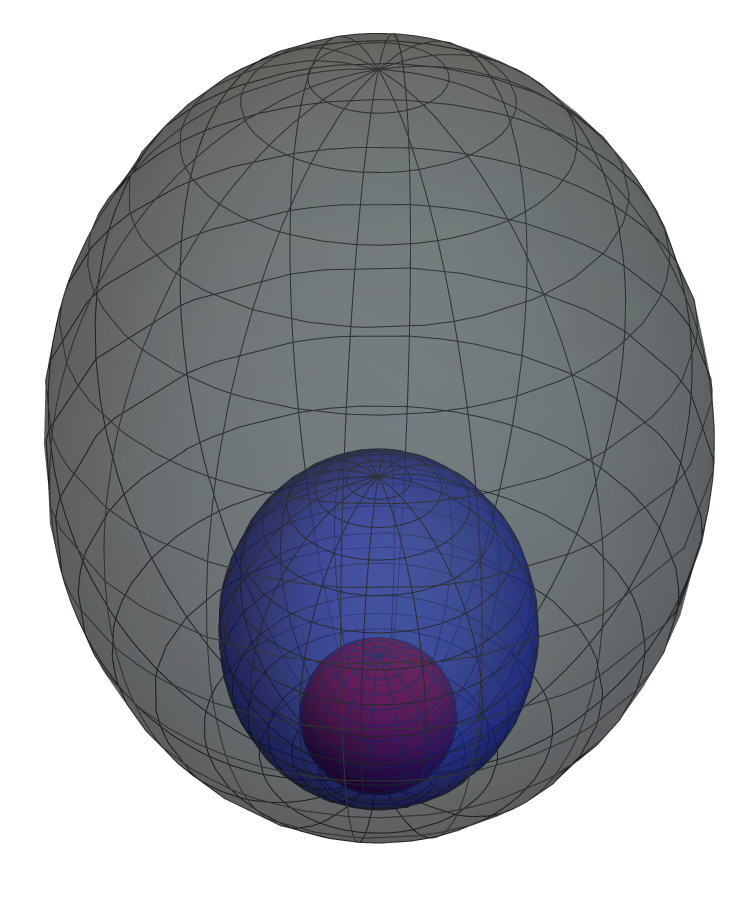} 
    \caption{Schematic evolution of a comoving perfect sphere (red) in an axially symmetric Bianchi I spacetime described by Eq.~\eqref{Eq:BianchiIMetric}. Differential expansion rates distort the sphere into an ellipsoid (blue, black) over time.}
    \label{fig:BianchiIExample} 
\end{figure}
\subsubsection{Challenges to Isotropy: Observational Anomalies}
\label{subsubsec:Anomalies} 

While the $\Lambda$CDM model assumes perfect large-scale isotropy and homogeneity (FLRW metric, Eq.~\eqref{eq:FLRWMetric}) and Gaussian random initial fluctuations, certain features observed primarily in the CMB maps present potential challenges or ``anomalies'' relative to these assumptions. These include unexpected alignments of the lowest multipole moments (quadrupole and octopole), a hemispherical power asymmetry, anomalous cold spots, and perhaps unusual patterns in polarization data~\cite{Planck:2019evm, Schwarz:2015cma, Jones:2023ncn}. Some studies also suggest potential large-scale anisotropies in galaxy distributions or flows~\cite{Kashlinsky:2008ut, Wiltshire:2012uh, Bengaly:2017slg}.

The statistical significance and origin of these anomalies are subjects of ongoing debate – they might be residual systematic effects, foreground contamination, or simply rare statistical fluctuations within the standard model. However, their persistence motivates considering deviations from perfect isotropy. Anisotropic cosmological models, such as the Bianchi spacetimes, offer a theoretical framework for such investigations. Their general metric is described by
\begin{equation}
    ds^2 = - dt^2 + a^2(t)\gamma_{ij}dx^idx^j. \label{Eq:BianchiMetric}
\end{equation}
In this equation, $\gamma_{ij}=e^{2\sigma_i}\delta_{ij}$, and the condition $\sum_{i=1}^3\sigma_i=0$ guarantees that $a$ describes the average expansion. The Bianchi I model, for instance, describes a spatially flat but anisotropically expanding universe, whose line element can be written (in axially symmetric form) as:
\begin{equation}
    ds^2 = - dt^2 + a^2(t) \left[ e^{-4\sigma(t)} dx^2 + e^{2\sigma(t)} \left( dy^2 + dz^2 \right) \right], \label{Eq:BianchiIMetric}
\end{equation}
where $\sigma(t)$ quantifies the shear or deviation from isotropy (see Figure~\ref{fig:BianchiIExample}). Although CMB observations strongly constrain any residual anisotropy today ($\sigma(t_0) \ll 1$)~\cite{Planck:2018jri}, studying these models helps test the robustness of the Cosmological Principle and explore potential explanations for observed large-scale anomalies.

These outstanding problems – the theoretical puzzle of $\Lambda$, the observational tensions in $H_0$ and $S_8$, and the potential CMB anomalies – collectively provide strong motivation for exploring cosmological models beyond the standard $\Lambda$CDM paradigm.

\section{Observational Constraints and Dark Energy}

\begin{figure}[t!]
 \centering
 \includegraphics[width=13cm]{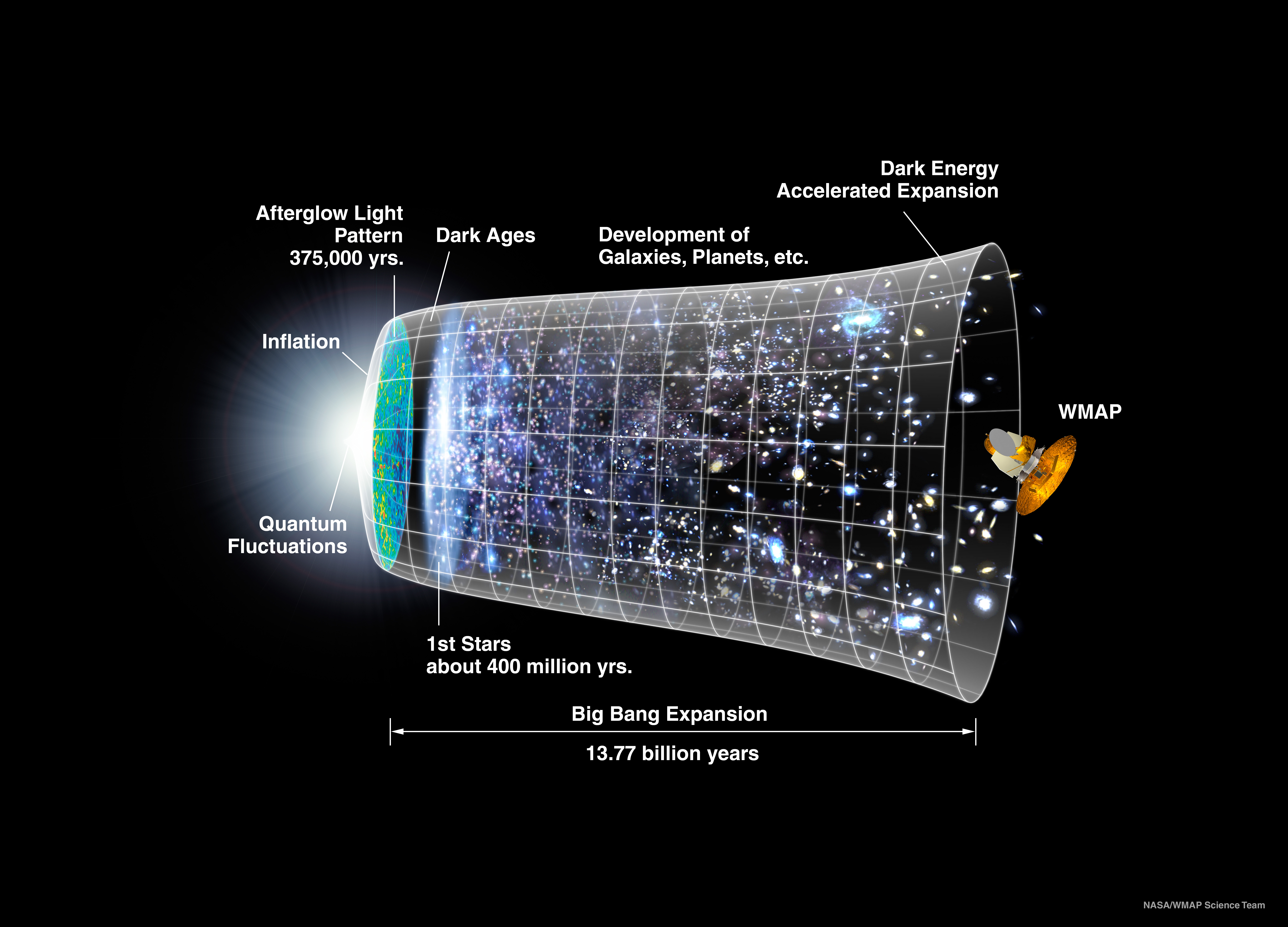} 
 \caption{Timeline of the Universe according to the $\Lambda$CDM model including inflation. Inflation occurs at the very beginning, followed by the hot Big Bang evolution (radiation domination, matter domination, recombination/CMB emission $\sim$380,000 yrs), and late-time acceleration driven by dark energy ($\Lambda$). (Image credit: NASA/WMAP Science Team~\cite{WMAP_Timeline}).}
 \label{fig:History} 
\end{figure}

 As has been discussed, the Universe today is experiencing accelerated expansion~\cite{Perlmutter_1999,SupernovaSearchTeam:1998fmf}. Similar to inflation, this accelerated expansion can be explained by the fluid known as dark energy, whose main characteristic is that its equation of state satisfies $w<-1/3$, meaning it can produce accelerated expansion. The first attempt to introduce dark energy in cosmological models was $\Lambda$. However, due to previously mentioned issues, such as the nature of $\Lambda$, the Hubble tension~\cite{DiValentino:2020vnx, Weinberg:2013agg, Planck:2018vyg, Riess:2021jrx}, the $S_8$ tension~\cite{DES:2017myr, Kazantzidis:2018rnb, Benevento:2020fev, Skara:2019usd}, anomalies in the CMB~\cite{Planck:2019evm, Schwarz:2015cma}, and the cosmic dipole misalignment~\cite{Kashlinsky:2008ut, Wiltshire:2012uh, Bengaly:2017slg}, among others~\cite{Bowman:2018yin, Perivolaropoulos:2021jda, Banerjee:2020bjq}. It is convenient to describe dark energy with additional degrees of freedom. These can include tachyonic fields~\cite{Copeland:2006wr,Nozari:2013mba}, quintessence fields~\cite{Copeland:2006wr}, vector fields~\cite{Armendariz-Picon:2004say,Koivisto:2008ig,Thorsrud:2012mu,Koivisto:2008xf,Landim:2016dxh,Gomez:2020sfz,Gomez:2021jbo}, p-forms~\cite{BeltranAlmeida:2019fou,Guarnizo:2020pkj,BeltranAlmeida:2018nin}, among other possible descriptions that can generate a viable model of dark energy.\\

Additionally, the fluid describing dark energy must comply with all observational constraints, such as the history of the Universe, illustrated in Figure~\ref{fig:History}. More specifically, it must meet constraints related to the processes occurring within this history. For instance, at the epoch where there is equality between the amount of matter and radiation, at $z_r=3200$, the Big Bang Nucleosynthesis (BBN) must be completed adequately, implying that the dark energy density parameter satisfies $\Omega_{\Lambda}<0.045$. At $z_r=50$, dark energy must satisfy $\Omega_{\Lambda}<0.02$ due to CMB constraints~\cite{dodelson2020modern,Abdalla:2022yfr,liddle2015introduction}. Additionally, the values of the density parameters today are $\Omega_{\Lambda}\sim 0.69$, $\Omega_m\sim 0.31$, and $\Omega_r\sim 10^{-4}$~\cite{Planck:2018vyg}.

\section{Dynamical System Approach\label{sec: dynamicalanalysis}}
Previously, we introduced the idea of defining new variables to apply the dynamical systems technique, which is useful for characterizing cosmological models. This approach is effective for analyzing the parameter space of theoretical models~\cite{Copeland:2006wr, Wainwright2009, Garcia-Salcedo:2015ora, Wimberger2014}. This technique is particularly powerful in cosmology, where the equations of motion for the fields are reformulated using dimensionless variables, typically derived from the first Friedman equation. This transformation results in a set of autonomous first-order differential equations, with their stationary solutions residing in the phase space of the dynamical variables.

In general, an autonomous system can be schematically written as:
\begin{equation}
\{ x'_i = f_i (x_1, \ldots, x_n; a_1, \ldots, a_k)\, | \ i = 1, \ldots, n\},
\label{ec:AutonomousEcuation}
\end{equation}
where the prime denotes differentiation with respect to the number of $e$-folds, and $f_i$ represents algebraic functions of the $n$ variables $x_j$ and the $k$ parameters $a_j$ of the model. 

The stationary solutions of this system, defined by:
\begin{equation}
\{ x'_i = 0 \ | \ i = 1, \ldots, n \},
\end{equation}
are known as fixed points. These fixed points are determined by solving the algebraic equations:
\begin{equation}
\{ f_i (x_1, \ldots, x_n ; a_1, \ldots, a_k) = 0 \ | \ i = 1, \ldots, n \}.
\end{equation}
In many cases, these equations are analytically solvable, allowing the fixed points to be expressed as functions of the model parameters. This enables a comprehensive characterization of the parameter space and its corresponding asymptotic behaviors.

However, in more complex scenarios, analytical solutions may not exist. For example, if any equation $f_i = 0$ contains a polynomial of degree greater than four, the fundamental theorem of Galois theory~\cite{ribes2010free} guarantees the absence of closed-form solutions. In such cases, numerical methods become indispensable for investigating the parameter space and analyzing the behavior of the system near its fixed points.

This dynamical systems approach has found extensive applications in cosmology (see, e.g., Refs.~\cite{Bahamonde:2017ize, Alvarez:2019ues, Guarnizo:2020pkj, Motoa-Manzano:2020mwe, Orjuela-Quintana:2020klr, DeFelice:2016yws, Basilakos:2019dof}), enabling the study of critical phenomena such as cosmic acceleration, transitions between different cosmological epochs, and the asymptotic dynamics encoded in the properties of each fixed point and its stability. This provides a foundation for deeper exploration of the system's evolution and its physical implications.

\subsubsection{Stability}
Now, let us analyze the stability of the fixed points, specifically investigating whether the system's evolution in their vicinity converges towards or diverges away from them. To achieve this, we consider Eq.~(\ref{ec:AutonomousEcuation}) in a component-wise manner near a fixed point, where $f(\vec{x_0}) = 0$. In this context, we proceed as follows:

\begin{equation}
 \dot{x}^i = \left. \frac{\partial f^i}{\partial x^j} \right|_{\vec{x}_0} \left(x^j - x_0^j\right) + \mathcal{O}\left(\left(x^j - x_0^j\right)^2\right),
\end{equation}
where $J^i_j = \left. \frac{\partial f^i}{\partial x^j} \right|_{\vec{x}_0}$ is known as the Jacobian matrix, which is not necessarily diagonal. However, it is always possible to diagonalize the system by finding the eigenvalues and eigenvectors, transforming the system into the eigenbasis such that the equations take the form

\begin{equation}
 \dot{u}^i = D^i_j \left(u^j - u_0^j\right) = \lambda_i \left(u^i - u_0^i\right),
\end{equation}
where $\lambda_i$ are the eigenvalues of the matrix $J$. Thus, in the variables $u$, which correspond to the eigenbasis, the problem is decoupled, and the solutions near the fixed point $u_0^j$ take the form

\begin{equation}
 u^i(t) = A^i e^{\lambda_i t} + u_0^i,
 \label{ec:EvolutionFixedPoints}
\end{equation}
where $A^i$ is a constant depending on the initial conditions. Depending on the sign of $\text{Re}(\lambda_i)$, the fixed point can be classified as an attractor or a repeller in the direction of $u^i$ axe, such that:

\begin{itemize}
 \item If $\text{Re}(\lambda_i) < 0$, $u_0$ is an attractor since, in Eq. (\ref{ec:EvolutionFixedPoints}), as $t \to \infty$, $|u^i(t) - u^i_0| \to 0$, meaning it gets closer to the fixed point.
 \item If $\text{Re}(\lambda_i) > 0$, $u_0$ is a repeller since, in Eq. (\ref{ec:EvolutionFixedPoints}), as $t \to \infty$, $|u^i(t) - u^i_0| \to \infty$, meaning it moves away from the fixed point.
 \item If $\text{Re}(\lambda_i) = 0$, the point exhibits periodic solutions around the fixed point $u^i$, without necessarily converging to it.
\end{itemize}

Thus, a fixed point is classified as an \textbf{attractor} if the real part of all eigenvalues satisfies $\text{Re}(\lambda_i) < 0$, a \textbf{repeller} if for all eigenvalues $\text{Re}(\lambda_i) > 0$, and if the real parts of the eigenvalues include both positive and negative values, it is classified as a \textbf{saddle} point~\cite{Ghaffarnejad:2017hqt, Wimberger2014, azamov2018differential}. When the system is not compact, the existence of multiple attractors can significantly influence the dynamical behavior, necessitating the study of nullclines or pseudo fixed points. Unlike traditional fixed points, pseudo fixed points do not require $ \dot{x}_j = 0 $ for all $ j $; however, the approach for analyzing the dynamics remains similar to that used for traditional fixed points\cite{Ott:2002}.

\subsection{Numerical approach}
\label{Sec: Dynamical systems: numerical approach}

As we mentioned, there are cases where finding the fixed points becomes an impossible task. In this context, we present a general framework to analyze the stability properties of regions in the parameter space of models without analytical fixed points. Assuming an autonomous system of $n$ differential equations, as described in Eq.~\eqref{ec:AutonomousEcuation}, our numerical approach follows these key steps:

\begin{itemize}
\item[1.] Selection of a representative region within the parameter space of the model.
\end{itemize}

This involves defining an interval for each parameter, such that $a_{i_\text{min}} < a_i < a_{i_\text{max}}$ for $i = 1, \ldots, k$. The Cartesian product of these intervals forms a subsection of the parameter space. The length of each interval should account for the model's physical constraints and the symmetries of the autonomous system. 

\begin{itemize}
\item[2.] Application of physical constraints to differentiate between cosmologically viable and non-viable fixed points.
\end{itemize}

These constraints ensure that the numerical solutions of the autonomous equations align with physically meaningful properties of the model. For instance, imposing the condition $w_{\text{eff}} < -1/3$, where $w_{\text{eff}}$ represents the effective equation of state, ensures that any resulting solution describes an accelerated expansion of the universe. Other constraints may include non-negativity of density parameters for consistency with physical laws. Solutions meeting these criteria are labeled as ``cosmologically viable'', while those failing to satisfy them are categorized as ``non-viable'' and excluded. It is important to note that all cosmologically viable solutions derived from the same parameter set are indistinguishable in terms of physical interpretation.

\begin{itemize}
\item[3.] Stochastic exploration of cosmologically viable solutions within the chosen parameter space region.
\end{itemize}

\begin{figure}[t!]
\centering
\includegraphics[width = 10cm]{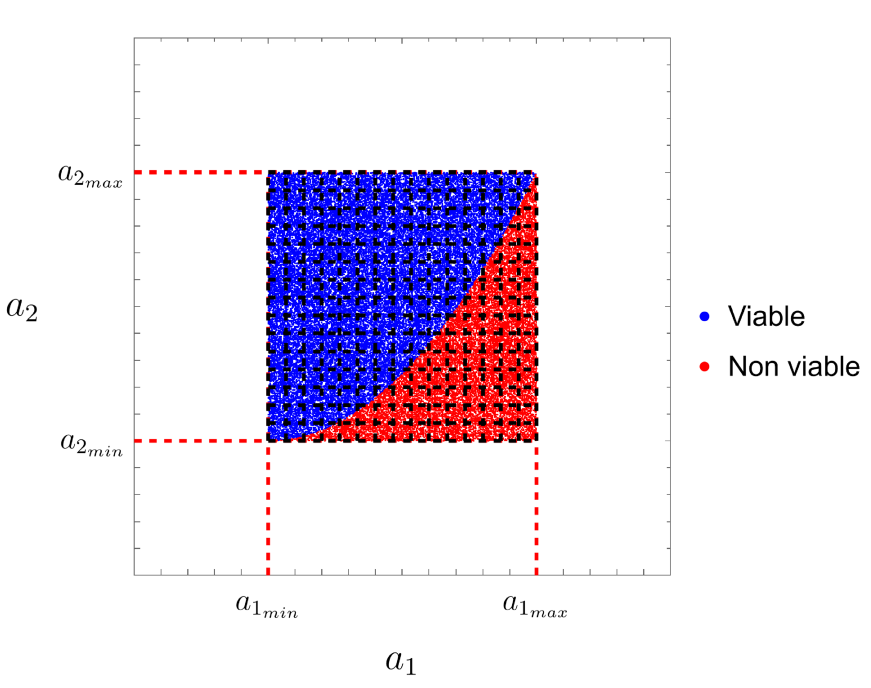}
\caption{Parameter space where a stochastic search for viable and non-viable points has been performed.}
\label{Fig: Stochastic Search}
\end{figure}

A large set of $N$ random points in the parameter space $\{a_1, \ldots, a_k\}$ is generated. For each set of parameters, fixed points are determined by numerically solving the system of algebraic equations $\{ f_i(x_1, \ldots, x_n) = 0 \ | \ i = 1, \ldots, n \}$\footnote{This can be achieved using standard numerical methods, such as \texttt{NSolve} in \texttt{Mathematica}.}. Subsequently, the physical constraints from the prior step are applied to classify the solutions. Points that fulfill the constraints are marked as ``cosmologically viable'', whereas those that do not are labeled ``non-viable'' and removed from further consideration. For models involving two parameters, this stochastic exploration may result in outcomes similar to those illustrated schematically in Figure~\ref{Fig: Stochastic Search}. In this context, all cosmologically viable solutions correspond to accelerated states of the system, such as Dark Energy (DE)-dominated fixed points.

\begin{itemize}
\item[4.] Stability analysis of the region of viable points. 
\end{itemize}

In general, the stability of a fixed point is assessed by calculating the real parts of the eigenvalues of the Jacobian matrix evaluated at that point, as discussed in the previous section. In the context of cosmology, a consistent expansion history necessitates the presence of at least three distinct fixed points: $i)$ a radiation-dominated point, which may function as either a saddle or a source; $ii)$ a matter-dominated saddle point; and $iii)$ a dark energy DE dominated point, which typically acts as an attractor of the system.\\

Note that for a given point in the parameter space, there could exist several fixed points with their own stability; for example, the three stages of dominance (radiation, matter, and DE epochs) should exist for a given set of parameters in order for the model to be able to reproduce a correct expansion history. Therefore, it is necessary to compute all the available Jacobian matrices (and their eigenvalues) for each set of parameters yielding cosmologically viable solutions. Since viable solutions from a given set of parameters have the same physical characteristics, the stability of the point in the parameter space is determined by the most stable point, that is:

\begin{itemize}
\item If all eigenvalues of at least one of the matrices are negative, then this point is an attractor and the other possible fixed points are discarded.
\item If all eigenvalues of all matrices are positive, then the stability corresponds to a repeller.
\item If the stability is not an attractor neither a repeller in the sense described above, one or more matrices have a mix of positive and negative eigenvalues, and the stability is of a saddle point.
\end{itemize}

Here, we are mainly interested in DE dominated attractor points. Therefore, once an attractor is found, the computation of the Jacobian matrices for the given set of parameters is stopped, just to proceed to another point in the parameter space. In the chapter~\ref{Ch: anisotrpic Tachyon field}, we will apply the numerical method to analyze the asymptotic behavior of DE model in an anisotropic background, prioritizing the search of DE domination points.

\section{Alternative Framework: The Generalized SU(2) Proca Theory}
\label{Sec: GSU2P_Intro}

The previously discussed challenges facing the standard $\Lambda$CDM model, including the theoretical puzzles surrounding the cosmological constant and the observational tensions in $H_0$ and $S_8$, motivate the exploration of alternative cosmological paradigms. These alternatives often involve modifying gravity itself or introducing new dynamical fields beyond the standard model components. Vector fields, particularly those associated with gauge symmetries like SU(2), represent an intriguing possibility, potentially arising from more fundamental physics frameworks and offering richer dynamics than simple scalar fields.

One such well-motivated framework is the Generalized SU(2) Proca (GSU2P) theory. This theory describes the dynamics of interacting, massive, non-Abelian vector fields ($B^a_\mu$) belonging to the Lie algebra of the SU(2) group, coupled to gravity. As detailed in Refs.~\cite{GallegoCadavid:2020dho, GallegoCadavid:2022uzn} (building on earlier work~\cite{BeltranJimenez:2016afo, Allys:2016kbq} and extended in~\cite{GallegoCadavid:2021ljh}), the action for GSU2P includes specific derivative self-interactions and couplings to curvature designed to maintain the global SU(2) symmetry and, crucially, to ensure the theory propagates the correct number of degrees of freedom (avoiding ghostly Ostrogradski instabilities often associated with higher-derivative theories~\cite{ErrastiDiez:2019trb, Woodard:2006nt, Ostrogradsky:1850fid, Woodard:2015zca}, although see~\cite{ErrastiDiez:2023gme, Janaun:2023nxz} for further discussion). These vector fields can potentially act as a source of dark energy or modify gravity, offering new avenues to address cosmological puzzles~\cite{Garnica:2021fuu, Rodriguez:2017wkg}. The following section details the specific structure of the GSU2P action.

\subsection{General Framework} 
\label{subsec:GSU2P_Framework}

The generalized SU(2) Proca theory considers the dynamics of a vector field belonging to the Lie algebra of the SU(2) group. The corresponding action, as presented in Refs.~\cite{GallegoCadavid:2020dho, GallegoCadavid:2022uzn} (see also Refs.~\cite{BeltranJimenez:2016afo, Allys:2016kbq} for older constructions and Ref.~\cite{GallegoCadavid:2021ljh} for an extended version), is designed to respect global invariance under this group of transformations and to propagate the right number of degrees of freedom~\cite{ErrastiDiez:2019trb} (see, anyway, Refs.~\cite{ErrastiDiez:2023gme, Janaun:2023nxz}), thereby circumventing Ostrogradski instabilities~\cite{Woodard:2006nt,Ostrogradsky:1850fid, Woodard:2015zca}. The action of the GSU2P theory is given by:
\begin{align}
\label{Eq: Action GSU2P}
S \equiv \int \text{d}x^4 \sqrt{-g} & \left\{ \mathcal{L}_{\text{EH}} + \mathcal{L}_{\text{YM}} + \sum_{i=1}^{2} \chi_i \mathcal{L}_2^i \right. \left.+ \sum_{i=3}^{7} \frac{\chi_i}{m_{\text{P}}^2} \mathcal{L}_{2}^i + \sum_{i=1}^{6} \frac{\alpha_i}{m_{\text{P}}^2} \mathcal{L}_{4,2}^i \right\}, 
\end{align}
where
\begin{equation}
 \mathcal{L}_{\text{EH}} \equiv \frac{m_{\text{P}}^2}{2}R, \qquad
 \mathcal{L}_{\text{YM}} \equiv-\frac{1}{4}F^a_{\ \mu\nu}F_a^{\ \mu\nu},
\end{equation}
denote the Einstein-Hilbert and the Yang-Mills Lagrangians, respectively, while the different vector-tensor interactions are:
\begin{align}
 \mathcal{L}_2^1 &\equiv B^a_{\ \mu} B_a^{\ \mu} B^b_{\ \nu} B_b^{\ \nu}, \quad
 \mathcal{L}_2^2 \equiv B^a_{\ \mu} B^{b\mu} B_a^{\ \nu} B_{b\nu}, \quad
 \mathcal{L}_2^3 \equiv A_a^{\ \mu \nu} A^\rho{ }_\nu{ }^a B_{\ \mu}^b B_{\rho b}, \\
 \mathcal{L}_2^4 &\equiv A_a^{\ \mu \nu} A^\rho{ }_\nu{ }^b B_{\mu b} B_{\ \rho}^a, \quad
 \mathcal{L}_2^5 \equiv A_a^{\ \mu \nu} A^\rho{ }_\nu{ }^b B_{\ \mu}^a B_{b\rho}, \quad
 \mathcal{L}_2^6 \equiv A_a^{\ \mu \nu} A_{\ \mu \nu}^a{ } B_b^{\ \rho} B^b_{\ \rho}, \\
 \mathcal{L}_2^7 &\equiv A_a^{\mu \nu} A_{\mu \nu}^b B^a_{\ \rho} B_b^{\ \rho},
\end{align}
and
\begin{align}
 \mathcal{L}_{4, 2}^1 & \equiv B_b^{\ \rho} B^b_{\ \rho} \left[S_{\ \ \mu}^{a\mu} S_{a \nu}^{\ \ \nu} - S_{\ \ \nu}^{a \mu} S_{\mu a}^{\ \ \nu} \right] + 2 B_a^{\ \rho} B_{b\rho} \left[ S_{\ \ \mu}^{a\mu} S_{\ \ \nu}^{b\nu} - S_{\ \ \nu}^{a\mu} S_{\ \ \mu}^{b\nu}\right], \\
 \mathcal{L}_{4,2}^2 & \equiv A_{\ \mu \nu}^a S_{\ \ \sigma}^{b\mu} B_a^{\ \nu} B_b^{\ \sigma} - A_{\ \mu \nu}^a S_{\ \ \sigma}^{b\mu} B_b^{\ \nu} B_a^{\ \sigma} + A_{\ \mu \nu}^a S_{\ \ \rho}^{b\rho} B_a^{\ \mu} B_b^{\ \nu}, \\
 \mathcal{L}_{4,2}^3 & \equiv B^{\mu a} R^\alpha{ }_{\sigma \rho \mu} B_{\alpha a} B^{\rho b} B_b^{\ \sigma} + \frac{3}{4} B_b^{\ \mu} B^b_{\ \mu} B^a_{\ \nu} B_a^{\ \nu} R, \\
 \mathcal{L}_{4,2}^4 & \equiv \left(B_b^{\ \mu} B^b_{\ \mu} B^a_{\ \nu} B_a^{\ \nu} + 2 B_a^{\ \mu} B_{b\mu} B^a_{\ \nu} B^{b\nu}\right) R, \\
 \mathcal{L}_{4,2}^5 & \equiv G_{\mu \nu} B^{a\mu} B_a^{\ \nu} B^b_{\ \rho} B_b^{\ \rho}, \\
 \mathcal{L}_{4,2}^6 &\equiv G_{\mu \nu} B^{a\mu} B^{b\nu} B_a^{\ \rho} B_{b\rho}.
\end{align}

In the previous expressions, $g$ is the determinant of the metric, $G_{\mu\nu}$ is the Einstein tensor, $R^{\alpha}_{\ \sigma\rho\mu}$ is the Riemann tensor, $\epsilon_{abc}$ is the Levi-Civita symbol, $B^a_{\ \mu}$ is the vector field within the Lie algebraic structure of SU(2) whose strength tensor is $F^{a}_{\ \mu\nu}\equiv\nabla_{\mu}B^a_{\ \nu}-\nabla_{\nu}B^a_{\ \mu}+\tilde{g}{\epsilon^a}_{bc}B^{b}_{\ \mu}B^{c}_{\ \nu}$, where $\tilde{g}$ is the SU(2) coupling constant, and we define the symmetric and antisymmetric tensors $S^{a}_{\ \mu\nu}\equiv\nabla_{\mu}B^a_{\ \nu}+\nabla_{\nu}B^a_{\ \mu}$ and $A^{a}_{\ \mu\nu}\equiv\nabla_{\mu}B^a_{\ \nu}-\nabla_{\nu}B^a_{\ \mu}$, respectively. All the $\alpha_i$ and $\chi_i$ are arbitrary dimensionless constants. Hereinafter, Greek indices denote space-time indices that run from $0$ to $3$ while Latin indices run from $1$ to $3$ and denote space indices and/or SU(2) group indices. 

It is worth mentioning that the only pieces of the GSU2P theory that have been considered are those that are relevant for the cosmic acceleration mechanism discussed in Refs.~\cite{Garnica:2021fuu, Rodriguez:2017wkg}, i.e., those that, prior to covariantization, involve two derivatives and two vector fields, or four vector fields (see Refs.~\cite{GallegoCadavid:2020dho, GallegoCadavid:2022uzn}).


\chapter{Numerical Exploration of Dynamical Systems: Anisotropic Tachyon Field in Cosmology}
\label{Ch: anisotrpic Tachyon field}

\section{Introduction} 
\label{Introduction}
Cosmological models have expanded significantly beyond the standard $\Lambda$CDM paradigm in response to persistent observational tensions and theoretical challenges~\cite{Perivolaropoulos:2021jda, Abdalla:2022yfr}. Among the proposed alternatives, scalar and vector fields have emerged as essential tools for describing phenomena such as inflation, dark energy, and deviations from isotropy. Scalar-tensor frameworks, including those involving tachyon fields~\cite{Aguirregabiria:2004xd, Abramo:2003cp, Bagla:2002yn, Nozari:2013mba, Hussain:2022dhp,Pace:2024vcd}, and vector-tensor extensions, such as Proca theory and its generalizations~\cite{Tasinato:2014eka, Heisenberg:2014rta, Allys:2015sht, BeltranJimenez:2016rff, Allys:2016jaq, GallegoCadavid:2019zke}, provide robust frameworks to enrich the gravitational dynamics of the universe. These models open new avenues to address discrepancies within $\Lambda$CDM, such as the Hubble and $S_8$ tensions, while also offering tools to explore anisotropic cosmologies.\\

Tachyon scalar fields have been widely studied for their role in cosmological evolution, particularly in isotropic spacetimes where analytical solutions are more readily obtained~\cite{Aguirregabiria:2004xd, Abramo:2003cp}. However, the complexity increases significantly in scenarios involving non-canonical scalar fields in anisotropic Bianchi-I spacetimes~\cite{Ohashi:2013pca, Orjuela-Quintana:2021zoe}, where analytical solutions often become intractable. Such cases require numerical methods to analyze the behavior of the system and assess its stability.\\

In this chapter, we examine a model that couples a scalar tachyon field to a vector field in an anisotropic Bianchi-I background. The tachyon field predominantly governs the evolution of dark energy, while the vector field primary role is to keep the anisotropic nature of the background. This coupling introduces structural parallels to the Dirac-Born-Infeld (DBI) Lagrangian, which further complicates the dynamical equations and precludes a full analytical characterization of the parameter space~\cite{Orjuela-Quintana:2021zoe}. To overcome these challenges, we utilize the dynamical systems approach to numerically reconstruct the parameter space and explore the asymptotic behavior of the solutions, as detailed in Section~\ref{Sec: Dynamical systems: numerical approach}.

\section{Illustration of numerical Method}
\label{Sec: Illustration of the Method}

\subsection{Tachyon field in an anisotropic background}

The Lagrangian of our template model is given by 
\begin{equation}
S \equiv \int \text{d}^4 x \sqrt{- g} \left( \mathcal{L}_\text{EH} + \mathcal{L} + \mathcal{L}_m + \mathcal{L}_r \right),
\label{Eq: Tachyon Field Action}
\end{equation}
where $\mathcal{L}_\text{EH} \equiv m_\text{P}^2 R/2$, $m_\text{P}$ is the reduced Planck mass, $R$ is the Ricci scalar, $\mathcal{L}_m$ and $\mathcal{L}_r$ are the Lagrangians for matter and radiation, respectively, and 
\begin{eqnarray} 
\mathcal{L} \equiv - V(\phi)\sqrt{1+\partial_{\mu}\phi\partial^{\mu}\phi} - \frac{1}{4} f(\phi) F^{\mu\nu} F_{\mu\nu},
\label{Eq: Tachyonic Lagrangian}
\end{eqnarray}
where $\phi$ is the scalar tachyon field, $V(\phi)$ is its potential, $F_{\mu\nu}\equiv \nabla_{\mu}A_{\nu}-\nabla_{\nu}A_{\mu}$ is the strength tensor associated to the vector field $A_{\mu}$, and $f(\phi)$ is a coupling function between $\phi$ and $A_\mu$. Next, we assume that the background geometry is described by a Bianchi I metric with rotational symmetry in the $y$-$z$ plane, Eq.~\eqref{Eq:BianchiIMetric}. In order to preserve the symmetries of the background, we choose the field profiles as
\begin{equation}
\phi \equiv \phi(t), \quad A_{\mu} \equiv \left(0, A(t), 0, 0\right),
\label{Eq: Field Profiles}
\end{equation}
being $A(t)$ a scalar field and the unique component of the vector field.

\subsection{Dynamical System} 

Following the standard procedure~\cite{Garcia-Salcedo:2015ora}, we derive the evolution equations of the model by varying the action in Eq.~\eqref{Eq: Tachyon Field Action} with respect to the metric, the scalar field and the vector field. After these variations, we replace the Bianchi I metric given in Eq.~\eqref{Eq:BianchiIMetric}, and the ansatz for the fields in Eq.~\eqref{Eq: Field Profiles} obtaining:
\begin{align}
3 m_\text{P}^2 H^2 &= \frac{1}{2} f \frac{e^{4\sigma} \dot{A}^2}{a^2} +\frac{ V }{ \displaystyle\sqrt{1-\dot{\phi}^2}} +\rho_m + \rho_r \label{Eq: 1º Friedman}+ 3 m_\text{p}^2 \dot{\sigma}^2, 
\\
- 2 m_\text{P}^2 \dot{H} &= \dot{\phi}^2 \frac{ V }{ \displaystyle\sqrt{1-\dot{\phi}^2} } + \frac{2}{3} f \frac{e^{4\sigma} \dot{A}^2}{a^2} + \frac{4}{3} \rho_r \label{Eq: 2º Friedman} + \rho_m + 6 m_\text{p}^2 \dot{\sigma}^2,
\\
\ddot{\sigma} + 3H \dot{\sigma} &= \frac{e^{4 \sigma} f \dot{A}^2 }{ 3 a^2 m_\text{p}^2 },
\label{Eq: sigma} \\
\frac{\ddot{\phi}}{1 - \dot{\phi}^2 } &= \frac{ \displaystyle\sqrt{1-\dot{\phi}^2} f_{,\phi} }{ 2 V a^2 } e^{4\sigma} \dot{A}^2 -\frac{V_{,\phi}}{V}-3H\dot{\phi}, \label{Eq: Eq phi} \quad
\frac{\ddot{A}}{\dot{A}} = -\frac{\text{d}}{\text{d} t}\ln{\left(a f e^{4\sigma}\right)},
\end{align}
where $\rho_r$ and $\rho_m$ are the densities of matter and radiation, respectively, and a dot indicates a derivative with respect to $t$. Equations~\eqref{Eq: 1º Friedman} and~\eqref{Eq: 2º Friedman} correspond to the first and second Friedman equations, Eq.~\eqref{Eq: sigma} is the evolution equation for the geometrical shear, and Eqs.~\eqref{Eq: Eq phi} are the equations of motion for the scalar and vector fields, respectively. The above equations can be recast in terms of the following dimensionless variables
\begin{gather}
x \equiv \dot{\phi}\text{,\hspace{0.4cm}}y^2 \equiv \frac{V(\phi)}{3m_\text{p}^2H^2}\text{,\hspace{0.4cm}}z^2 \equiv \frac{1}{2}f(\phi)\frac{e^{4\sigma}\dot{A}^2}{3m_\text{p}^2H^2a^2}\text{,\hspace{0.4cm}}
\Omega_m \equiv \frac{\rho_m}{3m_\text{p}^2H^2}\text{,\hspace{0.4cm}}\Omega_r \equiv \frac{\rho_r}{3m_\text{p}^2H^2}\text{,\hspace{0.4cm}} \Sigma \equiv \frac{\dot{\sigma}}{H}.
\label{Eq: Dynamical Variables}
\end{gather}
The first Friedman equation in Eq.~\eqref{Eq: 1º Friedman} becomes the constraint
\begin{equation}
1 = \frac{y^2}{\sqrt{1 - x^2}} + z^2 + \Sigma^2 + \Omega_m + \Omega_r,
\end{equation}
while from the second Friedman equation in Eq.~\eqref{Eq: 2º Friedman} we can compute the deceleration parameter $q \equiv - 1 - \dot{H}/H^2$ obtaining
\begin{equation}
 q = \frac{1}{2} \left[1 - 3 \sqrt{1 - x^2} y^2 + z^2 + \Omega_r + 3 \Sigma^2 \right].\label{Eq: q deceleration parameter}
\end{equation}
We can easily integrate the right hand Eq.~\eqref{Eq: Eq phi} such that for the vector field degree of freedom we have 
\begin{equation}
 \dot{A} = c \frac{e^{-4 \sigma}}{a f},
\end{equation}
where $c$ is a constant.
From the Friedman equations~\eqref{Eq: 1º Friedman} and~\eqref{Eq: 2º Friedman}, we can identify the density and pressure of dark energy, which can be written in terms of the dimensionless variables as
\begin{align}
\rho_{\text{DE}} \equiv\text{ } 3m_\text{P}^2 H^2 \left( \frac{y^2}{\sqrt{1 - x^2}} + z^2 + \Sigma^2 \right), \quad
p_{\text{DE}} \equiv\text{ } 3m_\text{p}^2H^2\left(-y^2\sqrt{1-x^2}+\frac{1}{3}z^2+\Sigma^2\right),
\end{align}
respectively. Now, the equation of state of DE, $w_{\text{DE}} \equiv p_{\text{DE}} / \rho_{\text{DE}}$, is given by
\begin{equation}
w_{\text{DE}} = -1 + \frac{2}{3} \frac{\frac{3}{2} \frac{x^2 y^2}{\sqrt{1 - x^2}} + 2 z^2 + 3 \Sigma^2}{\frac{y^2}{\sqrt{1 - x^2}} + z^2 + \Sigma^2}.
\end{equation}
Note that we included the geometrical shear $\Sigma$ in the definition of $\rho_{\text{DE}}$ and $p_{\text{DE}}$. This choice allows us to write the DE continuity equation as 
\begin{equation}
\dot{\rho}_{\text{DE}} + 3 H (\rho_{\text{DE}} + p_{\text{DE}}) = 0,
\end{equation}
which is the usual form of the continuity equation of an uncoupled fluid, as it was described on the Section~\ref{sec:LCDM_Dynamics_v2}, stressing that we are considering that DE can be an anisotropic fluid.

In order to calculate the evolution equations for the dimensionless variables, we differentiate each variable in Eq.~\eqref{Eq: Dynamical Variables} with respect to the number of $e$-folds.\footnote{Remember, the relation between the number of $e$-folds and the scale factor is $N \equiv \ln a$.} We get
\begin{align}
 x' &= \sqrt{3} \left(1 - x^{2} \right) \left[\frac{\sqrt{1 - x^{2}}}{y} \beta z^{2} - \sqrt{3} x - \alpha y \right], \label{Eq: x prime} \\
 y' &= y \left[\frac{\sqrt{3}}{2} \alpha y x + q + 1 \right], \\
 z' &= z (q - 1) - 2 z \Sigma - \frac{\sqrt{3}}{2} \beta x y z, \label{zprime} \\
 \Sigma' &= \Sigma(q - 2) + 2 z^{2}, \label{Eq: Sigma prime}\\
 \Omega_r' &= 2 \Omega_r (q - 1),\label{Eq: Omega prime}
\end{align}
where
\begin{equation}
\alpha(t) \equiv m_\text{P} \frac{V_{, \phi}}{V^{3/2}}, \quad 
\beta(t) \equiv m_\text{P} \frac{f_{,\phi}}{f \sqrt{V}}.
\label{Eq: alpha and beta}
\end{equation} 
The last equations can be easily integrated when $\alpha$ and $\beta$ are constants. Under this assumption, we get that the potential $V(\phi)$ and the coupling function $f(\phi)$ are given by expressions of the form\footnote{In the case of varying $\alpha$ and $\beta$, the autonomous set is not closed and we would be forced to introduce more dimensionless variables to close the system. Thus, we choose $\alpha$ and $\beta$ constant to keep our presentation simple.}
\begin{equation}
V(\phi) \propto 1/(\alpha \phi)^{2}, \quad f(\phi) \propto (\alpha \phi)^{2\beta\phi/\left|\alpha\phi\right|},
\end{equation}
which correspond to common power law expressions typically used in the literature~\cite{Copeland:2006wr,Yang:2012ht,Nozari:2013mba}.

\subsection{Analytical Fixed points}
\label{Section: AFP}

The asymptotic behavior of the autonomous system in Eqs.~\eqref{Eq: x prime}-\eqref{Eq: Omega prime} is encoded in its fixed points, which can be found by setting $x' = 0$, $y' = 0$, $z' = 0$, $\Sigma' = 0$, and $\Omega'_r = 0$. This yields a set of algebraic equations given by
\begin{align}
0 &=\left( -1 + x^2 \right) \big\{ \alpha^2 y^4 + 2 \sqrt{3} \alpha x y^3 + 3 x^2 y^2 \label{Eq: x prime polynomial} + \beta^2 x^2 z^4 - \beta^2 z^4 \big\}, \\
0&=y\Big\{9 \left(x^2-1\right) y^4+\left(3 \Sigma ^2+\Omega_r+z^2+3\right)^2+2 \sqrt{3} \alpha x y \left(3 \Sigma ^2+\Omega_r+z^2+3\right)\label{Eq: y prime polynomial} 
+3 \alpha ^2 x^2 y^2\Big\},\\
0&=z\Big\{3 \beta ^2 x^2 y^2+\left[ \Sigma(3\Sigma-4) +\Omega_r+z^2-1\right]^2\label{Eq: z prime polynomial} -2 \sqrt{3} \beta x y \left[\Sigma(3\Sigma-4)+\Omega_r+z^2-1\right]\\
&+9 \left(x^2-1\right) y^4\Big\},\nonumber\\
0&=9 \Sigma ^6+\Sigma ^4 \left(6 \Omega_r+6 z^2-18\right)+24 \Sigma ^3 z^2\label{Eq: Sigma prime polynomial}+\Sigma ^2 \Big\{9 \left(x^2-1\right) y^4+\Omega_r^2-6 \Omega_r+z^4+2 \Omega_r z^2\\
&-6 z^2+9\Big\}+16 z^4+\Sigma \left(8 z^4+8 \Omega_r z^2-24 z^2\right),\nonumber\\
0&=\Omega_r \left\{9 \left(x^2-1\right) y^4+\left(3 \Sigma ^2+\Omega_r+z^2-1\right)^2\right\} \label{Eq: Omega_r prime polynomial},
\end{align}
where we have replaced the expression for $q$ in terms of the dynamical variables given in Eq.~\eqref{Eq: q deceleration parameter}, and some algebraic manipulations have been performed. Notice that these equations compose a polynomial system of degree greater than $4$, since Eq.~\eqref{Eq: Sigma prime polynomial} is an equation of degree $6$ in the variable $\Sigma$. This means that the system is not analytically solvable in general due to the fundamental theorem of Galois theory~\cite{ribes2010free}. However, in the specific case where $\Sigma = 0$, Eq.~\eqref{Eq: Sigma prime polynomial} reduces to $z = 0$, and Eq.~\eqref{Eq: z prime polynomial} is trivially satisfied. This is consistent with the fact that the vector field is the source of the anisotropy, as can be seen in Eq.~\eqref{Eq: sigma}. Under this simplification, the reduced system, consisting of Eqs.~\eqref{Eq: x prime polynomial},~\eqref{Eq: y prime polynomial}, and~\eqref{Eq: Omega_r prime polynomial}, becomes a polynomial system of degree $4$. In such cases, analytical solutions exist.

In order to do some analytical progress, in the following we neglect anisotropy, i.e., we consider $\Sigma = 0$ and $z = 0$, and study the fixed points of the simplified system in Eqs.~\eqref{Eq: x prime polynomial},~\eqref{Eq: y prime polynomial} and~\eqref{Eq: Omega_r prime polynomial} relevant for the radiation era ($\Omega_r \simeq 1$, $ w_{\text{eff}} \simeq 1/3$), the matter era, ($\Omega_m\simeq 1$, $w_{\text{eff}} \simeq 0 $) and an isotropic DE era ($\Omega_{\text{DE}}\equiv\rho_{\text{DE}}/3m_\text{p}^2H^2\simeq 1$, ~$w_{\text{eff}}<-1/3$).

\begin{itemize}
\item (\emph{R}) Radiation dominance:
\end{itemize}
\begin{equation}
\label{Eq: R fixed point}
x = 0, \ y = 0, \ z = 0, \ \Sigma = 0, \ \Omega_r = 1,
\end{equation} 
with $\Omega_{\text{DE}} = 0$, $w_{\text{DE}}=-1$, $\Omega_m = 0$ and $w_{\text{eff}} = 1/3$. The eigenvalues of the Jacobian evaluated in this point are
\begin{equation}
\lambda_1=2, \quad \lambda_2=-1, \quad \lambda_3=1, \quad \lambda_4=0, \quad \lambda_5=-3.
\end{equation}
Therefore, (\emph{R}) is a saddle. 

\begin{itemize}
\item (\emph{M}) Matter dominance:
\end{itemize}
\begin{equation}
x = 0, \ y = 0, \ z = 0, \ \Sigma = 0, \ \Omega_r = 0,\label{Eq: Matter dominance FP}
\end{equation}
with $\Omega_{\text{DE}} = 0$, $w_{\text{DE}}=-1$, $\Omega_m = 1$ and $w_{\text{eff}} = 0$. The eigenvalues of the Jacobian evaluated in this point are
\begin{equation}
\lambda_1=-\frac{3}{2}, \quad \lambda_2=\frac{3}{2}, \quad \lambda_3=-1, \quad \lambda_4=-\frac{1}{2}, \quad \lambda_5=-3.
\end{equation}
Then, this point is a saddle. Note that the Jacobian has $4$ negative eigenvalues in this point, in contrast with the $2$ negative eigenvalues in the point (\emph{R}). This difference is crucial, since a correct expansion history requires a radiation dominated epoch followed by a matter dominated epoch. These eigenvalues reflect the fact that cosmological trajectories passing around the radiation dominated point (\emph{R}) can go to the ``more stable'' matter dominated point (\emph{M}).

\begin{itemize}
\item (\emph{DE-I}) Isotropic DE dominance:
\end{itemize}
\begin{equation*}
x = \mp \frac{\alpha \sqrt{\sqrt{36 + \alpha^{4}} -\alpha^{2}}}{3 \sqrt{2}}, \quad \Sigma = 0,
\end{equation*}
\begin{equation}
y = \pm \frac{\sqrt{\sqrt{36 + \alpha^{4}} - \alpha^{2}}}{\sqrt{6}}, \quad z = 0, \label{Eq: DE-I dominance FP}\\
\end{equation}
with $\Omega_{\text{DE}} = 1$, $\Omega_m = 0$, $\Omega_r = 0$ and 
\begin{equation}
w_{\text{DE}} = -1 + \frac{1}{18} \alpha^{2} \left(- \alpha^{2} + \sqrt{36 + \alpha^{4}}\right).
\end{equation}
Since $w_{\text{DE}} = w_\text{eff}$ in this point, the condition for accelerated expansion ($w_{\text{eff}} < -1/3$) is satisfied when 
\begin{equation}
|\alpha| < 3^{1 / 4} \sqrt{2} \approx 1.86121.
\end{equation}

\begin{figure*}
\centering
\begin{minipage}[b]{.4\textwidth}
\includegraphics[width=\textwidth]{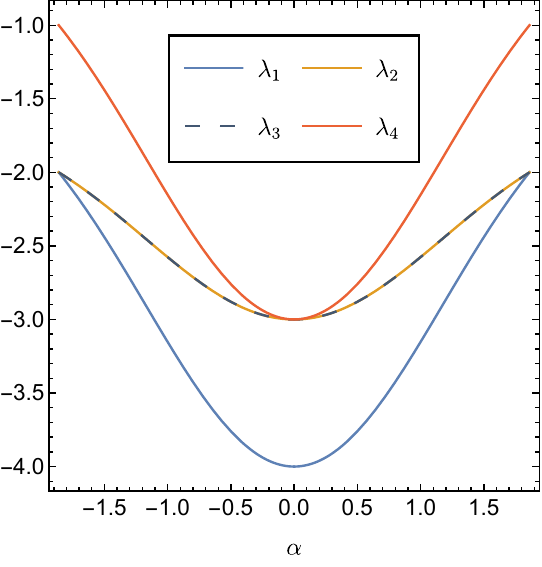}
\end{minipage} \hfill
\begin{minipage}[b]{.5\textwidth}
\includegraphics[width=\textwidth]{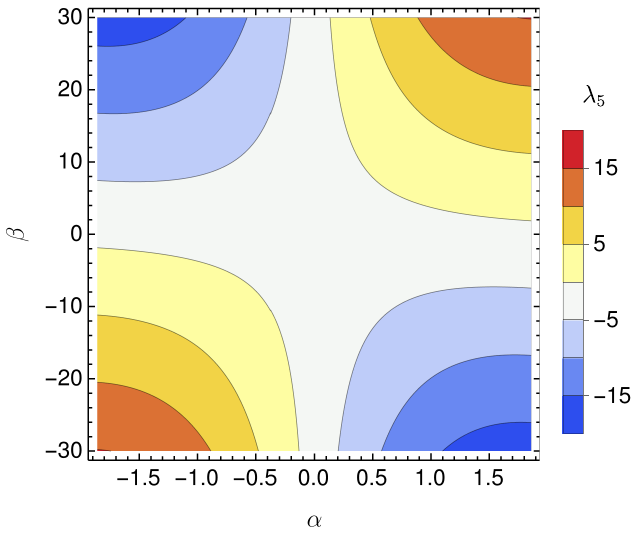}
\end{minipage}
\caption{Eigenvalues of the Jacobian matrix evaluated at ({DE-I}). (Left) Note that $\lambda_{1, 2, 3, 4} < 0$ in the whole region where ({DE-I}) is an accelerated solution, i.e., $w_{\text{eff}} < -1/3$. (Right) The eigenvalue $\lambda_5$ is negative in the regions given in Eqs.~\eqref{Eq: DE-I Attractor 1} and~\eqref{Eq: DE-I Attractor 2}.}
\label{Fig: Eigenvalues DEI}
\end{figure*}

In this case, the eigenvalues are given by large expressions which we present in Appendix~\ref{App: DEI Eigenvalues}. From those expressions, we see that just one eigenvalue depends on $\alpha$ and $\beta$, while the remaining four eigenvalues depend only on $\alpha$. In the left panel of Figure~\ref{Fig: Eigenvalues DEI}, we see that $\lambda_{1, 2, 3, 4} < 0$ in the region where (\emph{DE-I}) is an accelerated solution. In the right panel, we plot $\lambda_5$ in the same interval for $\alpha$, as in the left-panel, and we choose $\beta \in [-30, 30]$ as a representative region for this parameter. We find that $\lambda_5$ is negative when $\alpha = 0$ or in the regions
\begin{align}
0 < \alpha \quad &\land \quad \beta < -\frac{\alpha}{3} + \frac{2}{3} \sqrt{\frac{36 + \alpha^4}{\alpha^2}}, \label{Eq: DE-I Attractor 1} \\
\alpha < 0 \quad &\land \quad \beta > -\frac{\alpha}{3} - \frac{2}{3} \sqrt{\frac{36 + \alpha^4}{\alpha^2}}. \label{Eq: DE-I Attractor 2}
\end{align}
Therefore, (\emph{DE-I}) is an attractor in these regions. By looking at Eqs.~\eqref{Eq: x prime}-\eqref{Eq: Omega prime}, we note that the autonomous system is invariant under the transformation $\{x \rightarrow -x, \alpha \rightarrow -\alpha, \beta \rightarrow -\beta \}$. This symmetry is reflected in Figure~\ref{Fig: Eigenvalues DEI} and in the regions in Eqs.~\eqref{Eq: DE-I Attractor 1} and~\eqref{Eq: DE-I Attractor 2}.

\subsection{Numerical Fixed points}

As shown in the last section, the set of algebraic equations in Eqs.~\eqref{Eq: x prime polynomial}-\eqref{Eq: Omega_r prime polynomial} does not have analytical solutions when $\Sigma \neq 0$. Therefore, our numerical setup can be useful for further analysis of the model. 

Despite early anisotropies being expected to be insignificant given the homogeneity of the CMB~\cite{Planck:2019evm}, late-time anisotropies sourced by a dark energy component are not discarded by observations~\cite{Campanelli:2010zx, Amirhashchi:2018nxl}. In the last section, we analytically found isotropic radiation, matter, and dark energy dominated points (\emph{R}), (\emph{M}), and (\emph{DE-I}). Then, our numerical search will be focused on anisotropic accelerated solutions. Moreover, it is possible to find initial conditions, in the deep radiation epoch, ensuring a proper radiation era followed by a standard matter era, given that the point (\emph{M}) is ``more stable'' than the point (\emph{R}). Hence, in the following, we will neglect the radiation component, i.e., $\Omega_r = 0$. This assumption will simplify our numerical treatment.

Now, we proceed with the implementation of the numerical setup explained in Sec.\ref{Sec: Dynamical systems: numerical approach}. Firstly, we have to choose a specific window parameter where the stochastic search will be performed. Having in mind that $|\alpha| < 1.86121$ for (\emph{DE-I}) being an accelerated solution, we choose $\alpha \in [-30,30]$ and $\beta \in [0,30]$; more precisely,
\begin{equation}
\{\alpha, \beta\} \in [-30,30]\times[0,30].
\label{Eq: search region}
\end{equation} 
Note that unlike $\alpha$, $\beta$ takes on nonnegative values since the autonomous system enjoys the symmetry $\{x \rightarrow -x, \alpha \rightarrow -\alpha, \beta \rightarrow -\beta \}$. Secondly, we impose the following physically motivated conditions
\begin{equation}
0 \leq \Omega_m \leq 1, \quad 0 \leq \Omega_\text{DE} \leq1.
\label{Eq: Physical Constraint}
\end{equation}
Therefore, any point in the parameter space is cataloged as a ``non-viable solution'' if all the found fixed points do not obey these physical constraints. If at least one of the corresponding fixed points meets all the conditions, the point is cataloged as a ``viable solution''. Now, we generate $N \sim 10^4$ random points in the region specified in Eq.~\eqref{Eq: search region}. For each of this points, the set of algebraic equations in Eqs.~\eqref{Eq: x prime polynomial}-\eqref{Eq: Sigma prime polynomial} is solved, considering $\Omega_r = 0$, using the \texttt{NSolve} command of \texttt{Mathematica}.\footnote{Note that our numerical scheme is not tied to the \texttt{Mathematica} software or to the \texttt{NSolve} command. This step, namely, the solution of the algebraic equations, can be done using any other coding language or algebraic system solver.} The code for solving $\sim 10^4$ algebraic systems takes approximately 45 minutes running on 136 parallelized mixed physical cores. This code is available on \href{https://github.com/sagaser/Numerical-dynamical-analysys}{GitHub}.\footnote{\href{https://github.com/sagaser/Numerical-dynamical-analysys}{https://github.com/sagaser/Numerical-dynamical-analysys}}

\begin{figure}[h!]
\centering
\begin{tikzpicture}
 \node at (0,0) {\includegraphics[width=10cm]{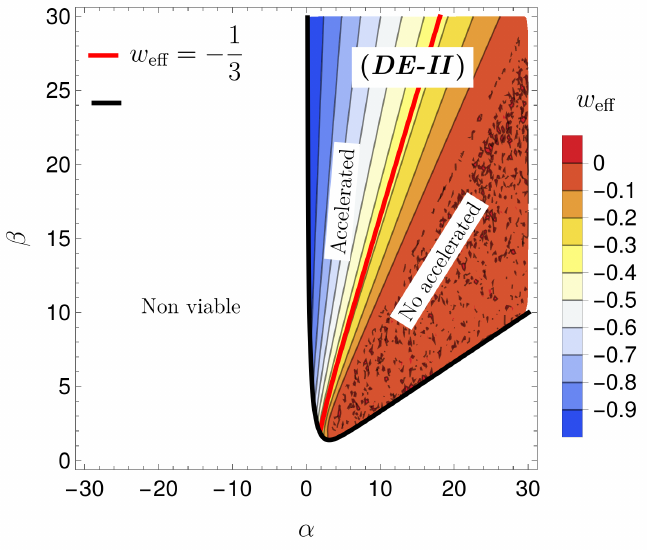}};
 \node at (-2.1,2.7) {Eq.~\eqref{Eq: Curve Viable Non-Viable}};
\end{tikzpicture}
\caption{Region in the parameter space where the stochastic search was performed. This region is divided in ``non-viable'' solutions and ({DE-II}). Points in ({DE-II}) match the physical constraints in Eq.~\eqref{Eq: Physical Constraint}. These regions are separated by the black curve whose expression is given in Eq.~\eqref{Eq: Curve Viable Non-Viable}. The red line represents the points for which $w_{\text{eff}} = -1/3$, which further divides ({DE-II}) into two regions: accelerated solutions and non-accelerated solutions.}
\label{Fig: Region Accelerated Solutions}
\end{figure}

The result of code runs are depicted in Figure~\ref{Fig: Region Accelerated Solutions}, which we explain in the following. If all the solutions for a given point in the region in Eq.~\eqref{Eq: search region} do not satisfy the physical conditions in Eq.~\eqref{Eq: Physical Constraint}, this point in the parameter space is allocated in the white region called ``non viable''. The points filling this region are cosmologically irrelevant. In turn, if at least one of the solutions for a given point matches all the physical constraints, this point is allocated in the ``viable region'' which is called (\emph{DE-II}). We found that these regions, non viable and (\emph{DE-II}), are separated by the curve
\begin{equation}
\beta = - \frac{\alpha}{3} + \frac{2}{3} \sqrt{\frac{36 + \alpha^4}{\alpha^2}},
\label{Eq: Curve Viable Non-Viable}
\end{equation}
which is the same curve dividing the regions where (\emph{DE-I}) is an accelerated solution [see Eqs.~\eqref{Eq: DE-I Attractor 1} and~\eqref{Eq: DE-I Attractor 2}]. Region (\emph{DE-II}) can be further divided into ``accelerated'' and ``non accelerated'' solutions by the line $w_{\text{eff}}(\alpha,\beta) = -1/3$. The color code in Figure~\ref{Fig: Region Accelerated Solutions} shows that it is possible to get $w_\text{eff} \sim -1$ for some values of $\alpha$ and $\beta$. The existence of (\emph{DE-I}) and the ``accelerated'' region in (\emph{DE-II}) ensures an expanding universe at an accelerated rate, which can be isotropic or anisotropic.
 
\begin{figure}[t!]
\centering
\includegraphics[width=10cm]{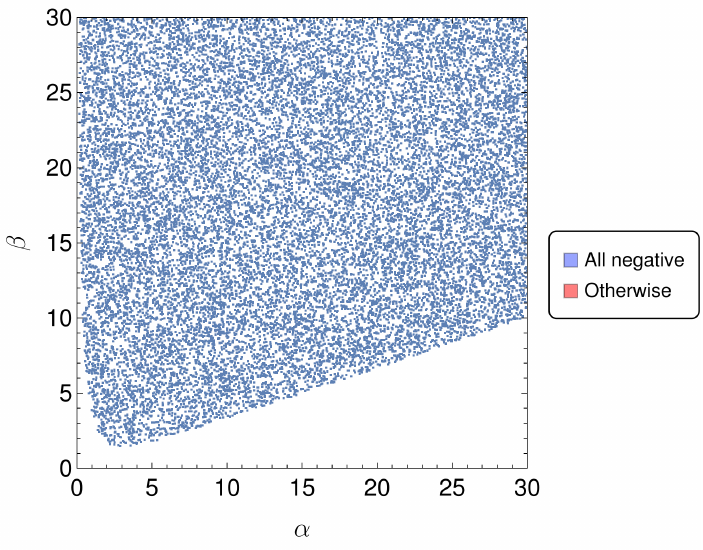}
\caption{Determination of the stability of the numerical fixed points conforming the region ({DE-II}). A blue point represents an attractor point, which means that all the corresponding eigenvalues of at least one of the available Jacobian matrices are negative. Red points represent non-attractor points, namely, saddles or sources. Interestingly, there are no red points, meaning that ({DE-II}) is an attractor in its own region of existence.} 
\label{Fig: Eigenvalues DE-II}
\end{figure}

\begin{figure}[t!]
\centering
\includegraphics[width = 8cm]{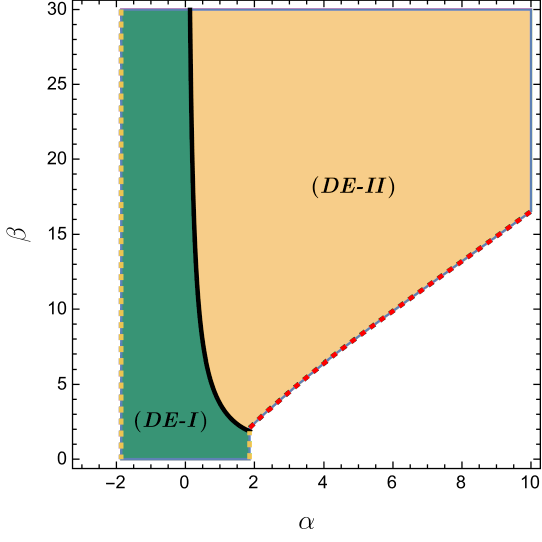}
\caption{Regions where ({DE-I}) and ({DE-II}) are attractors and accelerated solutions. The black line is the bifurcation curve given in Eq.~\eqref{Eq: Curve Viable Non-Viable}.} 
\label{Fig: DEI and DEII}
\end{figure}

The stability of (\emph{DE-II}) can be determined following the step 4 in Section~\ref{Sec: Dynamical systems: numerical approach}. We compute all the available Jacobian matrices (and their eigenvalues) for each of the points in (\emph{DE-II}). We expect that at least a portion of the ``accelerated'' region in (\emph{DE-II}) could be an attractor of the system. As mentioned in Section~\ref{Sec: Dynamical systems: numerical approach}, a point in the parameter space yield attractor solutions if all the eigenvalues of at least one of the corresponding Jacobian matrices are negative. As shown in Figure~\ref{Fig: Eigenvalues DE-II}, this attractor condition is obeyed by all the points in (\emph{DE-II}). Therefore, (\emph{DE-II}) is an attractor inside its own region of existence. We summarize the possible attractors of the system in Figure~\ref{Fig: DEI and DEII}. In this figure, we find out that Eq.~\eqref{Eq: Curve Viable Non-Viable} is the bifurcation curve that separates both attractors.

\section{Numerical Integration of the Autonomous Set}
\label{Sec: Numerical Integration of the Autonomous Set}

In order to check our claims about the asymptotic behavior of the system, in this section we numerically solve the full autonomous system in Eqs.~\eqref{Eq: x prime}-\eqref{Eq: Omega prime} for specific values of $\alpha$ and $\beta$. The initial conditions are set at a very high redshift, $z_r = 6.57 \times 10^7$, ensuring that cosmological trajectories start in the deep radiation epoch. Moreover, we assume that possible anisotropies can be sourced only at late-times when the contribution of DE is significant to the energy budget, hence $\Sigma_i = 0$. For the remaining variables we choose
\begin{equation}
x_i = 10^{-25}, \quad z_i = 10^{-15}, \quad \Omega_{ri} = 0.99995. 
\label{Eq: Initial Conditions}
\end{equation}
The value for $y_i$ is constrained by Eq.~\eqref{Eq: 2º Friedman}. For $0 \leq \Omega_{mi} \leq 1$, $y_i$ have to satisfy $|y_i| \leq 0.00707$, so we choose 
\begin{equation}
 y_i = 2.01 \times 10^{-14},
\end{equation}
such that $\Omega_{mi} = 5.0 \times 10^{-5}$. We would like to point out that the values of $x_i$, $y_i$, and $z_i$ are chosen so small to avoid possible large contributions of dark energy during the radiation dominated epoch.

\subsection{Isotropic Dark Energy Attractor}
\label{Sec: Isotropic Dark Energy Attractor}
Firstly, we choose $\alpha$ and $\beta$ such that (\emph{DE-I}) is the attractor of the system:
\begin{equation}
\alpha = 0.5, \quad \beta = 0.1.
\end{equation}

\begin{figure}[t!]
\centering
\includegraphics[width = 8cm]{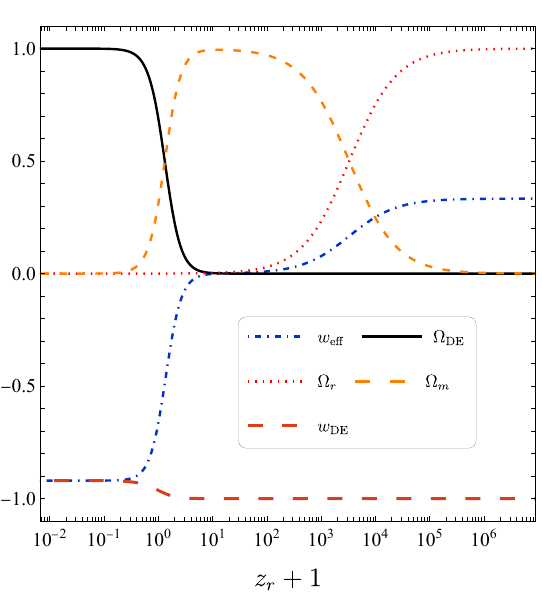}
\caption{Evolution of the density parameters, $w_\text{eff}$ and $w_\text{DE}$ during the whole expansion history. The initial conditions, Eq.~\eqref{Eq: Initial Conditions}, were chosen in the deep radiation era at the redshift $z_r = 6.57 \times 10^7$. The universe passes through radiation dominance at early times (red dotted line), followed by a matter dominance (light brown dashed line), and ends in the DE dominance (black solid line) characterized by $w_{\text{eff}} = w_\text{DE} \approx -0.92$ (blue dot-dashed line and tangelo dashed line, respectively).} 
 \label{Fig: Isotropic Evolution}
\end{figure}

\begin{figure}[t!]
\centering
\includegraphics[width = 8cm]{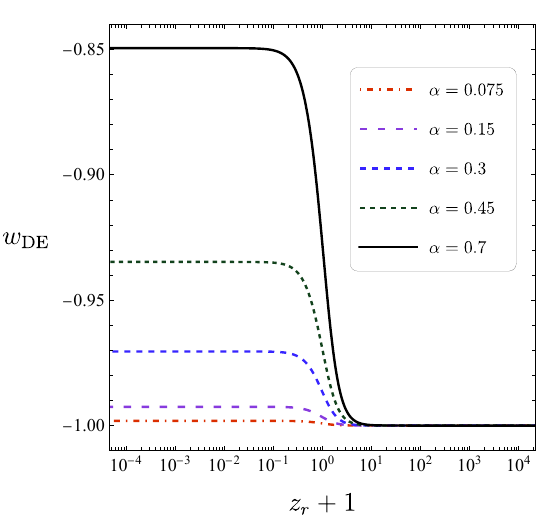}
 \caption{Time evolution of $w_{\text{DE}}$ for different values of the parameter $\alpha$, while $\beta = -90$. The initial conditions are the same given in Eq.~\eqref{Eq: Initial Conditions}.} 
\label{Fig: DE-I Attractor}
\end{figure}

In Figure~\ref{Fig: Isotropic Evolution}, we plot the evolution of the density parameters, the effective equation of state of the universe, and the equation of state of dark energy (DE) as functions of the redshift $z_r$. The early universe ($z_r > 10^6$) is dominated by radiation, as indicated by the red dotted line. At $z_r \approx 3200$, the radiation-matter equality occurs, where $\Omega_m \simeq \Omega_r$. Following this transition, the DE contribution is $\Omega_{\text{DE}} \approx 9.89 \times 10^{-10}$, which satisfies the Big Bang Nucleosynthesis (BBN) constraint, $\Omega_{\text{DE}} < 0.045$ at $z_r = 1200$~\cite{Bean:2001wt}. From this transition until $z_r \approx 0.3$, the universe is matter-dominated, as shown by the light brown dashed line. At $z_r = 50$, the DE contribution to the energy budget is $\Omega_{\text{DE}} = 1.74 \times 10^{-5}$, consistent with the Cosmic Microwave Background (CMB) constraint, $\Omega_{\text{DE}} < 0.02$ at this redshift~\cite{Planck:2018vyg}. Subsequently, the Universe transitions to a DE-dominated phase, represented by the black solid line, during which the expansion accelerates as $w_\text{eff} < -1/3$ (blue dot-dashed line). The behavior of the DE equation of state matches theoretical expectations: it remains $w_{\text{DE}} = -1$ during the radiation and matter-dominated epochs but takes on a different value during the DE-dominated era. This value depends solely on $\alpha$ and, in this case, is given by $w_{\text{eff}} = w_{\text{DE}} \simeq -0.92$. Additionally, in Figure~\ref{Fig: DE-I Attractor}, we plot $w_\text{DE}$ for several values of $\alpha$ while keeping $\beta$ fixed, demonstrating that $w_\text{DE} \rightarrow -1$ as $\alpha \rightarrow 0$, consistent with theoretical predictions.\\

We want to point out that we verified that $w_\text{DE}$ does not depend on $\beta$ and also that $\Sigma = 0$ during the whole expansion history. This had to be so, since (\emph{DE-II}) does not exist when (\emph{DE-I}) is an attractor. However, the conversely is not true. In general, when (\emph{DE-II}) is the attractor of the system, (\emph{DE-I}) exist as a saddle, and the Universe could expand isotropically during a brief period of time. 

\subsection{Anisotropic Dark Energy Attractor}

For (\emph{DE-II}) to be the attractor of the system, we choose the following parameters
\begin{equation}
\alpha=0.5, \quad \beta=80,
\end{equation}
and we use the same initial conditions as in Eq.~\eqref{Eq: Initial Conditions}.

\begin{figure}[h!]
\centering
\includegraphics[width = 8cm]{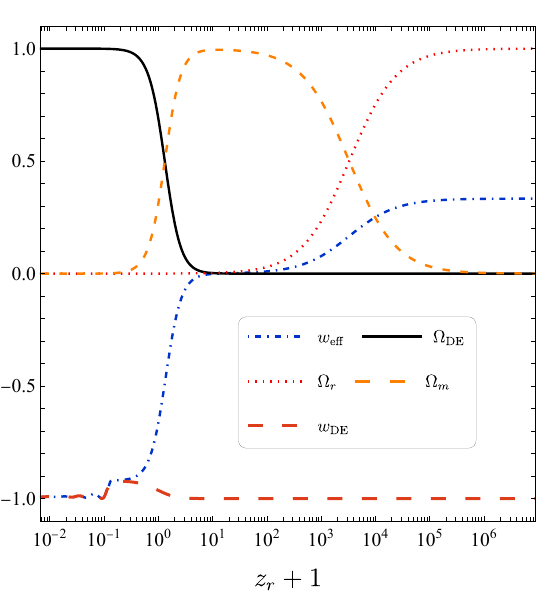}
\caption{Evolution of the density parameters, $w_\text{eff}$ and $w_\text{DE}$ during the whole expansion history for an anisotropic accelerated attractor. The initial conditions are the same as in Figure~\ref{Fig: Isotropic Evolution}. The most appreciable difference with respect to the isotropic accelerated case (shown in Figure~\ref{Fig: Isotropic Evolution}) is that $w_\text{DE} \approx -1$ oscillates at late times.} 
 \label{Fig: Anisotropic Evolution}
\end{figure}

In Figure~\ref{Fig: Anisotropic Evolution}, we plot the evolution of the density parameters, the effective equation of state and the equation of state of the DE as function of the redshift $z_r$. The most prominent difference between this case and the (\emph{DE-I}) attractor case (shown in Figure~\ref{Fig: Isotropic Evolution}) is the oscillatory behavior of $w_{\text{DE}}$ at late times. As mentioned in Ref.~\cite{Orjuela-Quintana:2021zoe}, these oscillations occur when the kinetic term of the scalar field is comparable to the vector density, turning the equation of motion of the scalar field (Eq.~\eqref{Eq: Eq phi} in this case) into an equation describing a damped harmonic oscillator. 

\begin{figure*}[h!]
\centering
\begin{minipage}[b]{.472\textwidth}
\includegraphics[width=\textwidth]{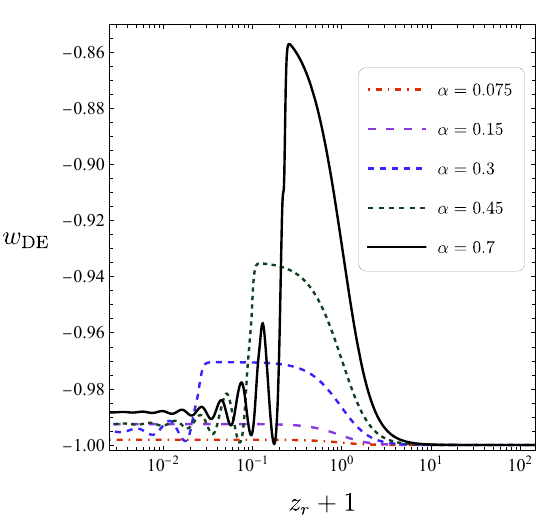}
\end{minipage} \hfill
\begin{minipage}[t]{.445\textwidth}
\includegraphics[width=\textwidth]{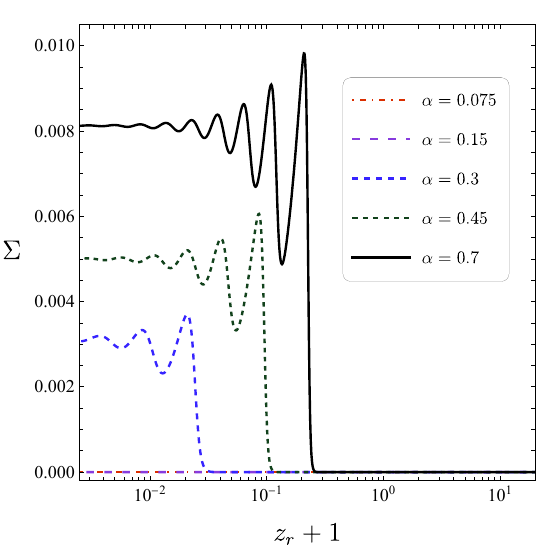}
\end{minipage}
\caption{Late-time evolution ($z_r < 100$) of the equations of state of DE $w_{\text{DE}}$ and the shear $\Sigma$ for different values of the parameter $\alpha$, while $\beta = 80$. The initial conditions are the same given in Eq.~\eqref{Eq: Initial Conditions}. Note that the amplitude of the oscillations grows with $\alpha$. As $w_\text{DE}$ as $\Sigma$ behave as damped oscillators until they reach the asymptotical value predicted by the numerical fixed point ({DE-II}).}
\label{Fig: Dark Energy Alpha}
\end{figure*}

\begin{figure*}[h!]
\centering
\begin{minipage}[b]{.472\textwidth}
\includegraphics[width=\textwidth]{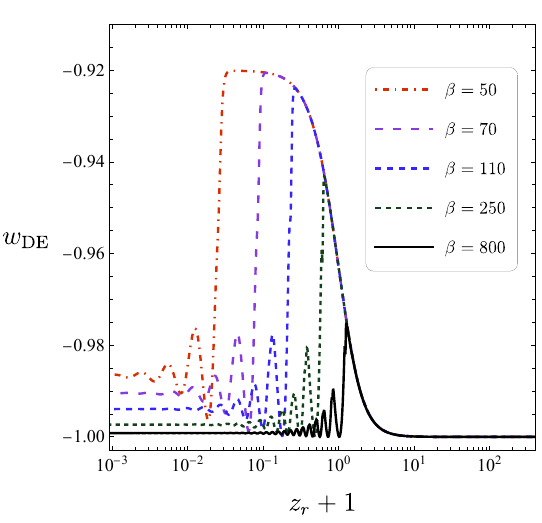}
\end{minipage} \hfill
\begin{minipage}[b]{.445\textwidth}
\includegraphics[width=\textwidth]{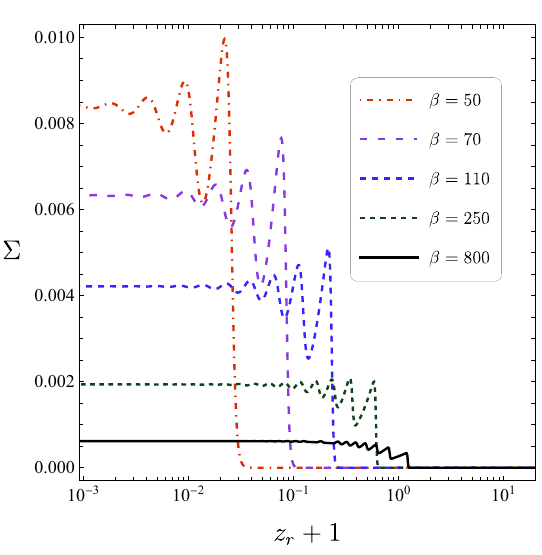}
\end{minipage}
\caption{Late-time evolution ($z_r < 100$) of the equations of state of DE (left) $w_{\text{DE}}$ and the shear $\Sigma$ (right) for different values of the parameter $\beta$, while $\alpha = 0.5$. The initial conditions are the same given in Eq.~\eqref{Eq: Initial Conditions}. Note that the amplitude of the oscillations is greater for smaller $\beta$. As $w_\text{DE}$ as $\Sigma$ oscillate until they reach the asymptotical value predicted by the numerical fixed point ({DE-II}).}
\label{Fig: Dark Energy Beta}
\end{figure*}

In Figure~\ref{Fig: Dark Energy Alpha}, we plot the late-time evolution of $w_\text{DE}$ and $\Sigma$ for fixed $\beta = 80$ and varying the parameter $\alpha$. We observe that both $w_\text{DE}$ and $\Sigma$ oscillate until they stabilize to the value predicted in the numerical fixed point (\emph{DE-II}), as shown in Table~\ref{tab: Predicted values}. Note that the amplitude of these oscillations grows with $\alpha$. We also investigate the behavior of $w_\text{DE}$ and $\Sigma$ when $\alpha$ is fixed and $\beta$ varies, as shown in Figure~\ref{Fig: Dark Energy Beta}. In this case, the amplitude of the oscillations grows while $\beta$ decreases. Note in Figure~\ref{Fig: Dark Energy Alpha} that when $w_\text{DE}$ deviates from $-1$ it grows until a value which mainly depends on $\alpha$. This indicates that the Universe is crossing the point (\emph{DE-I}) which is a saddle. When $\alpha$ is fixed, as in Figure~\ref{Fig: Dark Energy Beta}, the time spent by the Universe crossing (\emph{DE-I}) depends on $\beta$. For larger values of $\beta$, the coupling between the tachyon field and the vector field is stronger, and thus oscillations start earlier. 

\begin{table}[h!]
 \centering
 \caption{Predicted values of $\Sigma$ and $w_{\text{DE}}$ as a function of each $\alpha$ and $\beta$ in the region where the numerical approach was applied and solved with a mean error of $20\%$ in $\Sigma$.}
\begin{tabular}{l|l|l|l}
\hline 
\hline
$\alpha$ & $\beta$ & $\Sigma\times 10^{-3}$ & $w_{\text {DE }}$ \\
\hline 
$0.5$ & $50$ & $5.58 \pm 2.70 $ & $-0.9867$ \\
$0.5$ & $70$ & $4.21 \pm 0.46 $ & $-0.9905$ \\
$0.5$ & $110$ & $2.81 \pm 0.56 $ & $-0.9939$ \\
$0.5$ & $250$ & $1.29 \pm 0.61 $ & $-0.9973$ \\
$0.5$ & $800$ & $41.2 \pm 25.1 $ & $-0.9992$ \\
\hline
$0.075$ & $80$ & $20.8 \pm 0.1 $ & $-0.9988$ \\
$0.15$ & $80$ & $83.3 \pm 0.1 $ & $-0.9975$ \\
$0.3$ & $80$ & $2.08 \pm 0.56 $ & $-0.9950$ \\
$0.45$ & $80$ & $3.33 \pm 0.46 $ & $-0.9925$ \\
$0.7$ & $80$ & $5.40 \pm 2.70 $ & $-0.9884$
\end{tabular}

 \label{tab: Predicted values}
\end{table}

\hspace{-1cm}

In summary, we have shown that the following cosmological trajectories exist:
\begin{center}
(\emph{R}) \ $\rightarrow$ \ (\emph{M}) \ $\rightarrow$ \ (\emph{DE-I})/(\emph{DE-II}),
\end{center}
where (\emph{DE-II}) is a region in the parameter space with no analytical description. At this point, we would like to stress that the analysis presented in this section serves as a check for the asymptotic behavior predicted from the numerical fixed point (\emph{DE-II}). Hence, it establishes the consistency of the numerical scheme proposed in this work, and its usefulness when portions of the parameter space of a cosmological model is unreachable through analytical means.

\section{Conclusions} 

In this chapter, we have put forward a numerical method to explore the parameter space of a cosmological model when no analytical fixed points are available. We applied our method to a specific model of anisotropic dark energy based on the interaction between a scalar tachyon field and a vector field in a Bianchi I background. We have explicitly shown that the anisotropic attractor of the system has no analytical description, given that the degree of the algebraic system from which this point must be computed is greater than four [see Eqs.~\eqref{Eq: x prime polynomial}-\eqref{Eq: Sigma prime polynomial}]. However, when the anisotropy is neglected, the system is reduced to a system of degree 4 and thus analytical solutions exist, which we presented in Eqs.~\eqref{Eq: R fixed point},~\eqref{Eq: Matter dominance FP} and~\eqref{Eq: DE-I dominance FP}. In particular, our method allowed us to find the parameter space of the model where anisotropic accelerated solutions exist as attractors of the system, which we plot in Figure~\ref{Fig: DEI and DEII}. Then, we checked the consistency of the method by numerically solving the full autonomous system in Eqs.~\eqref{Eq: x prime}-\eqref{Eq: Omega prime} for a particular set of initial conditions. As a last remark, we would like to stress on the generality of our method, as explained in Section~\ref{Sec: Dynamical systems: numerical approach}, i.e., in principle, it can be applied to any DE scenario.

\chapter{Challenges to Cosmic Acceleration in Generalized SU(2) Proca Dynamics
\label{Ch: GSU2P}}

\section{Introduction} 
\label{Sec: Introduction}

In Chapter~\ref{Ch: ModernCosmology}, we presented an overview of cosmological studies, emphasizing the $\Lambda$CDM model, illustrating the application of a simple dynamical systems approach to a toy model, and providing a brief introduction to the Generalized SU(2) Proca theory. We demonstrated how even minor modifications to a model can significantly increase its complexity, necessitating advanced methods to derive meaningful insights and reconstruct the cosmic history. Chapter~\ref{Ch: anisotrpic Tachyon field} further developed these concepts by focusing on an anisotropic Tachyon field, showcasing the use of numerical techniques within the dynamical systems framework to analyze its behavior and explore its implications for cosmic evolution.

Building on these foundations, we now extend our analysis to a prominent avenue for modifying gravity: the introduction of additional dynamical fields. Horndeski’s seminal work on scalar-tensor theories established the most general second-order scalar-tensor framework, now widely recognized as Horndeski theory~\cite{Horndeski:1974wa} or generalized Galileon theory~\cite{Deffayet:2011gz,Deffayet:2009wt}. This framework has inspired numerous extensions~\cite{Rodriguez:2017ckc}, including vector-tensor theories such as the generalized Proca (GP) theory~\cite{Tasinato:2014eka,Heisenberg:2014rta,Allys:2015sht,BeltranJimenez:2016rff,Allys:2016jaq,GallegoCadavid:2019zke}, scalar-vector-tensor (SVT) theories~\cite{Heisenberg:2018acv}, and the generalized SU(2) Proca (GSU2P) theory~\cite{GallegoCadavid:2020dho,GallegoCadavid:2022uzn,BeltranJimenez:2016afo,Allys:2016kbq}. These theories enrich gravitational dynamics by incorporating vector fields, scalar fields, or combinations thereof, offering new pathways to address the limitations of $\Lambda$CDM and General Relativity (GR).

Among these extensions, the GSU2P theory stands out as particularly intriguing. By introducing a vector field with a global SU(2) symmetry in the action, this theory provides a natural framework for exploring anisotropic cosmological evolutions and alternative mechanisms for cosmic acceleration. While Horndeski~\cite{Kobayashi:2019hrl,Kreisch:2017uet,Kobayashi:2011nu}, GP~\cite{DeFelice:2020sdq,DeFelice:2016yws,Cardona:2023gzq,Heisenberg:2020xak,Gomez:2020sfz,Gomez:2022okq}, and SVT theories~\cite{Heisenberg:2018vsk,Heisenberg:2018mxx,Heisenberg:2018vti,Cardona:2022lcz,Gonzalez-Espinoza:2023qba} have been extensively studied, the GSU2P theory remains relatively underexplored. Existing research has addressed specific aspects, such as stability issues~\cite{Gomez:2019tbj}, black hole and neutron star solutions~\cite{Martinez:2022wsy,Gomez:2023wei,Martinez:2024gsj}, inflationary scenarios~\cite{Garnica:2021fuu}, and late-time cosmic acceleration~\cite{Rodriguez:2017wkg}. However, a comprehensive analysis of the full cosmological implications of the theory, including its ability to reproduce the expansion history of the Universe, is still lacking.

This work aims to bridge that gap by conducting a detailed investigation of the GSU2P theory in a flat, homogeneous, and isotropic background. Specifically, we employ the dynamical systems approach~\cite{Bahamonde:2017ize} to identify the conditions under which the theory can drive cosmic acceleration, either during the early inflationary phase or the late-time accelerated expansion. Our analysis reveals the presence of a fixed point corresponding to de-Sitter expansion, which may represent a stable or transient state in the evolution of the universe. Additionally, we identify "pseudo-stationary" states at distinct scales in the two-dimensional phase space, corresponding to accelerated expansion and radiation-like behavior.

In the regime of large field values, the GSU2P model predicts a constant-roll evolution~\cite{Motohashi:2017vdc,Motohashi:2014ppa,Motohashi:2019tyj}, which can correspond to either an inflationary phase or late-time cosmic acceleration. Conversely, in the regime of small field values, the system exhibits oscillatory behavior between two pseudo-stationary states, resembling a radiation-like fluid. This behavior may signify either the graceful exit from an inflationary epoch or a precursor phase leading to late-time acceleration. However, as we will demonstrate, the theory encounters significant challenges during the transition between these phases, notably the emergence of a non-physical expansion rate. Although the autonomous system can be regularized, this process reveals both physical singularities, such as undefined observable quantities like the Hubble parameter, and numerical singularities that render the system non-integrable. These issues further cast doubt on the viability of the GSU2P model as a consistent cosmological framework.

\subsection{Stability Conditions and Gravitational Wave-Speed Constraint}

The action defined in Eq.~\eqref{Eq: Action GSU2P} has been meticulously constructed to circumvent Ostrogradski's instability, thereby ensuring the correct number of propagating degrees of freedom~\cite{ErrastiDiez:2019trb} (see, however, Refs.~\cite{ErrastiDiez:2023gme, Janaun:2023nxz}). Nonetheless, for the theory to be considered physically viable, it is imperative that it remains free from other forms of instabilities which could undermine its consistency. Notable pathologies that must be avoided include ghost instabilities and gradient or Laplacian instabilities.

Ghost instabilities manifest when the linearized perturbations exhibit negative kinetic energy terms, resulting in nonphysical behavior when the ghost field interacts with other fields. In contrast, Laplacian instabilities occur when the propagation speed of perturbations is imaginary, leading to the uncontrollable, often exponential, growth of initially small perturbations~\cite{Sbisa:2014pzo}. Furthermore, the detection of gravitational waves (GW) by LIGO~\cite{LIGOScientific:2017vwq} and the subsequent determination of their speed, which has been shown to be equal to the speed of light with astonishing precision~\cite{Liu:2020slm, Baker:2022eiz}, have imposed stringent constraints on several modified gravity theories~\cite{Ezquiaga:2017ekz, Sakstein:2017xjx, Creminelli:2017sry, Kreisch:2017uet, Baker:2017hug, Jana:2018djs}.

In the context of the GSU2P theory, it has been demonstrated that to preclude ghost and Laplacian instabilities from appear and to ensure tensor modes to propagate at the speed of light, the parameters in the action~\eqref{Eq: Action GSU2P} must satisfy the following conditions~\cite{Garnica:2021fuu}:
\begin{align}
 \chi_3 &= 0, &\alpha_4 &= -2 \alpha_1+\frac{7}{20} \alpha_3, \\
 \chi_7 &= 5 \alpha_1+\alpha_3-\frac{1}{2}\chi_4 -3 \chi_6, &\alpha_5 &= \frac{14}{3} \alpha_3 - \frac{20}{3} \alpha_1, \\
 \alpha_2 &= 2 \alpha_3,  &\alpha_6 &= -20 \alpha_1+6 \alpha_3-3 \alpha_5.
\end{align}
Thus, the final form of the GSU2P theory that satisfies these stability requirements is given by:
\begin{align}
\label{Eq: Stable Action}
S &= \int \text{d}^4 x \, \sqrt{-g} \Bigg[ \mathcal{L}_\text{EH} + \mathcal{L}_\text{YM} + \chi_1 \mathcal{L}^1_2 + \chi_2 \mathcal{L}^2_2+ \frac{\chi_4}{m_\text{P}^2} \left( \mathcal{L}_2^4 - \frac{\mathcal{L}_2^7}{2} \right) \\
 &\left. + \frac{\chi_5}{m_\text{P}^2} \mathcal{L}_2^5 + \frac{\chi_6}{m_\text{P}^2} \left( \mathcal{L}_2^6 - 3 \mathcal{L}_2^7 \right) \nonumber \right. \left.+ \frac{\alpha_1}{m_\text{P}^2} \left(\mathcal{L}_{4,2}^1- 2 \mathcal{L}_{4,2}^4 - \frac{20}{3} \mathcal{L}_{4,2}^5 + 5 \mathcal{L}_2^7 \right) \nonumber \right.\\
 & + \frac{\alpha_3}{m_\text{P}^2} \left(2 \mathcal{L}_{4,2}^2 + \mathcal{L}_{4,2}^3 + \frac{7}{20} \mathcal{L}_{4,2}^4 + \frac{14}{3} \mathcal{L}_{4,2}^5 - 8 \mathcal{L}_{4,2}^6 + \mathcal{L}_2^7 \right) \Bigg]. \nonumber
\end{align}
In the following sections, we will focus on the cosmological dynamics encoded in this final action, exploring its implications for the evolution of the universe.

\subsection{Homogeneous and Isotropic Configuration}

Observational evidence has pointed out that the universe is largely homogeneous and isotropic on cosmological scales~\cite{Planck:2018vyg}.\footnote{There exists controversy around this~\cite{Aluri:2022hzs}. However, we will neglect a possible anisotropic expansion as a first approximation.} This allows one to describe the geometry of the universe by the flat (FLRW) metric, Eq.~\eqref{eq:FLRWMetric}.

The symmetries of this metric impose significant constraints on the dynamics of cosmological fields. For example, vector fields inherently break rotational invariance, potentially introducing substantial anisotropy into the dynamics of the expansion of the Universe, as discussed for the anisotropic tachyon field in Chapter~\ref{Ch: anisotrpic Tachyon field}. In theories involving vector fields without internal global symmetry in the action, this challenge can be addressed by either introducing three identical and orthogonal vector fields, known as the \emph{cosmic triad}~\cite{Armendariz-Picon:2004say, Emami:2016ldl}, or by restricting the analysis to time-like vector fields (see, e.g., Refs.~\cite{Koivisto:2008xf, DeFelice:2016yws}).

In contrast, within the framework of the GSU2P theory, rotational invariance can be preserved by compensating for spatial rotations of the vector field with internal rotations in the isospin space. The most general configuration of the vector field consistent with spatial isotropy has been demonstrated to be given by~\cite{Witten:1976ck, Forgacs:1979zs, Sivers:1986kq}:
\begin{align}
 B_{0a} &= b_0(t)\hat{\bar{r}}_a, \quad
 B_{ia} = b_1(t)\hat{r}_i\hat{\bar{r}}_a+b_2(t)[\delta_{ia}-\hat{r}_i\hat{\bar{r}}_a]+b_3(t)\epsilon_{ia}{}^k\hat{r}_k,
\end{align}
where $b_0(t)$, $b_1(t)$, $b_2(t)$, and $b_3(t)$ are arbitrary functions of time only, $\hat{\bar{r}}$ is the unit vector in the isospin space pointing in the direction of $\vec{B}_i$, and $\hat{r}$ is the unit vector in physical space pointing in the direction of $\vec{B}_a$. The cosmic triad can be naturally accommodated within this general configuration by assuming $b_0(t) = b_3(t) = 0$ and $b_1(t) = b_2(t)$, such that:
\begin{equation}
\label{Eq: triad cosmic}
 B_{0a}(t)=0, \qquad B_{ia}(t) = a(t)\psi(t)\delta_{ia},
\end{equation}
where $b_2 (t) \equiv a(t) \psi(t)$, being $\psi(t)$ the norm of the physical 3D vector fields. 

\subsection{Cosmological Dynamics in the GSU2P Theory}

The cosmological dynamics encoded in the action~\eqref{Eq: Stable Action} can be revealed through the application of the variational principle. By varying this action with respect to the metric, we derive the gravitational field equations:
\begin{equation}
 m_\text{P}^2 G_{\mu \nu} = T_{\mu\nu}^{(B)} + T_{\mu\nu}^{(m)},
\end{equation}
where $T_{\mu\nu}^{(B)}$ represents the energy-momentum tensor containing contributions from the SU(2) vector field $B^a_{\ \mu}$, and $T_{\mu\nu}^{(m)}$ is the energy-momentum tensor for the rest of matter fluids in the cosmic budget.

Due to its length, we refrain from presenting the full form of $T_{\mu\nu}^{(B)}$ here.\footnote{The full calculation is available in a \texttt{Mathematica} notebook accessible on the \texttt{GitHub} repository: \href{https://github.com/sagaser/GSU2P}{sagaser/GSU2P}. See also Ref.~\cite{Martinez:2022wsy}.} Substituting the FLRW metric from Eq.~\eqref{eq:FLRWMetric} and the cosmic triad configuration from Eq.~\eqref{Eq: triad cosmic} into the gravitational field equations yields the Friedman equations:
\begin{align}
 3m_{\text{P}}^2H^2 &= \rho_{B} + \rho_m, \quad
 -2m_{\text{P}}^2\dot{H} = p_{B} + \rho_{B} + \rho_m, \label{Eq: Friedman GSU2P} 
\end{align}
where the density $\rho_B$ and pressure $p_B$ considering contributions from the vector field are given by:
\begin{align}
 \rho_B &\equiv \left(\dot{\psi} + H \psi \right)^2 \left[ \frac{3}{2} - 9 c_2 \frac{\psi^2}{m_\text{P}^2} \right] + \frac{3}{2} \hat{g}^2\psi^4 \label{Eq: Density} + 6H (c_1 - c_2) \frac{\psi^3 \dot{\psi}}{m_\text{P}^2}, \\
 p_B &\equiv \left(\dot{\psi} + H \psi \right)^2\left[\frac{1}{2} + 3 c_2 \frac{\psi^2}{m_\text{P}^2}\right] + \frac{1}{2} \hat{g}^2\psi^4 \label{Eq: Pressure} + 6 \frac{\psi^2}{m_\text{P}^2} (c_2 - c_1)\left\{\dot{\psi}^2-\psi^2\left(H^2+\frac{\dot{H}}{3}\right)+\frac{1}{3}\psi\ddot{\psi}\right\}.
\end{align}
In these expressions, an over-dot denotes differentiation with respect to cosmic time. The new constants $c_1$ and $c_2$ are defined through:
\begin{equation}
\label{Eq: New Constants}
 \alpha_3 \equiv \alpha_1 + \frac{1}{20}(c_2 - c_1), \quad \chi_5 \equiv - 2 \alpha_1 + \frac{1}{10}(c_1 + 9c_2),
\end{equation}
and we have defined a generalized version of the SU(2) coupling constant as:
\begin{equation}
\label{Eq: Generalized Charge}
 \hat{g}^2 \equiv \tilde{g}^2 - 6\chi_1 - 2\chi_2.
\end{equation}
Finally, varying the action in Eq.~\eqref{Eq: Stable Action} with respect to $B^{a}_{\ \mu}$, and substituting the homogeneous and isotropic configurations for the metric and the field, yields the equation of motion for the sole dynamical degree of freedom representing the vector field:
\begin{align}
 0 &= \ddot{\psi} + 3H \dot{\psi} + \psi\left(2 H^2 + \dot{H} - 6 c_2 \frac{\dot{\psi}^2}{m_\text{P}^2}\right) -6 c_2 \frac{\psi^2}{m_\text{P}^2} \left(\ddot{\psi} + 3 H \dot{\psi}\right)\nonumber \\ 
 &+ 2 \psi^3\left[\hat{g}^2+3\left(c_1-2 c_2\right) \frac{H^2}{m_\text{P}^2}+\left(c_1-4 c_2\right) \frac{\dot{H}}{m_\text{P}^2} \right]. \label{Eq: Vector EoM}
\end{align}
In the subsequent section, we examine the dynamics of the universe's accelerated expansion as influenced by the vector field.

\section{Accelerated Expansion Driven by the SU(2) Vector Field}
\label{Sec: Accelerated Expansion}

According to the current cosmological paradigm, the universe has experienced two phases of accelerated expansion: an early inflationary phase preceding the radiation-dominated epoch and the present accelerated expansion, likely driven by dark energy. To gain insight into the system's asymptotic behavior, we initially neglect the matter sector, thereby isolating the dynamics of the vector field. This allows us to determine the conditions under which the vector field can induce accelerated expansion, whether in the primordial or late-time phases.

\subsection{Autonomous System}

To identify the conditions under which the vector field drives accelerated expansion, it is necessary to determine the parameter space where such solutions exist and assess their stability properties. The asymptotic behaviour of the model, encoded in its fixed points~\cite{Bahamonde:2017ize}, provides valuable insights. To facilitate this analysis, we reformulate the dynamical equations using the following dimensionless variables:
\begin{equation}
 x \equiv \frac{\dot{\psi}}{\sqrt{2} m_{\text{P}}H}, \quad y \equiv \frac{\psi}{\sqrt{2} m_{\text{P}}}, \quad z \equiv \sqrt{\frac{\hat{g}}{2m_\text{P} H}}\psi. \label{Eq: Dynamical System}
\end{equation}
Neglecting $\rho_m$, the first Friedman equation, left hand in Eq.~\eqref{Eq: Friedman GSU2P} simplifies to the following constraint:\footnote{Since the analysis spans the inflationary epoch in the distant past and the dark energy-dominated future, the effects of radiation are negligible during both periods.}
\begin{equation}
\label{Eq: Friedman Constraint}
 1 = (x + y)^2(1 - 12 c_2 y^2) + 8 (c_1 - c_2) x y^3 + 2 z^4, 
\end{equation}
which allows us to eliminate the variable $z$ from the dynamical system, expressing it in terms of the variables $x$ and $y$. Here, $x$ and $y$ unambiguously represent the speed and the magnitude of the vector field, respectively.

In terms of these variables, the evolution equations of the model reduce to an autonomous system governed by the following set of first-order differential equations:
\begin{align}
\label{Eq: Autonomous set}
x' &= \frac{p}{\sqrt{2}} + x\epsilon, \quad
y' = x,
\end{align}
where the prime denotes differentiation with respect to the number of $e$-folds, $N$, defined as $\text{d}N \equiv H \text{d}t$. The variables $p$ and $\epsilon$ are defined as:
\begin{align}
 p\equiv\frac{\Ddot{\psi}}{m_{\text{P}}H},\quad\epsilon\equiv-\frac{\dot{H}}{H^2},
\end{align}
with $p$ obtained from the vector field equation of motion in Eq.~\eqref{Eq: Vector EoM}, and $\epsilon$ from the second Friedman equation, right hand in Eq.~\eqref{Eq: Friedman GSU2P}, which in terms of the new variables read: 
\begin{align}
 \epsilon &= 2 + 12c_1 y^4 - 4y^3(c_1 - c_2)\left( \frac{p}{\sqrt{2}} + \epsilon y \right) - 4(c_1 - 7c_2)xy^3 - 12 (c_1 - 2c_2)x^2 y^2, \label{Eq: epsilon}\\
 \frac{p}{\sqrt{2}} &= 2 y^2 \left(2 x (4 c_1-7 c_2)+3 \sqrt{2} c_2 p\right) + \frac{2 \left(x^2-1\right)}{y} + x + y\left(\epsilon -12 c_2 x^2\right) + 4 y^3 (c_1 \epsilon -3 c_1-4 c_2 \epsilon) \label{Eq: p}.
\end{align}
In these expressions, notice that $\epsilon$ and $p$ are linearly coupled, allowing them to be expressed solely in terms of the variables $x$ and $y$.

\subsection{Fixed Points as Accelerated Solutions}

Finding the fixed points by solving the algebraic equations $x^\prime = y^\prime = 0$ in Eq.~\eqref{Eq: Autonomous set} leads to:
\begin{align}
A_\pm &= \left\{x \to 0,\, y \to \pm \sqrt[4]{-\frac{1}{6c_1}} \right\}, \\
B_\pm &= \left\{x \to 0,\, y \to \pm \frac{1}{2}\sqrt{\frac{1}{6c_2} - \frac{\sqrt{1 - 48c_2}}{6c_2}}\right\}, \\ 
C_\pm &= \left\{x \to 0,\, y \to \pm \frac{1}{2}\sqrt{\frac{1}{6c_2} + \frac{\sqrt{1 - 48c_2}}{6c_2}}\right\}. 
\end{align}

From these expressions, which only depend on the constants $c_1$ and $c_2$, we observe that the fixed points $A_\pm$, $B_\pm$, and $C_\pm$ assume real values under the following conditions:
\begin{align}
 A_\pm \in \mathbb{R}&: c_1 < 0, \label{Eq: Existence A} \\
 B_\pm \in \mathbb{R}&: c_2 < 0 \quad \lor \quad 0 < c_2 \leq 1/48, \\
 C_\pm \in \mathbb{R}&: 0 < c_2 \leq 1/48. \label{Eq: Existence C} 
\end{align}
For the vector field to drive an accelerated expansion phase, we must ensure that its equation of state $w_B$ satisfies:
\begin{equation}
 w_B \equiv \frac{p_B}{\rho_B} < -\frac{1}{3}.
\end{equation}
At the fixed points $A_\pm$, we find $w_B = -1$, indicating that these solutions correspond to de-Sitter points, where the universe undergoes exponential expansion. Evaluating $w_B$ at the fixed points $B_\pm$ shows that these correspond to accelerated expansion solutions as long as:
\begin{align} 
 B_\pm:\left( \ 1 + 8 c_1 + \sqrt{1 - 48 c_2} > 32 c_2 \right)
 \land \left( 1 + 16 c_1 + \sqrt{1 - 48 c_2} < 16 c_2\right),\label{Eq: Accelerated B} 
\end{align}
while for $C_\pm$, the condition for acceleration is: 
\begin{align}
 C_\pm: \left( \sqrt{1 - 48 c_2} + 16c_2 < 1 + 16 c_1 \right) \land \left( \sqrt{1 - 48 c_2} + 32 c_2 > 1 + 8 c_1\right). \label{Eq: Accelerated C}
\end{align} \\ 

The solutions reveal different cosmological phases associated with different values of the parameters $ c_1 $ and $ c_2 $. Notably, the fixed points $ A_\pm $ represent de-Sitter solutions corresponding to an exponentially expanding universe, while $ B_\pm $ and $ C_\pm $ describe more complex scenarios where the conditions for accelerated expansion are fulfilled under specific parameter ranges, which will be further studied in the next sections. 

\subsection{Stability Analysis}

Having established the conditions under which the fixed points yield accelerated expansion solutions, we now examine their stability properties to understand the system's asymptotic behavior.

As discussed in Section~\ref{sec: dynamicalanalysis}, the stability of fixed points can be determined by analyzing the eigenvalues of the Jacobian matrix, defined as:\footnote{Alternative techniques, such as Lyapunov exponents~\cite{Bahamonde:2017ize} or Malkin's criterion~\cite{Malkin_52}, may be required if linear stability analysis fails.}

\begin{equation}
J_{ij} \equiv \frac{\text{d} f_i}{\text{d}x_j},
\end{equation}
where $f_i \equiv x'_i$ represents the derivatives of the dynamical variables (e.g., $x'$ and $y'$), and $x_j$ represents the corresponding variables (e.g., $x$ and $y$). The sign of the real part of the eigenvalues of $J_{ij}$, evaluated at each fixed point, determines its stability.

The stability of the fixed points is crucial for determining the cosmological dynamics. As mentioned before, the universe has undergone two distinct accelerated phases: primordial inflation and late-time dark energy domination. To resolve the flatness, horizon, and unwanted relics problems of classical cosmology, inflation must last for a minimum of 60 $e$-folds, followed by a reheating phase that transitions the universe into the standard Big Bang evolution~\cite{dodelson2020modern,Ohashi:2013pca,Rodriguez:2017wkg,Guth:1980zm,Koivisto:2008xf,Planck:2018jri,Kobayashi:2011nu}. Given that inflation represents a transient period of accelerated expansion, it is expected, from a dynamical systems perspective, that inflationary solutions correspond to either a source or a saddle point. In contrast, the current accelerated expansion might persist indefinitely, implying that the corresponding solution might be an attractor point. 

Unlike the analysis in Chapter~\ref{Ch: anisotrpic Tachyon field}, the eigenvalues of the Jacobian matrix can be determined analytically in this case. However, the resulting expressions are too lengthy and complex to allow for an analytical description of the stability properties of the fixed points in the parameter space. As a result, we adopt the following strategy: we select a representative sector of the parameter space $\{c_1, c_2\}$ and numerically evaluate the eigenvalues, respecting the existence conditions in Eqs.~\eqref{Eq: Existence A}–\eqref{Eq: Existence C}, as well as the conditions for accelerated expansion given in Eqs.~\eqref{Eq: Accelerated B} and~\eqref{Eq: Accelerated C}. As demonstrated in subsequent sections, this representative region is sufficiently large to capture the essential dynamics of the system at the fixed points.

\begin{figure*}[h!]
\centering 
{\includegraphics[width=0.47\textwidth]{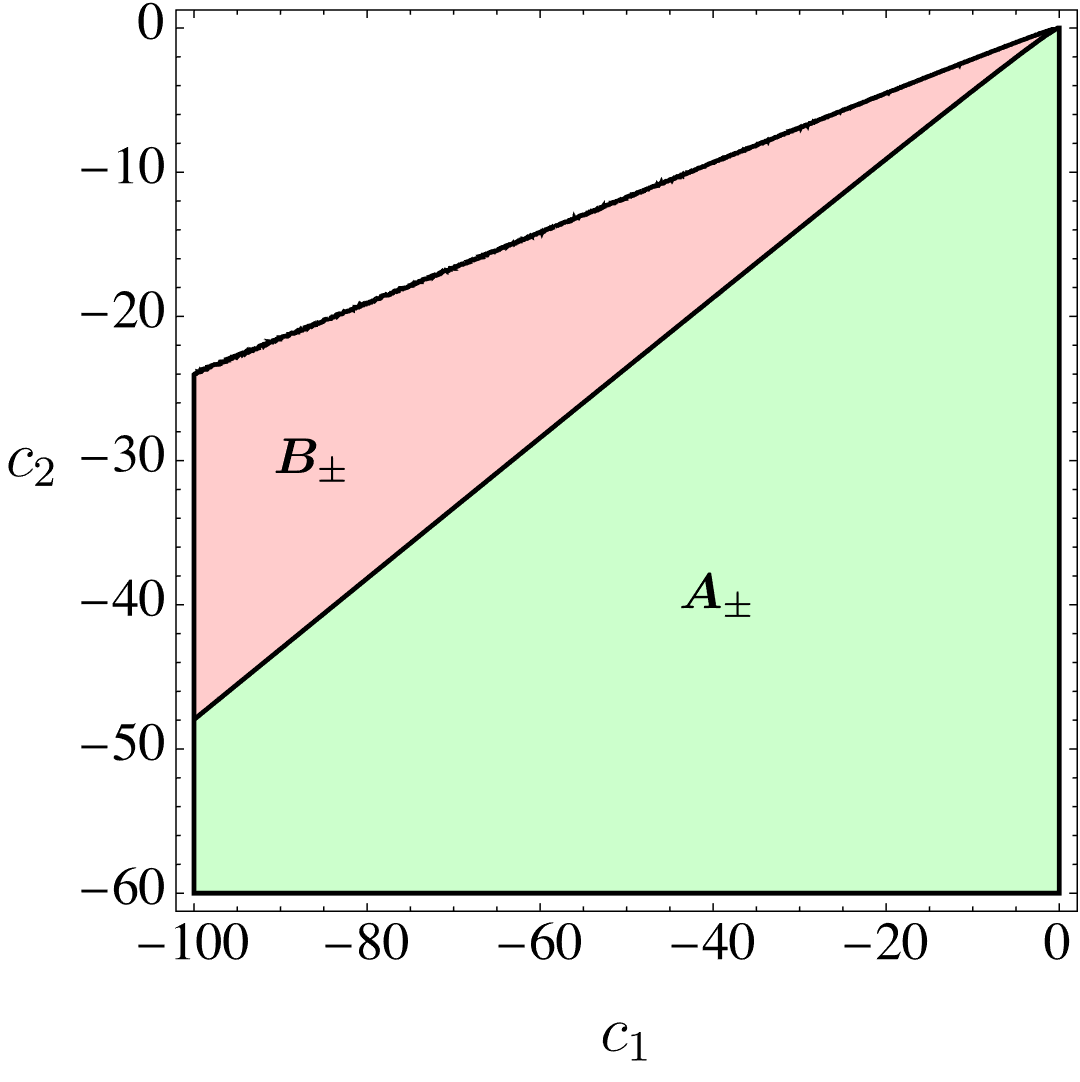}} \hfill
{\includegraphics[width=0.48\textwidth]{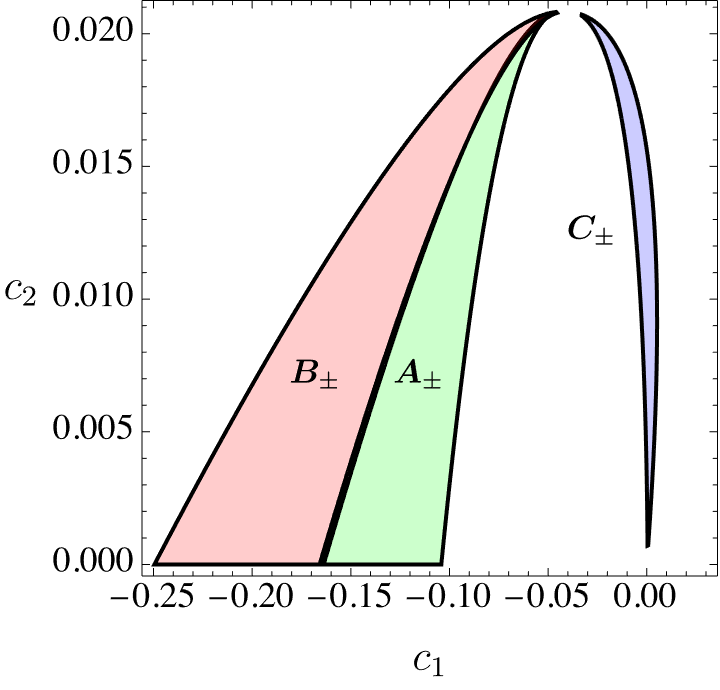}}
\caption{(Left) The regions in the parameter space $\{c_1, c_2\}$ where the fixed points $A_\pm$ (light green) and $B_\pm$ (light red) act as attractors are shown, with the two sets separated by a corresponding bifurcation curve and the attractors representing accelerated expansion. (Right) For small values of the parameters $c_1$ and $c_2$, all fixed points can serve as attractors representing accelerated expansion. However, it is important to note that these regions are considerably smaller compared to those in the left panel.}
\label{Fig: Stability}
\end{figure*}

The results are presented in Figure~\ref{Fig: Stability}. In the left panel, we display the regions where $A_\pm$ and $B_\pm$ are accelerated attractors, visibly separated by a bifurcation curve. In the right panel, we show smaller regions where $A_\pm$ and $B_\pm$ are attractors as well, along with the narrow region where $C_\pm$ serves as an attractor.

\subsection{Cosmological Viability of the Fixed Points}

For a model to be cosmologically viable, it must satisfy several physical conditions. For instance, the Hubble parameter must remain real, as it defines the horizon scale. In this section, we will investigate the ability of the fixed points to describe viable asymptotic cosmological scenarios.

The Hubble parameter, $H$, can be expressed in terms of the dynamical variables as: 
\begin{equation}
\label{Eq: Hubble in Variables}
 \frac{H^2}{m_\text{P}^2} = \hat{g}^2 \left(\frac{y}{z}\right)^4,
\end{equation}
where $z$ can be written in terms of $x$ and $y$ using the Friedman constraint in Eq.~\eqref{Eq: Friedman Constraint}. This formulation allows the energy scale set by $H$ to be tuned by adjusting the generalized coupling parameter $\hat{g}$, as defined in Eq.~\eqref{Eq: Generalized Charge}. This flexibility could be particularly relevant for determining the energy scale of primordial inflation or for adjusting the present-day value $H_0$ to match local observations, potentially alleviating the $H_0$ tension. 

On the other hand, since $y^4 > 0$ always holds, and $\hat{g}^2$ and $z^4$ are real and share the same sign dependence, $H^2$ is positive. This guarantees that $H$ is consistently real-valued once $z^4$ is determined at the fixed point. 

Using Eq.~\eqref{Eq: Hubble in Variables}, we can express the density and pressure of the vector field as: 
\begin{align}
 \frac{\rho_B}{m_\text{P}^4} &= 3 \hat{g}^2 \left(\frac{y}{z}\right)^4, \label{Eq: Rescaled Density} \\
 \frac{p_B}{m_\text{P}^4} &= \hat{g}^2 \left(\frac{y}{z}\right)^4 \Big[ x^2 + 2 x y + \left[1 + \left(36 c_2 - 24 c_1\right) x^2\right] y^2 + 4 \left[\sqrt{2} (c_2 - c_1) p + 6 c_2 x\right] y^3 \notag \\
 &+ 4 \left[ 6 c_1 - 3 c_2 + 2 (c_2 - c_1) \epsilon \right]y^4 + 2z^4 \Big]. \label{Eq: Rescaled Pressure}
\end{align}
Then, defining the re-scaled density as:
\begin{equation}
 \hat{\rho}_B \equiv \frac{\rho_B}{m_\text{P}^4 \hat{g}^2} = 3\left(\frac{y}{z}\right)^4,
\end{equation}
we note that $\rho_B > 0$ for all parameter values, as it depends on $\hat{g}^2$ and $z^4$, which share the same sign at a given fixed point. This further ensures that $H$ is a physical, real-valued quantity there.

Evaluating $\hat{\rho}_B$ at the fixed points $A_\pm$, we find: 
\begin{equation}
 \hat{\rho}_B (A_\pm) = \frac{6}{12 c_2 - 6c_1 - \sqrt{-6c_1}},
\end{equation}
which is positive under the condition: 
\begin{equation}
\label{Eq: Condition for positive density}
 c_1 \leq 0 \quad \land \quad c_2 > \frac{c_1}{2} + \sqrt{-\frac{c_1}{24}}.
\end{equation}
Conversely, $\hat{\rho}_B < 0$ if: 
\begin{equation}
\label{Eq: Condition for negative density}
 c_1 \leq 0 \quad \land \quad c_2 < \frac{c_1}{2} + \sqrt{-\frac{c_1}{24}}.
\end{equation}
A closer inspection of the left panel in Figure~\ref{Fig: Stability} reveals that the bifurcation curve between the attraction regions for points $A_\pm$ and $B_\pm$ is given exactly by:
\begin{equation}
 c_2 = \frac{c_1}{2} + \sqrt{-\frac{c_1}{24}},
\end{equation}
which leads to the conclusion, based on the second condition in Eq.~\eqref{Eq: Condition for negative density}, that $\hat{\rho}_B < 0$ on the attraction region of $A_\pm$ as depicted in the left panel of Figure~\ref{Fig: Stability}. This indicates that $z^4 < 0$, and consequently $\hat{g}^2 < 0$, in this region. Conversely, $\hat{\rho}_B > 0$ outside of this attraction region, leading to $z^4 > 0$ and $\hat{g}^2 > 0$. However, we stress that the physical energy density $\rho_B$ is positive on both regions of the parameter space. We want to clarify that effective negative energy densities are, in principle, possible in modified gravity theories, but they should not dominate the energy content, as this would lead, among other potential issues, to a negative squared Hubble parameter. For example, in a phantom dark energy model within Horndeski’s theory~\cite{Matsumoto:2017qil}, the effective energy density becomes negative only at high redshifts while exhibiting phantom behavior at low redshifts. 

Thus, from a dynamical systems perspective, the points $A_\pm$ represent viable asymptotic states describing a de-Sitter expansion phase of the universe. When acting as attractors, $A_\pm$ could describe the late-time accelerated expansion of the universe, whereas as saddle points, they may characterize the primordial inflationary phase. To validate these interpretations, we will further investigate the dynamics of selected cosmological trajectories in phase space in a subsequent section.


On the other hand, at the fixed points $B_\pm$ and $C_\pm$, we find that the denominator of $\hat{\rho}_B$ vanishes, while the numerator remains constant, causing $\hat{\rho}_B$ to diverge at these points. Consequently, we conclude that these fixed points do not correspond to viable scenarios of accelerated expansion.

Summarizing:
\begin{itemize}
 \item At the points $A_\pm$, the universe undergoes an exponential accelerated expansion, which could last forever or be a transient state.
 \item At the points $B_\pm$ and $C_\pm$, $\rho_B$ goes to infinity, which makes $H$ infinite, and thus cosmologically unreliable.
\end{itemize}

While the fixed points $A_\pm$ represent viable cosmological solutions, it is important to note that the phase space defined by the variables in Eq.~\eqref{Eq: Dynamical System} is not compact. Specifically, both $x$ and $y$ are unbounded, meaning the system may not have a well-defined global attractor~\cite{Coley:2003mj}. This opens up the possibility for other, potentially more complex, asymptotic behaviors that go beyond the fixed-point analysis.

In the absence of a global attractor, the dynamics could exhibit trajectories leading to different regimes or exhibit more complex structures such as limit cycles or chaotic behavior in certain sectors of the parameter space~\cite{Ott:2002}. To fully understand these potential outcomes, a more detailed exploration of the system’s trajectories in extended regions of the phase space is required. This could reveal additional solutions that might correspond to viable cosmological scenarios.

Next, we will delve into these possibilities by exploring the system's behavior at the boundaries and within regions where the fixed-point analysis does not capture the full dynamical complexity.

\section{Pseudo Stationary States}
\label{Sec: Pseudo States}

\subsection{The Stationary Straight Lines\label{SEC: TSSL}}

From the preceding analysis, we conclude that only the fixed points $A_\pm$, at which accelerated expansion occurs, serve as viable asymptotic states of the universe. In contrast, at $B_\pm$ and $C_\pm$, the field density is undetermined and thus these are not viable solutions.

As discussed in Ref.~\cite{Garnica:2021fuu}, the system may admit other asymptotic fates. Specifically, it could evolve towards ``pseudo-stationary states'' at distinct scales of $x$ and $y$. Specifically in Ref.~\cite{Garnica:2021fuu}, it is shown that for large values of $x$ and $y$, the system's behavior is governed by a linear relationship: 
\begin{equation} y = \beta x, 
\end{equation} 
where $\beta$ is a constant describing the slope of the line. In terms of the field $\psi$, this implies that the dynamics follows $\beta \dot{\psi} = H \psi$. During a de-Sitter phase, where $H$ is constant, the field enters a \textit{constant-roll regime}~\cite{Motohashi:2014ppa,Motohashi:2017vdc,Motohashi:2019tyj}, characterized by the equation: 
\begin{equation} 
\ddot{\psi} = \frac{1}{\beta} H \dot{\psi}. 
\end{equation}
In what follows, we will analyze in detail the existence and stability of this solution at large $x$.

To determine the slope of the straight line, we start by assuming $y = \beta x$. From the dynamical equation for $y$ in Eq.~\eqref{Eq: Autonomous set}, we obtain: \begin{equation} 
\label{Eq: Eq for beta} 
\frac{1}{\beta} - \frac{x'}{x} = 0. 
\end{equation} 
Next, using the equation for $x$ from Eq.~\eqref{Eq: Autonomous set}, we find: 
\begin{equation} 
\frac{x'}{x} = \epsilon + \frac{p}{\sqrt{2} x}, 
\end{equation} 
which depends only on $x$ under the assumption $y = \beta x$. In the limit $x \rightarrow \infty$, the dominant term depends solely on the constants $c_1$, $c_2$, and $\beta$. Thus, taking the limit, we obtain: 
\begin{equation} 
\lim_{x \rightarrow \infty} 
\left( \frac{1}{\beta} - \frac{x'}{x} \right) = 0, \end{equation} 
which yields the following cubic equation for $\beta$: \begin{align} 
 0 &= \left(- \frac{4}{3} + \frac{7}{3} \frac{c_2}{c_1}\right) + \left( - \frac{37}{9} + \frac{8}{9} \frac{c_1}{c_2} + \frac{56}{9} \frac{c_2}{c_1} \right) \beta + \left(\frac{4}{3} - \frac{2}{3} \frac{c_1}{c_2} + \frac{7}{3} \frac{c_2}{c_1} \right) \beta^2 + \beta^3.
\end{align} 
The roots of this equation are: 
\begin{align} 
 \beta_0 &= \frac{4}{3} - \frac{7}{3}\frac{c_2}{c_1}, \label{Eq: beta0} \\
 \beta_\pm &= - \frac{4}{3} + \frac{c_1}{3c_2} \pm \frac{\sqrt{(c_1 - c_2)(c_1 - 7c_2)}}{3c_2}. \label{Eq: beta+-} 
\end{align}
Therefore, for large $x$, the system evolves along a straight line defined by $y = \beta_i x$, where $\beta_i$ is one of the three slopes found in Eqs.~\eqref{Eq: beta0} and~\eqref{Eq: beta+-}.

Along these lines, we find that for $y = \beta_0 x$, the equation of state parameter is $w_B = -1$, corresponding to a de-Sitter phase. For the cases where $y = \beta_\pm x$, we find: 
\begin{equation} 
w_B = -\frac{23}{9} + \frac{8}{9} \frac{c_1}{c_2} \pm \frac{8}{9c_2} \sqrt{(c_1 - c_2)(c_1 - 7c_2)}. \label{Eq: Equation of state beta}\end{equation} 
Although all three lines may correspond to accelerated solutions, only the first one describes a de-Sitter phase. Thus, to keep our presentation simple, we will focus on the system's behavior around this line in the following sections and leave the discussion of the other lines to the Appendix~\ref{App: Straight Lines}.

As noted in Ref.~\cite{Garnica:2021fuu}, if $\beta_0$ is negative, the system evolves toward smaller values of $x$ and $y$, and as $y$ approaches zero, the system becomes dominated by the lower powers of the field, represented by the Yang-Mills Lagrangian. This leads to an exit from the accelerated phase into a decelerated expansion, where the vector field behaves like a radiation fluid. Conversely, a positive slope describes a system moving toward larger values of $x$ and $y$, potentially describing a phase of dark energy domination~\cite{Rodriguez:2017wkg}. However, this potential dark energy domination was not explored in sufficient detail in Ref.~\cite{Rodriguez:2017wkg}. Therefore, we will investigate this scenario further here. As a result, we will demonstrate that the inflationary phase, either primordial or late-time, exhibits several shortcomings leading to nonphysical outcomes, thereby challenging the viability of the GSU2P theory as a cosmological model.

\subsection{Existence of the Central Zone}

After analyzing the behavior of the system in the large-values regime of the variables $x$ and $y$, we now turn our attention to the opposite regime in which these variables are small. This regime is cosmologically relevant for two main reasons. First, in a primordial inflationary model, the field driving the accelerated expansion is expected to decay near the end of inflation, giving place to the reheating process. As a result, while trajectories may begin in the large-value regime of $x$ and $y$, they are expected to evolve towards smaller values of these variables as inflation concludes. Second, to accurately reproduce the post-Big Bang expansion history of the universe, the dark energy-dominated epoch must be preceded by a phase of decelerated expansion dominated by pressure-less matter. During this matter-dominated phase, the vector field should be subdominant, leading to trajectories in the phase space evolving with $y$ close to zero.

As pointed out in Ref.~\cite{Garnica:2021fuu}, small values of $y$ can lead to singularities in the system, as the denominator of $x'$ in Eq.~\eqref{Eq: Autonomous set} approaches zero, causing the system to diverge. This denominator is expressed as:
\begin{align}
\label{Eq: Denominator fx}
 D_{x'} &= 16 (c_1 - 7 c_2) (c_1 - c_2) y^7 + 8 (c_1 - c_2) y^5 - 12 c_2 y^3 + y.
\end{align} 
This singularity can occur in a region referred to as ``central zone'' of the phase space in which the system can enter provided that $D_{x'}$ does not vanish. However, the conditions under which this central zone arises, as well as its implications for the dynamics of the system, remain unexplored. In the subsequent analysis, we will investigate this issue in detail.

In non-compact phase spaces, the study of \textit{nullclines}—the geometric curves where $x_i' = 0$, for a given variable independent of the others—can reveal important asymptotic behaviors of the system. Although nullclines do not correspond to true fixed points, they can provide insight into the existence of ``pseudo-stationary states'' within the system~\cite{Ott:2002}. We will show that the existence of the central zone is guaranteed by two of such pseudo-stationary states, which arise from the nullcline of the variable $x$. 

Solving the equation $x' = 0$, when $y \rightarrow 0$, we find the points:
\begin{equation}
 U_\pm = \{\pm 1, 0\}.
\end{equation}
To analyze the stability of $U_\pm$, we compute the eigenvalues, $\lambda_i^\pm$, and eigenvectors, $\nu_i^\pm$, of the Jacobian matrix evaluated at $x = \pm 1$, obtaining the dominant terms for small $y$. For the point $U_+$, we find:
\begin{align}
U_+&: \quad \lambda_1^+ = - \frac{1}{2} y, \quad \lambda_2^+ = 3 + \frac{4}{y}, \nonumber \\
&\nu_1^+ = \{ 0, 1 \}, \quad \nu_2^+ = \left\{3 + \frac{4}{y}, 1 \right\},
\end{align}
while for $U_-$, the corresponding expressions are:
\begin{align}
U_-&: \quad \lambda_1^- = + \frac{1}{2} y, \quad \lambda_2^- = 3 - \frac{4}{y}, \nonumber \\
&\nu_1^- = \{ 0, 1 \}, \quad \nu_2^- = \left\{3 - \frac{4}{y}, 1 \right\},
\end{align}
These results indicate that $U_\pm$ are saddle-like points. A saddle point is characterized by some trajectories in phase space moving towards the fixed point while others move away from it.

Notice that $\nu_1^\pm$ are unitary vectors pointing towards the $y$-direction, whereas the $x$-component of the eigenvectors $\nu_2^\pm$ depends on the value of $y$. Notably, these $x$-components become large as $y$ approaches zero. For $U_\pm$, we must consider two cases: $y \rightarrow 0^+$ (from positive values), and $y \rightarrow 0^-$ (from negative values). When $y \rightarrow 0^+$, we have that $\lambda_2^-$ is negative and $\nu_2^- \approx \{-\infty, 1\}$, while $\lambda_2^+$ is positive and $\nu_2^+ \approx \{\infty, 1\}$. As a result, when $y$ takes on positive values, trajectories in phase space around $U_-$ converge towards it, while those near $U_+$ diverge from it. Conversely, when $y \rightarrow 0^-$, $\lambda_2^-$ becomes positive, and $\nu_2^- \approx \{\infty, 1\}$, causing trajectories to move away from $U_-$. Simultaneously, $\lambda_2^+$ turns negative, and $\nu_2^+ \approx \{-\infty, 1\}$, leading trajectories to converge towards $U_+$. In both cases, the remaining eigenvalue—$\lambda_1^+$ for $U_+$ and $\lambda_1^-$ for $U_-$—changes sign, confirming the saddle-like nature of these pseudo-critical points.

We conclude that when $y \rightarrow 0^+$, trajectories in phase space approach $U_-$ causing $y$ to become negative and move towards $U_+$. Upon reaching $U_+$, $y$ becomes positive again and is subsequently attracted back to $U_-$. This oscillating behavior, characterized by alternating repulsion and attraction between the points $U_\pm$ generates what we call the ``central zone''. This feature will be evidenced numerically.

\subsection{Issues in the Central Zone}

As previously discussed, the bouncing behavior between the points $U_\pm$ gives rise to what we call the central zone. However, it is important to emphasize that these points, $U_\pm$, are saddle-like pseudo-stationary states which arises from the nullcline for the variable $x$ when $y \rightarrow 0$. Consequently, the bouncing behavior does not necessarily lead to a limit cycle—self-sustained oscillations around a point, where the system follows a closed trajectory that repeats periodically, and any small perturbation causes the system to return to this trajectory~\cite{Ott:2002}. In the following, we will delve into these issues by numerically investigate the system's dynamics around the central zone. 

\begin{figure}[h!]
\centering 
\includegraphics[width=0.6\textwidth]{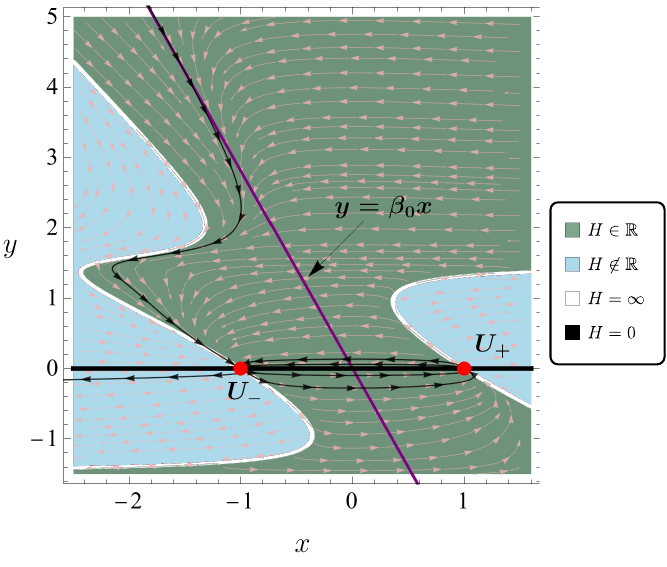}
\caption{Phase space evolution of a specific trajectory (line with arrows) using initial conditions and parameters from Figure 1(a) in Ref.~\cite{Garnica:2021fuu}: $x_i = 5 \times 10^9$, $y_i = 10^{10}$, $\alpha_1 = 1$, $\alpha_3 = 1.0008$, and $\chi_5 = -1.965$, corresponding to $c_1 = 0.0206$ and $c_2 = 0.0366$. The trajectory initially follows the attractor line $y = \beta_0 x$ (purple line), where $z^4 > 0$, fixing $\hat{g}^2 > 0$, and the Hubble parameter takes on real values (green region). As the system evolves, $z^4 \rightarrow 0$ and thus $H^2 \rightarrow \pm \infty$ (white line). Eventually, the trajectory enters the central zone, and oscillates between the points $U_\pm$ (red dots), which lies on a line where $H = 0$ (black line). Then, after some oscillations, it finally escapes from the central zone due to the saddle instability of $U_-$. When exiting from the central zone, $z^4$ flips sign causing the Hubble parameter to become complex (light blue region).}
\label{Fig: Example Phase Space}
\end{figure}

\subsubsection{\textbf{Issues for the Inflationary Scenario}}

To illustrate the system's dynamics within the central zone, we consider the phase space $\{x, y\}$ for a specific parameter set $\{c_1, c_2\}$. We focus on the trajectory represented by the line with arrows depicted in Figure~\ref{Fig: Example Phase Space}. This trajectory shows the system's evolution starting with initial conditions drawn from Figure 1(a) of Ref.~\cite{Garnica:2021fuu}:
\begin{equation}
\label{Eq: ICs Figure 1(a)}
 x_i = 5 \times 10^9, \quad y_i = 10^{10},
\end{equation}
with parameter values:
$$
\alpha_1 = 1, \quad \alpha_3 = 1.0008, \quad \chi_5 = -1.965.
$$
Using the relations from Eq.~\eqref{Eq: New Constants}, these parameters are translated into our system parameters as:
\begin{equation}
 c_1 = 0.0206, \quad c_2 = 0.0366.
\end{equation}
For these parameters, none of the fixed points $A_\pm$, $B_\pm$, or $C_\pm$ are real, leaving only the pseudo-stationary points $U_\pm$ visible. The system's trajectory initially follows the attractor line $y = \beta_0 x$ (depicted by the solid purple line) towards decreasing values of $y$. During this phase, $z^4 > 0$, fixing the sign of $\hat{g}^2$ as positive, and the Hubble parameter takes on real values (green region) [see Eq.~\eqref{Eq: Hubble in Variables}]. As $y$ decreases from positive values, the trajectory approaches $U_-$, $z^4 \rightarrow 0$, and thus $H$ becomes undetermined (white line). Note that the points $U_\pm$ are located at the intersection of the white lines (where $H \rightarrow \infty$) with the black line (where $H = 0$). Therefore, at each crossing through $U_\pm$, the expansion rate becomes undefined, highlighting a critical dynamical issue in the model.

As noted in Ref.~\cite{Garnica:2021fuu}, entering the central zone typically leads to oscillatory behavior, marking the end of the primordial inflationary phase. We argue that this access inherently introduces a dynamical inconsistency into the system. For trajectories approaching the point $U_-$ with $x \rightarrow - 1$ and $y \rightarrow 0$, the density and pressure of the vector field, according to Eqs.~\eqref{Eq: Rescaled Density} and~\eqref{Eq: Rescaled Pressure}, can be approximated as:
\begin{equation}
\label{Eq: Approximated density}
 \frac{\rho_B}{\hat{g}^2 m_\text{P}^4} \approx 3 y^3, \quad \frac{p_B}{\hat{g}^2 m_\text{P}^4} \approx y^3,
\end{equation}
indicating that the system behaves like a radiation fluid with $w_B \approx 1/3$ around this point. Then, upon reaching $U_-$, $y$ becomes negative, causing the trajectory to escape from $U_-$ and move towards $U_+$ as $y$ approaches 0 from the negative side. Once again, when the trajectory reaches $U_+$, the trajectory is repelled as $y$ becomes positive and subsequently it is attracted towards $U_-$ again since $y \rightarrow 0^+$. The system oscillates between $U_-$ and $U_+$ with $H \in \mathbb{R}$, but $H$ becomes undetermined at each crossing.\footnote{It is worth clarifying that the approximation presented in Eq.~\eqref{Eq: Approximated density} is valid in the vicinity of $U_-$ but just before the trajectory originating outside the central zone reaches it. Inside the central region, i.e., after the above mentioned trajectory reaches $U_-$, the behaviour of the density and pressure  are of the form $\rho_B \propto 6y^4$ and $p_B \propto 2y^4$ respectively, the proportionality factor being the same for both quantities.  The system, then, neither exhibits negative energy density nor negative pressure.}

This dynamics continue until the saddle point nature of $U_-$ eventually forces the trajectory out of the central zone, escaping towards infinity in the phase space, as no attractors exist in that region. When the trajectory escapes from the central zone, $z^4$ flips sign and thus $H$ becomes complex (blue region). This transition from real to complex values in $H$ signals a flaw in the theory.

Notably, this shortcoming arises near the end of the primordial inflationary phase. Following the constant-roll condition $y = \beta_0 x$, the amount of inflation can be calculated as:
\begin{equation}
\label{Eq: Expected N_inf}
 N_\text{inf} \equiv \int \text{d} t \ H = \int_{y_i}^{y_f} \beta_0 \frac{\text{d}y}{y} \approx - \beta_0 \ln y_i,
\end{equation}
where the magnitude of $y$ at the end of inflation, denoted as $y_f$, is neglected in comparison with its initial value $y_i$. Numerically, this translates into:
\begin{equation}
 N_\text{inf} \approx 64.75.
\end{equation}

\begin{figure*}[h!]
\centering 
{\includegraphics[width=0.48\textwidth]{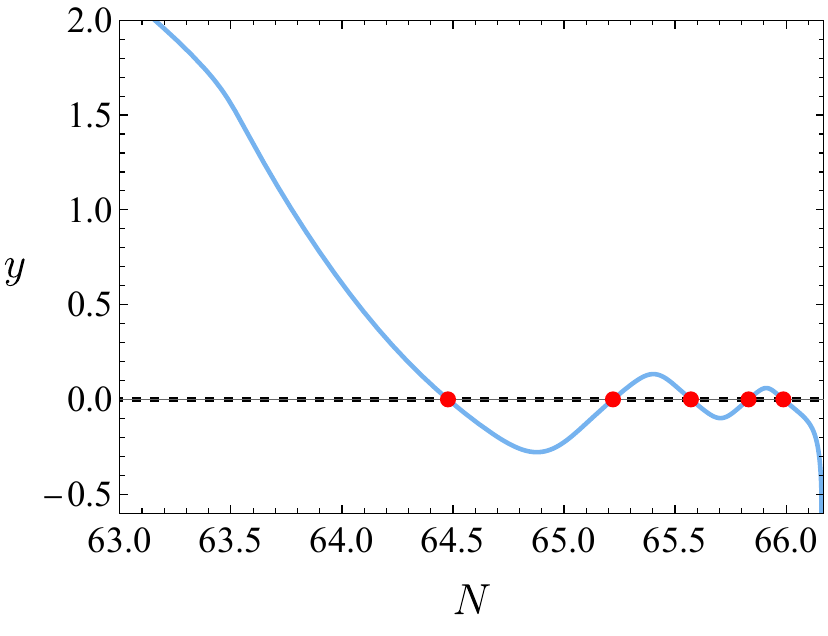}} \hfill
{\includegraphics[width=0.48\textwidth]{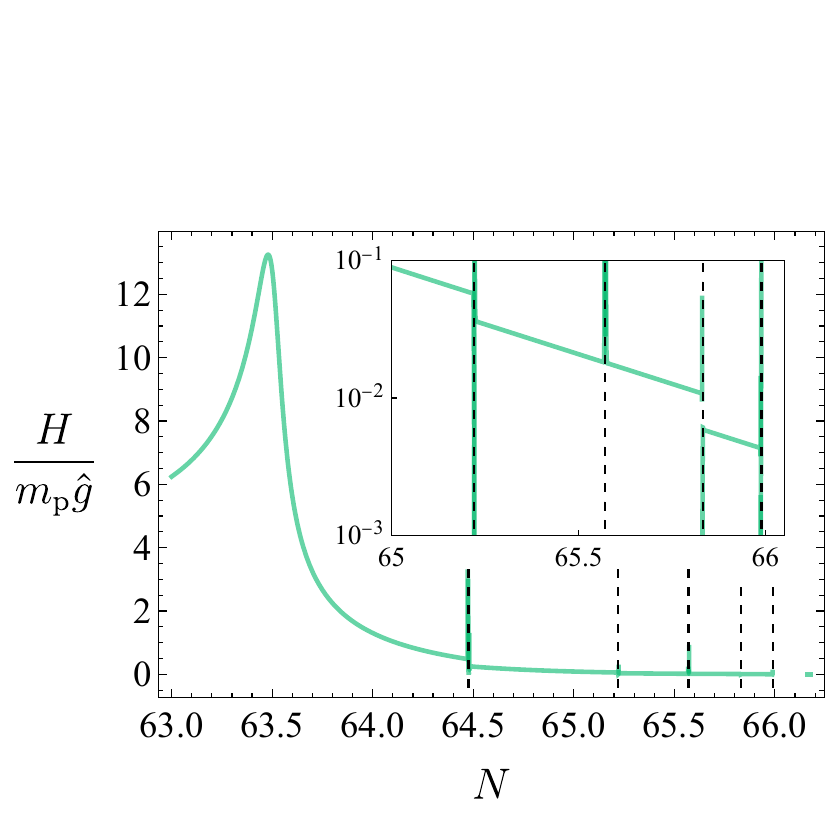}}
\caption{(Left) Time evolution of the variable $y$, with the red dots indicating the precise moments at which $y$ crosses cero, the first one occurring around $N = 64.5$. When $y$ crosses zero, $\hat{\rho}_B$ flips its sign as described by Eq.~\eqref{Eq: Approximated density}. (Right) Time evolution of the scaled Hubble parameter $H/(m_\text{p}\hat{g})$. At each crossing, $H$ exhibits undefined values, marked by the thin black dashed lines, corresponding to the physical singularities within the central zone.} 

\label{Fig: Instability}
\end{figure*}

For the initial conditions in Eq.~\eqref{Eq: ICs Figure 1(a)}, this issue is numerically confirmed in Figure~\ref{Fig: Instability}. The left panel shows the evolution of $y$ over time, with red dots marking the moments when $y$ crosses zero—corresponding to the system passing through the points $U_\pm$. The first crossing occurs approximately at $N = 64.5$. The right panel depicts the corresponding evolution of the rescaled expansion rate in Eq.~\eqref{Eq: Hubble in Variables}, which becomes undetermined precisely at these crossings.

\begin{figure}[h!]
\centering 
\includegraphics[width=0.6\textwidth]{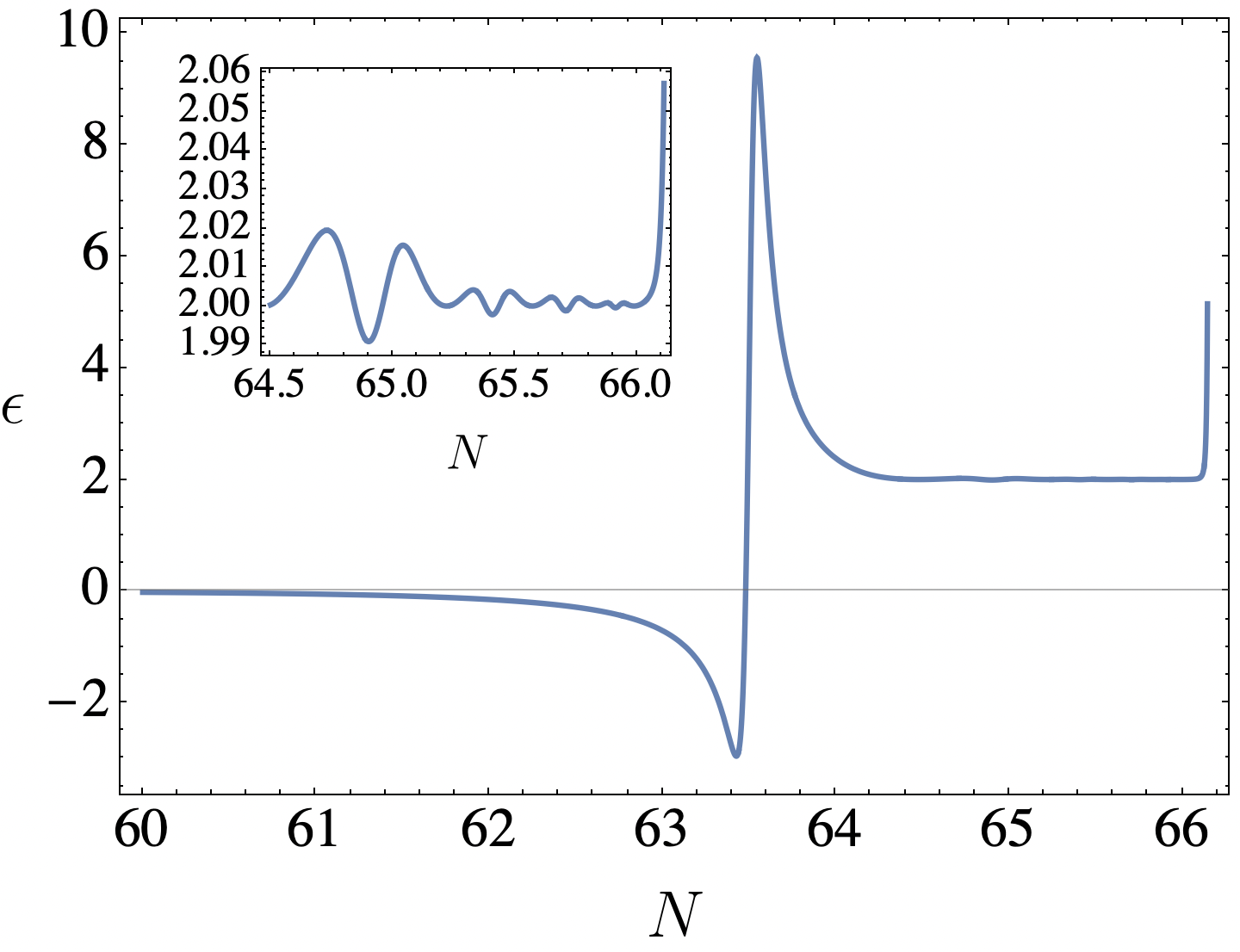}
\caption{Evolution of the parameter $\epsilon$. Two irregularities are evident in the evolution of $\epsilon$. The first occurs as the vector field's density flips sign upon entering the central zone. The second arises when the system exits the central zone, causing the field to grow uncontrollably. As highlighted in the inset, this uncontrolled growth ultimately results in a divergence of $\epsilon$.}
\label{Fig: Instability epsilon}
\end{figure}

The system exhibits another significant issue, stemming from the absence of a limit cycle in the central zone. As shown in Fig. \ref{Fig: Example Phase Space}, the trajectory oscillates between the points $U_\pm$ until, due to the saddle-like nature of $U_-$, eventually forces it to escape the central zone. Upon exiting this region, the lack of an attractor leads to an unbounded growth in the field magnitude and its velocity after approximately 66 $e$-folds, further highlighting the system's indeterminacy. 
This behaviour is directly reflected in the parameter $\epsilon$, as illustrated in Fig.~\ref{Fig: Instability epsilon}. Initially, $\epsilon \approx 0$ indicates that the system is in a constant-roll phase. Upon entering the central zone, $\epsilon \approx 2$, which reflects the system's behaviour as a radiation fluid. A natural interpretation—which is biased by the indeterminations in $H$ shown in the right-hand side of Fig.~\ref{Fig: Instability}—is to associate the transition from the accelerated phase to the decelerated one (corresponding to the first spike in $\epsilon$ in Fig.~\ref{Fig: Instability epsilon}) to some singularity in $H$. However, as shown by the behaviour of $H$ (right-hand side of Fig.~\ref{Fig: Instability}), the Hubble parameter remains well behaved during this transition, which begins at $N=63.5$ and ends near $N=64.5$. 

Furthermore, choosing $\epsilon$ as the main observable might conceal physical singularities because it behaves smoothly during the transition from the accelerated phase to the decelerated one. Afterwards, $\epsilon$ exhibits the typical oscillatory behaviour present in all viable models of inflation, without any of the singularities observed in $H$ in Fig.~\ref{Fig: Instability}. This behaviour suggests a graceful exit from the inflationary phase into the radiation-dominated era. However, this interpretation is delicate, as $H$ becomes indeterminate once the system reaches the $U_\pm$ points.  At this stage, the system may either escape the central zone or remain within it, depending on the tuning of the initial conditions and the integration interval. If the system exits the central zone, it experiences uncontrolled growth in $\epsilon$, ultimately leading to a divergence, as illustrated at the end of the inset plot of Fig.~\ref{Fig: Instability epsilon}.

In summary, trajectories that begin in the large-value regime of the variables $x$ and $y$ initially follow the attractor line $y = \beta_0 x$ until they approach the central zone. Upon entering this zone, the trajectories start oscillating between the pseudo-stationary states $U_\pm$, with the field behaving like a radiation fluid. However, during each crossing, the expansion rate becomes undefined, revealing a significant flaw in the model. After a limited number of oscillations, the trajectories inevitably exit the central zone, resulting in uncontrolled growth and divergence in the system. Moreover, as the trajectories escape, $z^4$ reverses sign, causing $H$ to acquire complex values.

\subsubsection{\textbf{Issues for the Dark Energy Scenario}}


\begin{figure*}[t!]
\centering 
{\includegraphics[width=0.43\textwidth]{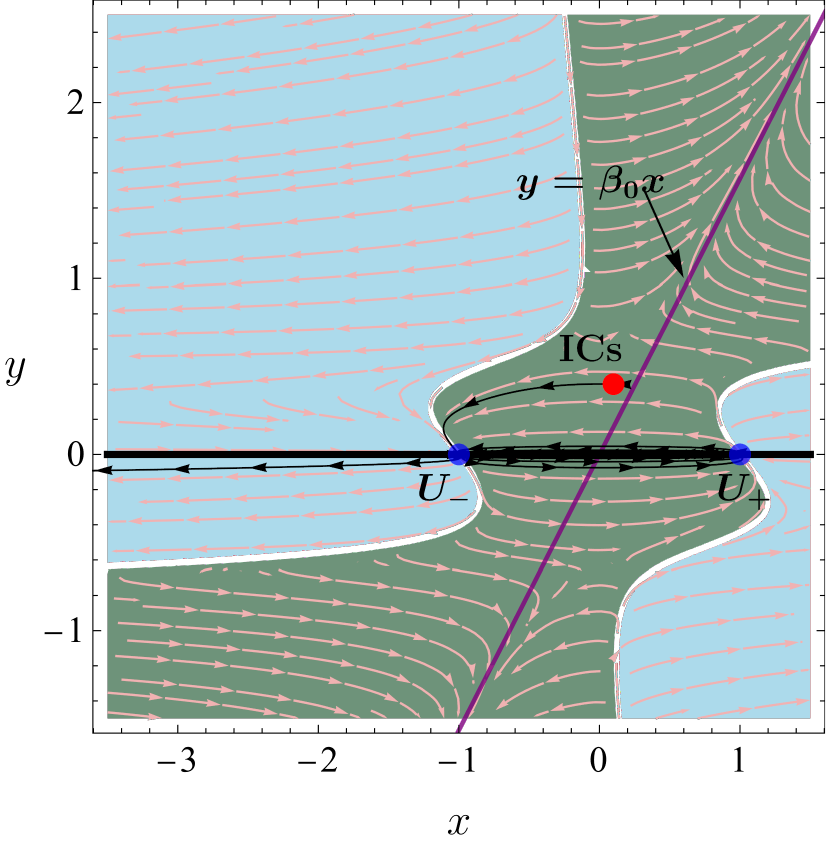}} \hfill
{\includegraphics[width=0.53\textwidth]{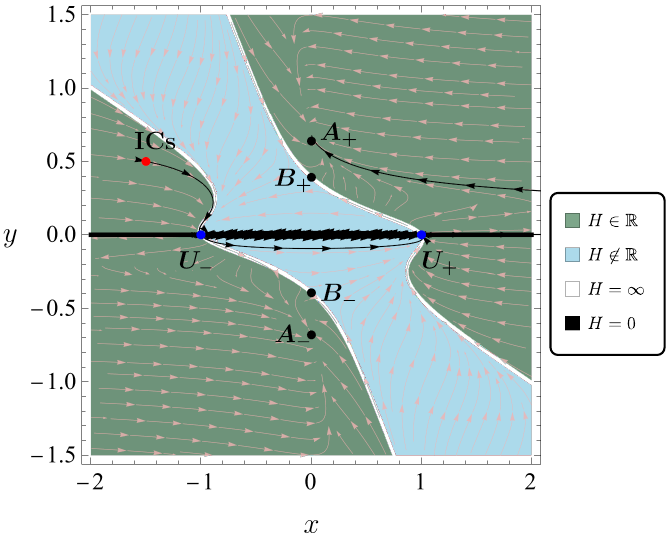}}
\caption{(Left) Evolution of a trajectory in phase space (line with arrows) starting within the central zone. The parameters are $c_1 = -1$ and $c_2 = 0.1$, with initial conditions are $x_i = 0.1$ and $y_i = 0.4$ (red dot). After a few oscillations, the system escapes the central zone and diverges, as the trajectory fails to align with the attractor line (purple line) with a positive slope. (Right) Phase space evolution of a specific trajectory with initial conditions $x_i = -1.5$, $y_i = 0.5$ and parameters $c_1 = -0.784$ and $c_2 = -2.99$, ensuring $A_+$ is an attractor. The trajectory initially follows the curve $z = 0$, enters the central region, and oscillates between the points $U_\pm$. Eventually, it escapes the central region due to the saddle point instability of $U_+$ and reaches the attractor point $A_+$. Although the evolution starts with a real-valued $H$ (green region), it becomes complex within the central zone due to a flip sign of $z^4$ (light blue region). Finally, $H$ becomes real once more when the trajectory escapes from the central zone to approach the attractor $A_+$. Moreover, at each crossing through the points $U_\pm$, $H$ becomes undetermined. This behavior of the system forbids the transition from a matter dominated epoch to a dark energy dominated epoch, and thus, the GSU2P theory is not suitable to describe the cosmic history of the universe.}
\label{Fig: Central Zone}
\end{figure*}

For the GSU2P theory to be a plausible candidate for describing the late-time accelerated expansion of the universe, the field must remain subdominant during earlier stages of cosmic history, as we got for the both examples in the Figures~\ref{Fig: Isotropic Evolution}  and ~\ref{Fig: Anisotropic Evolution}. Specifically, the domination of the vector field driving the accelerated expansion should be preceded by a matter-dominated epoch, ensuring that $\rho_B \ll \rho_m$ during that phase. This requires that trajectories in phase space originate within the central zone, where $x$ and $y$ are small. From there, the system must follow the straight line $y = \beta_0 x$ (with $\beta_0 > 0$) allowing the field to adhere to the constant-roll dynamics, or alternatively, approach the attractor points $A_\pm$.

In the case when there is no attractor point, when a trajectory escapes from the central zone, it diverges, as illustrated in the left panel of Figure~\ref{Fig: Central Zone}. This plot uses the parameters $c_1 = -1$, and $c_2 = 0.1$, such that $\beta_0 > 0$, with initial conditions set within the central zone: $x_i = -0.15$ and $y_i = 0.6$. The trajectory (line with arrows) starts from the initial conditions (marked by a red dot) fixing the expansion rate as real (green region), moves toward the pseudo-stationary state $U_-$, oscillates a few times between $U_\pm$ (blue points), where $H$ become undetermined at each crossing (intersection between the white lines and the black line), and then escapes from the central zone through $U_-$ to a region where there is no any attractor point, leading to divergence since the trajectory is not able to approach to the attractor line with positive slope (purple line). Moreover, after escaping the central zone, the expansion rate becomes complex (light blue region). Numerical integration of the autonomous set shows that these oscillations take around $1.24$ $e$-folds in total, after which the system becomes non-integrable due to the divergence.

From this numerical example, we conclude that although the field begins as a subdominant component of the cosmic budget, it rapidly becomes dominant after only a few $e$-folds.\footnote{As a reference, the time between photon decoupling and the present day corresponds to around 7 $e$-folds. Additionally, the radiation-dominated epoch is expected to last at least 15 $e$-folds~\cite{Alvarez:2019ue}.} Once the field transitions from its radiation-like behavior, the system diverges, making it impossible to achieve a viable cosmological scenario in which an early radiation-dominated epoch is followed by matter domination and eventually leads to an era dominated by the vector field driving the accelerated expansion of the universe.

In the case $A_\pm$ serve as attractors, the system does not diverge but other issues arise. For the right panel of Fig.~\ref{Fig: Central Zone}, we use the parameters $c_1 = -0.784$, and $c_2 = -2.99$, ensuring that $A_\pm$ are attractors. The initial conditions are chosen outside the central zone as $x_i = -1.5$ and $y_i = 0.4$ (red dot). The trajectory begins at these initial conditions, rapidly moves toward the pseudo-stationary state $U_-$, oscillates a few times between $U_\pm$ (blue points), and then escapes from the central zone through $U_+$, leading to the attractor $A_+$ (black point). However, the system exhibits a significant issue: while the trajectory starts with a real-valued expansion rate $H$ (green region) as it approaches $U_-$, $H$ becomes indeterminate (at the intersection of the black and white lines). Within the central zone, the Hubble parameter is no longer real. Finally, when the trajectory escapes from the central zone and reaches the attractor $A_+$, $H$ becomes real again. The entire process, from the initial conditions to reaching the attractor point $A_+$, spans approximately 4 $e$-folds, which is not long enough to cover the periods of radiation and matter domination. Therefore, although the attractor points $A_\pm$ represent viable cosmological solutions, the system cannot reach them from within the central zone, since $H$ is not real there. This implies that the system is not physically allowed to evolve within the central zone, which is crucial to ensure that dark energy remains subdominant before reaching its attractor, where it would drive perpetual cosmic acceleration. An alternative approach is either to choose initial conditions close to the attractor points $A_\pm$ or to adjust the parameters to avoid a divergence. However, this strategy poses significant challenges. If the initial conditions are set such that $y$ is close to $0$ and $x$ is large, the stability of the points $U_\pm$ remains independent of the parameters; this inevitably leads to uncontrolled growth of the field, causing it to dominate rapidly (within less than $3$ $e$-folds) over matter or radiation. Alternatively, selecting initial conditions after the uncontrolled growth has occurred allows the system to reach the attractor quickly without encountering a singularity; however, this requires fine-tuning the parameter $\hat{g}$ to slow the field's domination and allow matter and radiation to persist. In this case, the field rapidly stabilizes at the attractor, effectively behaving as a constant, which is indistinguishable from the $\Lambda$CDM model.

\subsection{Regularization of the Autonomous Set}

In the previous sections, we have analyzed the system's dynamics both outside the central zone, following the attractor line $y = \beta_0 x$ (where constant-roll occurs), and within the central zone when initial conditions are close to zero. We have demonstrated that the system is undetermined in both cases, as it eventually escapes the central zone after oscillating between the saddle-like pseudo-stationary states $U_\pm$ for a few $e$-folds. However, due to the non-compact nature of the phase space, it is also possible to encounter singularities caused by the presence of incomplete nullcline curves, i.e., points where a given dynamical equation becomes undetermined. In what follows, we will investigate the existence of such singularities in phase space.

As noted in Ref.~\cite{Garnica:2021fuu}, the system might not enter the central zone when approaching from the attractor line if the parameters are not properly chosen, leading to the vanishing of the denominator in the dynamical equation for the variable $x$ [see Eq.~\eqref{Eq: Denominator fx}]. The expression for this denominator, $D_{x'}$, can be written as a seventh-degree polynomial:
\begin{equation}
 D_{x'} = y(1 - \alpha y^2 + \gamma y^4 + \delta y^6),
 \label{Eq: Denominator}
\end{equation}
where
\begin{equation}
 \alpha \equiv 12c_2, \quad \gamma \equiv 8(c_1 - c_2), \quad \delta \equiv 2\gamma(c_1 - 7c_2).
\end{equation}
Since the expression inside the parentheses can be reduced to a cubic polynomial of $y^2$, we can find the seven roots of $D_{x'}$, which we present in Appendix~\ref{App: Regularization} to avoid overly long expressions here. One of these roots is $y = 0$. Notably, the system may still enter the central zone through the pseudo-stationary states $U_\pm$ at $x = \pm 1$ and $y = 0$.

This can be understood by noting that the dynamical equation for $x$ can be \textit{regularized}, where both the numerator and denominator of $x'$ vanish simultaneously. The numerator, $N_{x'}$, can be written as:
\begin{equation}
 N_{x'} = \sum_{i=0}^{3} f_i(c_1, c_2, y) x^i,\label{eq: Numerator}
\end{equation}
where $f_i(c_1, c_2, y)$ are polynomials in $y$, which we present in Appendix~\ref{App: Regularization}. Once we determine $y$ from the roots of $D_{x'}$, the expression for $N_{x'}$ becomes a third-degree polynomial in $x$, whose roots determine the location in phase space of singularities, i.e., points where the system is undetermined. For example, when $y = 0$, we find $N_{x'} = 2(x^2 - 1)$, whose roots are $x = \pm 1$. The analysis of other roots is more complex, so we focus on an illustrative case with specific initial conditions and parameters.

In Figure~5(b) of Ref.~\cite{Garnica:2021fuu}, it is demonstrated that a given trajectory in phase space is unable to enter the central zone as it reaches a region where the variable $z^4$ becomes negative. As we have shown, problems in the dynamical description arises not for $z^4$ to be negative, but for its flip of sign during the evolution. In this case, what happens is that the system find a regularization point, i.e., a singularity. In this analysis, we determine the precise point where this indeterminacy arises. Following the setup from Figure~5(b) in Ref.~\cite{Garnica:2021fuu}, we choose the initial conditions:
\begin{equation}
 x_i = -4 \times 10^9, \quad y_i = 10^{10},
\end{equation}
with parameters:
\begin{equation}
 \alpha_1 = 1, \quad \alpha_3 = 1.1, \quad \chi_5 = 0,
\end{equation}
which correspond to $c_1 = 0.2$ and $c_2 = 2.2$. Using Eq.~\eqref{Eq: Expected N_inf}, the inflationary phase is expected to last approximately 560 e-folds.

Solving for the roots of $D_{x'}$ gives one trivial solution $y = 0$ and six non-trivial solutions. After evaluating them for $c_1$ and $c_2$, we find two complex roots and four real roots: $y = \pm 0.479592$ and $y = \pm 0.194975$. Singularities occur at $y = \pm 0.479592$, the largest values of $|y|$, which will be firstly met when the trajectory comes from the attractor line. Solving for the roots of $N_{x'}$ at these values gives $x = \mp 0.653576$, yielding two singularities at the points:
\begin{align}
 S_+ = \{-0.653576, 0.479592\}, \quad
 S_- = \{0.653576, -0.479592\}.
\end{align}

\begin{figure}[h!]
\centering 
\includegraphics[width=0.45\textwidth]{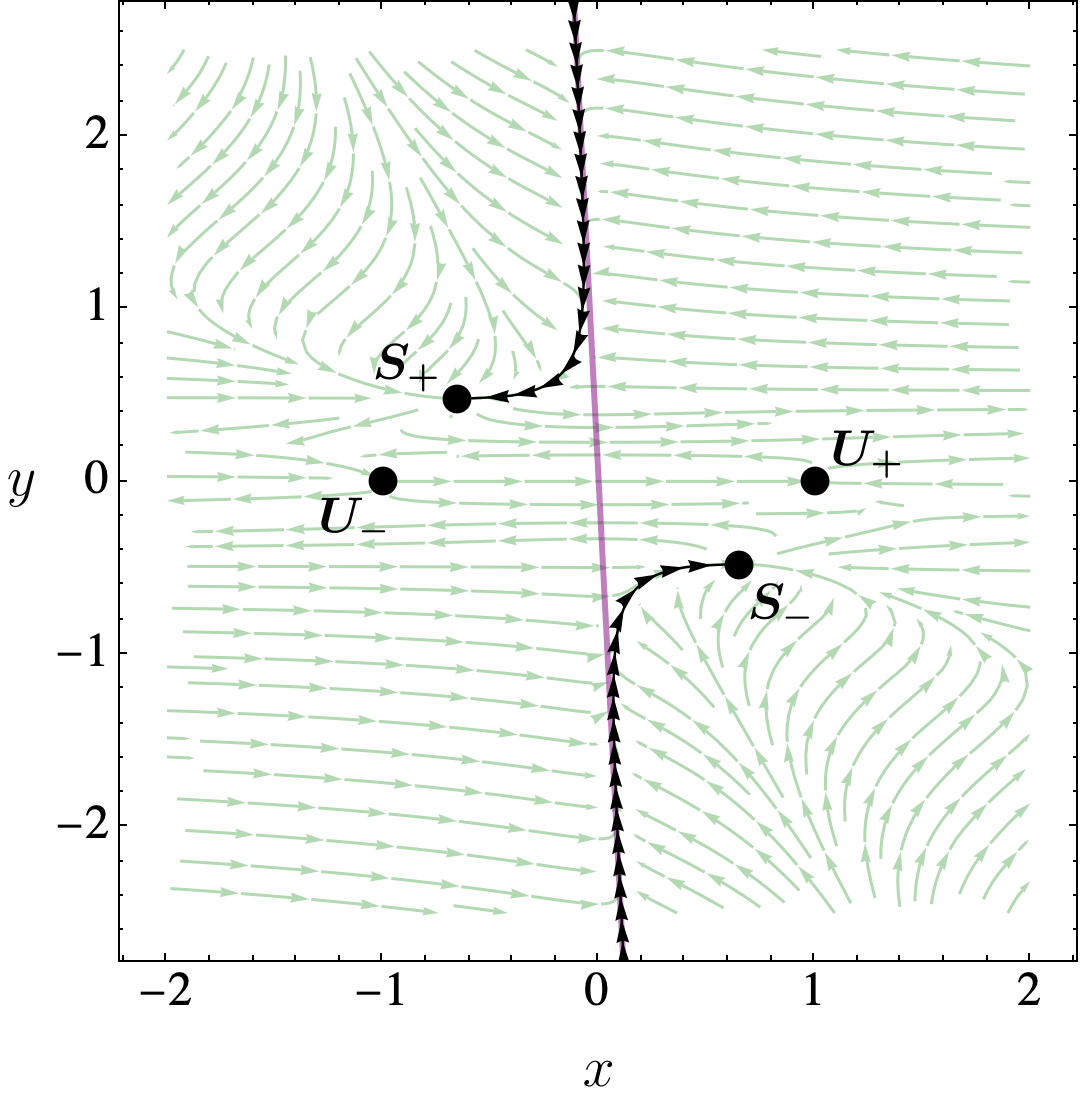}
\caption{Evolution of two trajectories in phase space with initial conditions $x_i = \pm 4 \times 10^9$ and $y_i = \mp 10^{10}$, using parameters $c_1 = 0.2$ and $c_2 = 2.2$. The trajectories initially follow the attractor line $y = \beta_0 x$, but eventually approach the singularity points $S_\pm$, determined by the regularization of the dynamical equation for $x$. After approximately 553.4 e-folds, the trajectories reach these singularities, at which point the system becomes non-integrable, causing the dynamics to cease.}
\label{Fig: No Central Zone}
\end{figure}

These results are confirmed numerically in Figure~\ref{Fig: No Central Zone}, which shows two trajectories approaching the singularity points $S_\pm$ from the attractor line $y = \beta_0 x$. After approximately 553.4 e-folds, the system diverges at these singularities, preventing access to the central zone spanned by $U_\pm$. Thus, the central zone can only be accessed if the initial conditions start within it.

In summary, further singularities may arise in the system due to the existence of regularization points, where the dynamical equation for $x$ becomes indeterminate, as both its denominator $D_{x'}$ and numerator $N_{x'}$ vanish simultaneously. A natural question then arises: if $U_\pm$ and $S_\pm$ are singularities resulting from the regularization of the autonomous system, why can one case be numerically integrated while the other cannot? The answer lies in the numerical evaluation of $x'$ during the integration process. Numerical solutions are computed in discrete steps, and singularities can sometimes be ``skipped'' if a sufficiently small neighborhood around the pseudo-fixed points is well-approximated. This enables the system to progress through regions near singularities, making numerical integration feasible, while preventing the trajectory from fully reaching the singular points.
\begin{figure*}[h!]
\centering 
{\includegraphics[width=0.46\textwidth]{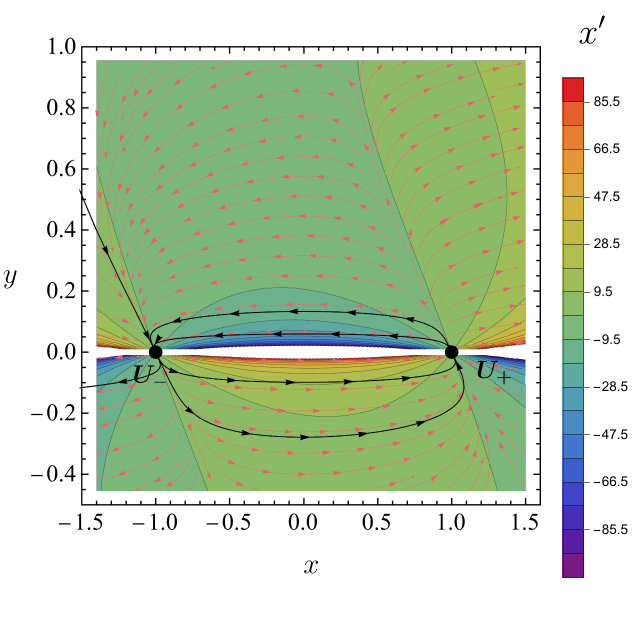}} \hfill
{\includegraphics[width=0.46\textwidth]{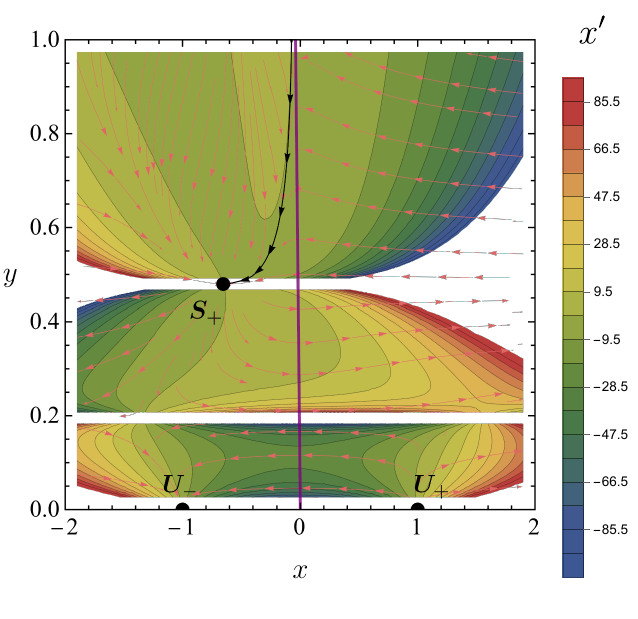}}
\caption{(Left) Evolution of a phase space trajectory using the same parameters and initial conditions as in Figure~\ref{Fig: Example Phase Space}. After a few oscillations, the trajectory exits the central zone, reaching the white zone and finally diverging. However, the regularization points $U_\pm$ can be bypassed during the integration process since the white region does not fully enclosed these points. (Right) Evolution of a phase space trajectory with the same parameters and initial conditions as in Figure~\ref{Fig: No Central Zone}. In this case, the regularization point $S_+$ is surrounded by a larger white region (where $x'$ is large), rendering numerical integration infeasible.}
\label{Fig: Numerical singularity}
\end{figure*}

In the left panel of Figure~\ref{Fig: Numerical singularity}, the phase space trajectory (using the same initial conditions and parameters as in Figure~\ref{Fig: Example Phase Space}) follows a well-behaved evolution. It enters the central zone through $U_-$ and then escapes. The color bar shows that as the trajectory approaches $U_-$ from positive $y$, $x'$ grows to large values, which could potentially cause numerical indeterminacy, marked by the white region.~\footnote{The large numerical values shown on the plot correspond to the white regions where $x^\prime > 100$, reaching maximum values of $x^\prime \sim 10^{60}$. These values are too large to be handled by our numerical integration method.} However, this white region does not fully enclose the point, allowing some trajectories to bypass the singularity.

Conversely, the right panel of Figure~\ref{Fig: Numerical singularity} shows the phase space trajectory (using the same initial conditions and parameters as in Figure~\ref{Fig: No Central Zone}). In this case, a wider white region around the regularization point $S_+$ prevents continuous integration of the system, leading to an indeterminacy.

\section{Conclusions} 
\label{Sec: Conclusions}

The Generalized SU(2) Proca (GSU2P) theory, widely studied for its applications in phenomena like inflation~\cite{Garnica:2021fuu}, late-time acceleration~\cite{Rodriguez:2017wkg}, and compact objects~\cite{Martinez:2022wsy, Gomez:2023wei, Martinez:2024gsj}, has not yet to provide a comprehensive cosmic history. To address this, we investigated the GSU2P theory within a flat FLRW spacetime using a cosmic triad configuration to assess its capacity to drive cosmic acceleration. Unlike the isotropic and anisotropic tachyon field scenarios discussed earlier, the dynamical system approach here revealed significant instabilities and inconsistent behaviors on the rate of expansion, ultimately challenging the GSU2P theory’s viability as a mechanism for cosmic acceleration.

Through a dynamical system approach, we have demonstrated that the fixed points of the system fail to yield viable cosmological scenarios. Specifically, the Hubble parameter was found to be either complex or undetermined at these points, precluding the existence of any stable accelerated attractors within the model. However, the chosen dynamical variables form a non-compact phase space, and additional stationary states could theoretically exist. We identified two pseudo-stationary states: the first corresponds to three straight lines that dominate the system’s behavior in the large $x$ and $y$ regime, where $x$ represents the velocity of the field and $y$ its magnitude. The second involves two points, $U_\pm$, which are solutions to the dynamical equation for $x$ when $y$ approaches zero, independent of how the parameters are chosen.

In the large $x$ and $y$ regime, we found that one of the attractor lines aligns with a de-Sitter-like expansion, consistent with constant-roll dynamics. In contrast, the pseudo-stationary points $U_\pm$ behave as saddles, creating what we refer to as the ``central zone''. When trajectories enter this zone, oscillations between the pseudo-stationary states $U_\pm$ occur, resembling radiation-like behavior of the field. However, these oscillations lead to instability, which undermines the Hubble parameter at each crossing. After several oscillations, the system escapes the central zone, experiencing uncontrolled growth in the field magnitude, ultimately causing a divergence in all observables. This instability renders the GSU2P model unviable for both inflation and late-time cosmic acceleration, as explained further below.

Regarding inflation, our findings suggest that the instability prevents a smooth transition from the inflationary phase to a radiation-dominated epoch, which would otherwise follow the attractor line $y = \beta_0 x$. The instability emerges each time the system passes through the singularities $U_\pm$, preventing the graceful exit from inflation. Furthermore, the lack of limit cycles within the central zone exacerbates the instability, as trajectories quickly escape this zone, disrupting the onset of the radiation-dominated era.

For late-time cosmic acceleration, the singularity points $U_\pm$ undetermined the Hubble parameter, regardless of the existence of attractors like $A_\pm$. The saddle-like behavior of $U_\pm$ prevents the system from maintaining periodic oscillations, pushing the trajectories out of the central zone. This results in the Hubble parameter becoming undetermined, transitioning between real and complex values, and disrupting the evolution. Once the system escapes the central zone, the lack of stable attractor points causes the trajectories to diverge, leading to rapid domination of the vector field in the cosmic budget. This behavior precludes the model from offering a viable cosmological history, as it fails to allow a smooth transition from a matter-dominated epoch to a phase of accelerated expansion driven by the vector field.

In addition to the physical instabilities previously discussed, we have also identified potential singularities in the phase space resulting from the regularization of the dynamical equation for $x$. These singularities arise when both the numerator and the denominator of the equation simultaneously vanish. Under certain conditions, such singularities prevent the system from entering the central zone, depending on the behavior of the numerical method during integration. Our simulations showed that trajectories following the attractor line can reach these singularities after a finite time, at which point the system becomes non-integrable, halting its evolution. These findings suggest that incomplete nullcline curves in the phase space contribute additional instabilities to the system’s dynamics, as illustrated in Figure~\ref{Fig: Numerical singularity}.

The distinction between numerically integrable singularities and those that cannot be integrated lies in the numerical evaluation of $x'$. Numerical solutions are computed in discrete steps, and singularities may be bypassed if a sufficiently small neighborhood around the pseudo-fixed points is well-approximated. This allows the system to progress through regions near singularities, making numerical integration possible, while preventing the trajectory from reaching the singular points entirely. For instance, in the left panel of Figure~\ref{Fig: Numerical singularity}, the trajectory bypasses the regularization points $U_\pm$ as it enters the central zone and then escapes, despite the large values of $x'$, which would normally cause numerical indeterminacy. In contrast, the right panel shows a trajectory that encounters a larger white region around the singularity $S_+$, making numerical integration infeasible due to the indeterminacy caused by large values of $x'$.

In conclusion, the GSU2P theory faces significant obstacles in providing a cosmologically viable explanation for both the exit of primordial inflation and late-time cosmic acceleration. While the theory can reproduce a constant-roll phase, the numerous numerical instabilities—the absence of limit cycles, and the occurrence of singularities—render the system unable to maintain a viable cosmological evolution across cosmic timescales. 

Are these results sufficient to conclusively rule out the theory? While not definitive, we can confidently state that choosing free parameters to ensure the theory behaves perturbatively (to second order in the action) like GR, which represents the simplest realization, is cosmologically unviable. Other trivial parameter choices, though largely unexplored due to their mathematical complexity, could either mitigate or exacerbate the instabilities already observed, highlighting the need for further investigation. These results point to the need for further refinement of the theory to reconcile it with the known expansion history of the universe.

        \chapter*{Summary\markboth{Summary}{Summary}}
\label{sec:Conclusions}
\addcontentsline{toc}{chapter}{Summary}
        
        In this thesis, we employed the dynamical systems approach to explore two different cosmological models, each addressing distinct challenges in understanding the expansion history of the Universe. In Chapter \ref{Ch: anisotrpic Tachyon field}, we developed a numerical framework for analyzing anisotropic dark energy models, specifically focusing on the interaction between a scalar tachyon field and a vector field in a Bianchi I background. This method, which does not rely on analytical fixed points, enabled the identification of regions within the parameter space where accelerated solutions could act as attractors of the system. The flexibility of this approach makes it applicable to any DE scenario, thereby demonstrating its general applicability.

In Chapter \ref{Ch: GSU2P}, we investigated the Generalized SU(2) Proca (GSU2P) theory, with particular focus on its potential to drive both primordial inflation and late-time cosmic acceleration. While the dynamical systems approach revealed regions where accelerated solutions could theoretically exist, it ultimately helped us rule out the GSU2P model as a viable candidate for a complete cosmological acceleration history. The model’s failure to produce stable attractors and maintain consistent cosmological evolution across different epochs prevented a smooth transition from inflation to a radiation-dominated phase, as well as from a matter-dominated phase to accelerated expansion. Thus, the dynamical systems approach served to discard this model as a viable candidate for cosmological history, highlighting its limitations in replicating the observed evolution of the Universe. In particular, the presence of singularities in phase space, such as those resulting from the regularization of dynamical equations, can disrupt the integrability and evolution of cosmological models. Depending on the behavior of the system during numerical integration, such singularities can halt its progression or introduce additional instabilities, ultimately limiting the model's ability to replicate a consistent and observationally accurate cosmological history. These challenges underscore the need for rigorous examination and refinement of models to ensure their robustness across different epochs of cosmic evolution.

Moreover, we note that although the GSU2P theory fails to produce a graceful exit from inflation, it could still serve as an effective theory for specific cases where the exit is not required. In such cases, the model satisfies all necessary constraints during the preceding evolution, as demonstrated in previous work \cite{Garnica:2021fuu}.

%
	\appendices
\chapter{Numerical Exploration of Dynamical Systems: Anisotropic Tachyon Field in Cosmology}
\section{(\emph{DE-I}) Eigenvalues}
\label{App: DEI Eigenvalues}

The eigenvalues for the isotropic DE, section \ref{Section: AFP} are given by
\begin{align}
 \lambda_1 &= \frac{1}{12} \left(-\sqrt{2} \sqrt{\alpha ^4+36} \Gamma +\sqrt{2} \alpha ^2 \Gamma -12\right)\\
 \lambda_2 &= \frac{1}{24} \left(-\sqrt{2} \sqrt{\alpha ^4+36} \Gamma +\sqrt{2} \alpha ^2 \Gamma -36\right)\\
 \lambda_3 &= \frac{1}{48}\left(-36+3 \sqrt{2} \alpha^2 \Gamma-3 \sqrt{2} \sqrt{36+\alpha^4} \Gamma\right. \\
&\notag -2\left(66 \alpha^8+\alpha^6\left(-66 \sqrt{36+\alpha^4}+32 \sqrt{2} \Gamma\right)+\alpha^2\left(-900 \sqrt{36+\alpha^4}+738 \sqrt{2} \Gamma\right)\right. \\
& \notag\left.\left.-162\left(-36+\sqrt{2} \sqrt{36+\alpha^4} \Gamma\right)-8 \alpha^4\left(-261+4 \sqrt{2} \sqrt{36+\alpha^4} \Gamma\right)\right)^{\frac{1}{2}}\right)\\
 \lambda_4 &= \frac{1}{48}\left(-36+3 \sqrt{2} \alpha^2 \Gamma-3 \sqrt{2} \sqrt{36+\alpha^4} \Gamma\right. \\
&\notag +2\left(66 \alpha^8+\alpha^6\left(-66 \sqrt{36+\alpha^4}+32 \sqrt{2} \Gamma\right)+\alpha^2\left(-900 \sqrt{36+\alpha^4}+738 \sqrt{2} \Gamma\right)\right. \\
& \notag \left.\left.-162\left(-36+\sqrt{2} \sqrt{36+\alpha^4} \Gamma\right)-8 \alpha^4\left(-261+4 \sqrt{2} \sqrt{36+\alpha^4} \Gamma\right)\right)^{\frac{1}{2}}\right)\\
 \lambda_5 &= \frac{1}{24} \left(2 \sqrt{\alpha ^4+36} \alpha  \beta -\sqrt{2} \sqrt{\alpha ^4+36} \Gamma -2 \alpha ^3 \beta +\sqrt{2} \alpha ^2 \Gamma -12\right)
\end{align}

where $\Gamma\equiv\sqrt{\alpha ^4-\sqrt{\alpha ^4+36} \alpha ^2+18}$, $\lambda_1, \ldots, \lambda_4$ are all negative if $\alpha$ is a real number, and $\lambda_5$ is negative when Eq. \eqref{Eq: DE-I Attractor 1} or Eq. \eqref{Eq: DE-I Attractor 2} are satisfied.

\chapter{Challenges to Cosmic Acceleration in Generalized SU(2) Proca Dynamics}

\section{Dynamical Analysis of the Pseudo-Stationary Straight Lines.}
\label{App: Straight Lines}

In Section \ref{SEC: TSSL}, we have introduced the existence of additional straight lines with slopes \( \beta_\pm \) [Eq.~\eqref{Eq: beta+-}] that influence cosmological dynamics. Unlike the line with slope \( \beta_0 \), which exists independently of \( c_1 \) and \( c_2 \) values, the lines with slopes \( \beta_\pm \) only arise if \( (c_1 - c_2)(c_1 - 7c_2) \geq 0 \).

Each line can dominate the dynamics when it has attractor stability. Reference~\cite{Garnica:2021fuu} shows that small perturbations around any point $x_s$ on these lines maintain the attractor condition if
\begin{equation}
    A_{\beta} = \frac{x_s}{x^{\prime}(x_s)} \frac{\partial x^{\prime}}{\partial x} \Big|_{x_s} > 1,
\end{equation}
where the attractor conditions for each line are found as:
\begin{align}
    A_{\beta_0} &= -6 + \frac{7 c_2}{c_1},\\
    A_{\beta_{\pm}} &= 2 + 2 \frac{c_1}{c_2} \pm 2 \frac{ \sqrt{(c_1 - 7 c_2)(c_1 - c_2)}}{c_2}.
\end{align}

Thus, the line with slope \( \beta_0 \) is an attractor when \( |c_1| < |c_2| \). Similarly, with \( (c_1 - c_2)(c_1 - 7c_2) \geq 0 \), the line with slope \( \beta_{-} \) becomes an attractor if
\begin{equation}
    c_2 < 0 \lor \left\{c_2 > 0 \land \left(\frac{3 c_2}{4} < c_1 \leq c_2 \lor c_1 \geq 7 c_2\right)\right\},
\end{equation}
while the line with slope \( \beta_+ \) is an attractor under
\begin{equation}
    c_2 > 0 \lor \left\{c_2 < 0 \land \left(c_2 \leq c_1 < \frac{3 c_2}{4} \lor c_1 \leq 7 c_2\right)\right\}.
\end{equation}
For large values of \( x \) and \( y \), the system naturally evolves along these lines. However, the lines with slopes \( \beta_\pm \) do not inherently generate accelerated expansion; specific parameter adjustments are required to satisfy \( w_B < -1/3 \) [Eq.~\eqref{Eq: Equation of state beta}]. When all lines are present, stability analysis shows that each line can potentially act as an attractor depending on the values $c_1$ and $c_2$. Nonetheless, the dynamics in the central zone exhibit the same physical shortcomings as those previously identified for the line with slope \( \beta_0 \).

\section{Long Expressions From Regularization}
\label{App: Regularization}

Previously, in the Eq.~\eqref{Eq: Denominator}, we have described the denominator of $x^\prime$ as a seventh-degree polynomial in $y$. We present the roots of this polynomial below:
\begin{align}
    y_0&=0,\\
    y_1^2&=-\frac{\gamma}{3\delta}+\frac{\sqrt[3]{2} \Delta }{3 \delta  \sqrt[3]{\sqrt{\Xi ^2-4 \Delta ^3}-\Xi }}+\frac{\sqrt[3]{\sqrt{\Xi ^2-4 \Delta ^3}-\Xi }}{3 \sqrt[3]{2} \delta },\\
    y_2^2&=-\frac{\gamma }{3 \delta }-\frac{\left(1-i \sqrt{3}\right) \Delta }{3\ 2^{2/3} \delta  \sqrt[3]{\sqrt{\Xi ^2-4 \Delta ^3}-\Xi }}-\frac{\left(1+i \sqrt{3}\right) \sqrt[3]{\sqrt{\Xi ^2-4 \Delta ^3}-\Xi }}{6 \sqrt[3]{2} \delta },\\
    y_3^2&=-\frac{\gamma }{3 \delta }-\frac{\sqrt[3]{-2} \Delta }{3 \delta  \sqrt[3]{3 \sqrt{3} \sqrt{\zeta}-\Xi }}-\frac{\left(1-i \sqrt{3}\right) \sqrt[3]{\sqrt{\Xi ^2-4 \Delta ^3}-\Xi }}{6 \sqrt[3]{2} \delta },
\end{align}
where 
\begin{align}
    \Delta&\equiv 3 \alpha  \delta +\gamma ^2,\\
    \Xi&\equiv 9 \alpha  \gamma  \delta +2 \gamma ^3+27 \delta ^2,\\
    \zeta&\equiv\delta ^2 \Big[\left(2 \gamma -\alpha ^2\right) (\alpha  \delta +\Delta )+\alpha  \gamma  \delta +\Xi \Big].
\end{align}
Additionally, the numerator of Eq.~\eqref{eq: Numerator} was described as third degree polynomial on the $x$ variable where the functions $f_i\equiv f_i(c_1,c_2,  y)$ are given by:
\begin{align}
    f_0&=-2 \left(6 c_1 y^4+1\right) \left(12 c_2 y^4-y^2+1\right),\\
    f_1&=y \Bigg[3+12 y^2 \Big\{4 y^4 \left(2 c_1^2-4 c_1 c_2-7 c_2^2\right)+y^2 (c_1+2 c_2)+2 c_1-5 c_2\Big\}\Bigg],\\
    f_2&=-2 \Bigg[8 y^6 \left(7 c_1^2-29 c_1 c_2+49 c_2^2\right)+6 y^4 (c_1-4 c_2)+6 c_2 y^2-1\Bigg],\\
    f_3&=4 y^3 \Big[12 c_2 y^2 (4 c_1-7 c_2)-5 c_1+8 c_2\Big].
\end{align}

%
	
%
\bibliography{susy.bib}
%


	\definecdlabeloffsets{0}{-0.65}{0}{0.55} 

	\createcdlabel{This is my MSc Thesis \\ Cosmology on the Generalized Proca Theory}{Santiago García Serna}{January}{2025}{1} 


	\createcdcover{This is my MSc Thesis \\ Cosmology on the Generalized Proca Theory}{Santiago García Serna}{January}{2025}{1} 

\end{document}